\documentclass[12pt]{report}

\usepackage{graphicx}
\usepackage{amsmath,amssymb}
\usepackage{bm}
\usepackage{siunitx}
\usepackage{enumitem}
\usepackage{hyperref}
\usepackage[margin=2.5cm]{geometry}

\hypersetup{colorlinks=true, linkcolor=blue, filecolor=blue, urlcolor=blue, citecolor=blue}

\newcommand{\bvec}[1]{{\bf #1}}
\newcommand{\bvecg}[1]{\bm{#1}}
\newcommand{\uvec}[1]{{\bf \hat{#1}}}
\newcommand{\uvecg}[1]{\bm{\hat{#1}}}
\newcommand{\norm}[1]{\tilde{#1}}

\begin{document}
\pagenumbering{roman}

\title{Plasma dynamics near the magnetic X-point\\ of the two-wire model: Theory and Simulation}

\author{Bin Ahn\\
Department of Nuclear and Quantum Engineering\\
Korea Advanced Institute of Science and Technology\\
Daejeon, South Korea\\
}

\maketitle

\renewcommand{\thefootnote}{}
\footnotetext{This paper is a revised version of the doctoral dissertation by Bin Ahn.}
\renewcommand{\thefootnote}{\arabic{footnote}}

\begin{abstract}
    The two-wire model (TWM) is a magnetic configuration generated by two parallel current-carrying wires, and it contains an X-point at its center and a separatrix.
    Since the TWM magnetic field is described by a closed-form analytic expression and contains no guide field, it offers a tractable setting for studying how a true magnetic null shapes plasma dynamics.
    This work investigates two complementary regimes: collisionless charged particle dynamics and collisional low temperature plasma transport.
    In the collisionless regime, a Lagrangian analysis identifies two particle motion invariants: the total kinetic energy and the base field line value, which is derived from the conserved axial canonical momentum.
    Collisionless test particle simulations show that the magnetic moment undergoes shifts when the particle traverses the large gradient region near the null.
    These shifts enable particles to migrate, the phenomenon in which a particle gyrating about one branch of a base field line jumps to the corresponding branch on the other side of the X-point.
    A threshold energy for migration is derived, and an empirical expression for the migration confinement time is formulated.
    In the collisional regime, reduced drift-diffusion models for low temperature plasmas are developed in a conformal field-aligned coordinate system, and they predict density plateau formation near the separatrix in the strongly magnetized regime.
    Self-consistent particle-in-cell simulations are performed to verify the predicted density plateau.
    The two complementary studies establish a fundamental understanding of plasma dynamics near a true magnetic null for both the collisionless and collisional regimes.
\end{abstract} 

\tableofcontents
\clearpage
\pagenumbering{arabic}


\chapter{Introduction}\label{chap.1}

Magnetic configurations containing X-points and separatrices appear across a wide range of plasma systems.
They are found in tokamak divertors,\cite{Umansky2016,Reiser2017,Wang2020,Eich2021} stellarator edge regions,\cite{Boozer2018,Pedersen2022} field-reversed configurations,\cite{Steinhauer2011} magnetic reconnection sites in astrophysical and laboratory plasmas,\cite{Ji2022,Yamada2014} and multipolar magnetic configurations used in low temperature plasma sources.\cite{Hagelaar2007,Lim2020}
An X-point is a location at which the in-plane field magnitude vanishes, and it lies on a separatrix that divides the domain into distinct magnetic regions.
Because magnetized plasma confinement depends on the ability of the magnetic field to constrain charged particle motion, the local reduction of the field magnitude near an X-point can weaken confinement and alter the local plasma dynamics.
Understanding how the presence of an X-point shapes plasma dynamics, from single particle orbits to macroscopic transport and density profiles, is therefore relevant to various plasma systems.

Most existing studies of X-point effects on plasma profiles, however, come from settings in which the role of the X-point is entangled with many other effects.
In tokamak edges, simulations and experiments report density shoulders and steepened or flattened gradients near the separatrix, with a notable example being the density shelf observed in DIII-D, attributed partially to $\mathbf{E}\times\mathbf{B}$ drift dynamics near the X-point.\cite{Wang2020, Eich2021}
However, this X-point is embedded in a strong toroidal magnetic field together with curvature, turbulence, and sheath effects, all of which influence the edge profile simultaneously.
Even in simpler toroidal X-point devices such as TORPEX, recent simulation efforts have not fully reproduced the experimental measurements described above.\cite{Sepulchre2025,Galassi2022}
In magnetic reconnection settings, where the X-point is central to the dynamics, the presence of a driving electric field perpendicular to the field lines is essential to the resulting chaotic single particle motion and the associated effective resistivity.\cite{Numata2003,Yamada2014}
In contrast to these configurations, the linear device MAXIMUS realizes a simpler X-point field topology using a pair of parallel current-carrying rods, with no superposed guide field; low temperature plasmas in that device are found to exhibit a transport barrier near the separatrix and a locally flattened pressure profile across the X-point.\cite{Lim2020}
Even here, the finite device length and localized plasma sources break the translational symmetry of the idealized configuration.
In order to directly address the role of the X-point alone in plasma dynamics without the complicating effects mentioned above, a clean theoretical setting is required.

Such a setting is provided by the two-wire model (TWM), an ideal X-point system formed by two parallel infinitely long current-carrying wires.
The TWM field contains an X-point at its center and a lemniscate separatrix passing through it, and the field is described by a closed-form analytic formula that allows an exact treatment of the magnetic geometry.
The TWM was originally introduced by Boozer and Rechester\cite{Boozer1978} as the simplest analytic example of a divertor field, with two parallel currents producing the X-point and a superposed axial field representing the tokamak toroidal field.
It has since been used to study classical cross-field diffusion in the presence of an X-point,\cite{Auerbach1980} divertor field line topology,\cite{Reiman1996} and stochastic layer scaling under magnetic perturbations.\cite{Ali2009}
Auerbach and Boozer,\cite{Auerbach1980} in particular, showed that with an axial guide field the classical cross-field diffusion coefficient remains finite at the separatrix and that cross-field flow concentrates near the X-point, but did not address the resulting density profile or the guide-field-free true null case.

A common feature of the prior TWM literature is that an axial guide field is included, so the configuration has a minimum in the field magnitude but not a true null.
The guide-field-free TWM, in which the only field is the in-plane field generated by the two wires and the field magnitude genuinely vanishes at the X-point, has not been investigated in detail.
This is the setting in which the X-point's effect on plasma dynamics is most extreme: charged particles are unmagnetized near the null, and the cross-field transport suppression characteristic of magnetized plasmas is locally relaxed.
This regime raises the following questions: how individual charged particles behave when their adiabatic invariants are violated near the X-point, and how the collisional cross-field transport and the resulting density profile of magnetized plasmas are affected by the presence of the X-point.

This work investigates the guide-field-free TWM as a clean setting for studying plasma dynamics near a true magnetic null, and addresses the two questions raised above in two complementary regimes.
In this work, the TWM refers to the guide-field-free case.
The first regime is the study of collisionless charged particle dynamics, in which individual charged particles move in the TWM field without the influence of additional fields induced by other particles.\cite{Ahn2024}
The second regime is the study of collisional dynamics of magnetized low temperature plasmas in the TWM field, in which the plasma can be described by classical drift-diffusion and complemented by self-consistent kinetic simulation.\cite{Ahn2026}

In the collisionless study, a Lagrangian analysis of single particle motion in the TWM field identifies two conserved quantities: the total kinetic energy, $W_\text{total}$, conserved because the Lorentz force does no work, and a base field line value, $\zeta_\text{base}$, derived from the axial canonical momentum, conserved by the translational symmetry of the field along the $z$-direction.
The base field line value labels the field line which particles gyrate about, distinct from the gyro-center, and consists of two disconnected branches symmetric about the X-point for particles inside the separatrix.
The total kinetic energy controls the speed and the Larmor radius of the particles, and therefore determines how deeply they travel into the large gradient region near the X-point.
Numerical simulations show that when a particle traverses this region, the magnetic moment undergoes abrupt shifts arising from the strongly inhomogeneous field.
These shifts are sensitive to initial conditions but statistically determined by the two conserved quantities, and they enable a phenomenon termed migration.
Migration refers to the phenomenon in which a particle, originally gyrating about one of the two branches of a base field line inside the separatrix, jumps to the other corresponding branch on the opposite side once its Larmor radius is large enough to cross the separatrix.
A threshold energy for migration, $\norm{W}_\text{total} > \zeta_\text{base}^2$, is derived from an effective potential analysis, and an empirical expression for the migration confinement time is formulated using simulation data over a wide range of invariant pairs.

In the collisional study, the drift-diffusion model (DD model) for a magnetized low temperature plasma is formulated in a field-aligned conformal curvilinear coordinate system derived from the TWM field.
Averaging the DD model along closed field lines, together with a fast parallel equilibration closure, yields a one-dimensional reduced description of the cross-field density profile.
The reduced models predict the formation of a density plateau near the separatrix in the strongly magnetized regime.
This plateau is a consequence of the structure of the magnetization reduction factor, $f = 1/(1+\beta^2)$, where $\beta = \omega_c/\nu$ is the ratio of the cyclotron frequency to the collision frequency.
Near the X-point, the magnetic field magnitude weakens and $f$ sharply rises to unity, so the conserved cross-field flux can be carried by a relaxed density gradient.
Fast parallel equilibration then extends this gradient relaxation across the entire field line, giving rise to a density plateau most strongly on the separatrix.
Two-dimensional electrostatic particle-in-cell (PIC) simulations of the TWM configuration are performed using the EDIPIC-2D code,\cite{EDIPIC2D} and they reproduce the predicted density plateau for the strongly magnetized electron population, with no such pronounced feature in the weakly magnetized ion population.
A scan of the wire current further confirms that the plateau strengthens with the magnetization.
The equilibration closure assumption underlying the reduced models is empirically validated in the PIC results through a direct fluctuation diagnostic.
Quantitative discrepancies between the reduced model and PIC results are attributed to numerical effects intrinsic to the PIC method and to the limitations of the fluid description near the magnetic null; however, these do not alter the central result that the presence of the magnetic X-point produces a density plateau near the separatrix in the strongly magnetized regime.

The remainder of the work is organized as follows.
Chapter~\ref{chap.2} introduces the TWM magnetic field, its normalization in terms of the various characteristic scales of the system, and the field-aligned conformal curvilinear coordinate system.
Chapter~\ref{chap.3} investigates collisionless charged particle dynamics.
This includes the invariants of single particle motion, the statistical magnetic moment shifts near the X-point, and the cross-separatrix migration phenomenon.
Chapter~\ref{chap.4} investigates collisional dynamics of magnetized low temperature plasmas.
This includes the reduced DD models in the TWM coordinate system, the prediction of density plateau formation near the separatrix, and its verification by self-consistent PIC simulations.
Chapter~\ref{chap.5} summarizes the principal findings, discusses the connection between the two regimes, and concludes the work.


\chapter{Two-Wire Model Magnetic Field}\label{chap.2}

The two-wire model (TWM) field in this work is defined as the superposition of the magnetic fields generated by two parallel wires carrying identical steady currents, and features a magnetic X-point and a separatrix.
The TWM is one of the simplest magnetic configurations that contain an X-point and have an exact analytical form, making it ideal for a focused study of the role of a true magnetic null in plasma dynamics.

The first three sections present the essential material used throughout the rest of the work.
In Sec.~\ref{sec.2.1}, we present the analytical expression of the TWM field and its normalization in terms of the characteristic scales of the system.
In Sec.~\ref{sec.2.2}, we construct the TWM coordinate system $(\zeta, \eta)$, a field-aligned curvilinear coordinate system in which $\zeta$ labels the field lines and $\eta$ parametrizes position along them.
In Sec.~\ref{sec.2.3}, we derive the scale factor and the differential operators in the TWM coordinate system.

The remaining sections present additional properties of the TWM field and coordinate system, which serve as background and reference material.
Section~\ref{sec.2.4} demonstrates the conformality of the coordinate transformation through complex analysis and the Jacobian.
Section~\ref{sec.2.5} collects useful coordinate identities and the backward transformation.
Section~\ref{sec.2.6} presents the connection coefficients and the gradient of the field magnitude.
Section~\ref{sec.2.7} examines the geometry of the TWM field lines, the Cassini ovals, including their spatial extent, arclength, and enclosed area.
Section~\ref{sec.2.8} summarizes the chapter.
In addition, Appendix~\ref{chap.A} briefly treats the TWM electric field generated by two uniformly charged wires.



\section{TWM magnetic field \& Normalization}\label{sec.2.1}

\begin{figure}
    \centering
    \includegraphics[width=13cm]{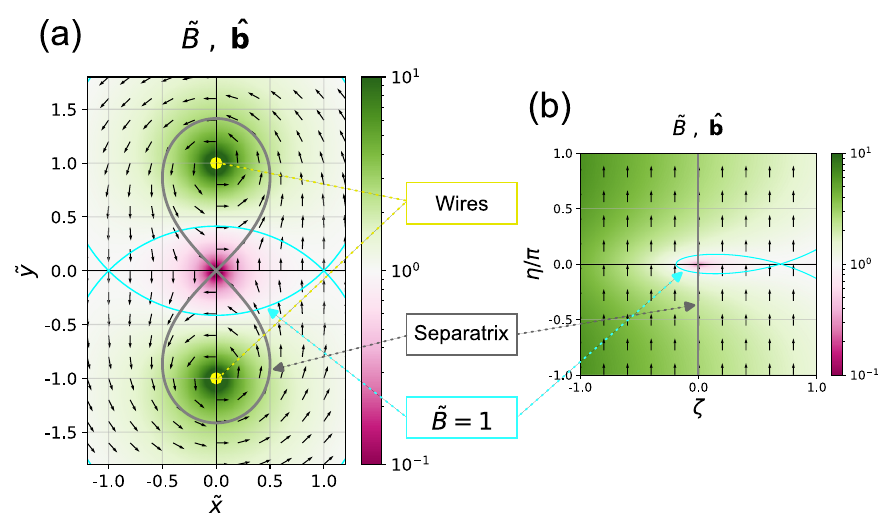}
    \caption[
    Normalized TWM field magnitude in the $\norm{x}-\norm{y}$ and $\zeta-\eta$ planes
    ]{
    Normalized TWM field magnitude, with the unit vector field shown as black arrows, the wire locations indicated by yellow circles, the separatrix ($s=1$, or equivalently $\zeta=0$) shown in gray, and the contour $\norm{B}=1$ shown in cyan, in (a) the $\norm{x}-\norm{y}$ plane and (b) the $\zeta-\eta$ plane.
    }\label{fig.2_1}
\end{figure}

The two-wire model (TWM) magnetic field is generated by two parallel infinitely long wires located at $(x,y) = (0, \pm \ell_0)$, each carrying a steady current, $I_0$, in the positive $z$-direction (see Fig.~\ref{fig.2_1}(a)).
To express the magnetic field in compact form, several quantities are defined as follows.
The quantities $y_\pm=y \pm \ell_0$ denote the signed vertical distances from each wire, $r_\pm^2 = x^2+y_\pm^2$ are the squared distances from each wire, and $A_0 = I_0 \mu_0/2\pi$ is the characteristic magnetic vector potential.
Then, the magnetic field is expressed as
\begin{equation}\label{eq.2_vecB}
    \bvec{B}(x, y)
    = B_x \uvec{x} + B_y \uvec{y}
    = A_0 \left[ \left( -\frac{y_+}{r_+^2}-\frac{y_-}{r_-^2} \right)\uvec{x}
    + \left( \frac{x}{r_+^2}+\frac{x}{r_-^2} \right)\uvec{y} \right].
\end{equation}
Note that a single current-carrying wire at $(x,y) = (0, 0)$ produces a purely azimuthal field $\bvec{B} = (A_0/r)\,\uvecg{\theta} = A_0\left[ (-y/r^2)\,\uvec{x} + (x/r^2)\,\uvec{y} \right]$, where $r^2 = x^2 + y^2$.\cite{Griffiths2013}
The TWM field is therefore the superposition of two such fields, with their centers shifted to $(x,y) = (0, \pm \ell_0)$.
Here, setting $y_+ = y + \ell_0 = 0$ gives $y = -\ell_0$, the location of the bottom wire.
Thus, the $+$ subscript denotes the field from the bottom wire, and vice versa.

We normalize physical quantities using the characteristic scales of the TWM.
A tilde over a symbol, e.g., $\norm{x}$, denotes a normalized quantity.
The vertical distance from the origin to each wire, $\ell_0$, is used as the characteristic length, so that $\norm{x} = x/\ell_0$, $\norm{y} = y/\ell_0$, $\norm{y}_\pm = y_\pm/\ell_0$, $\norm{r}_\pm = r_\pm/\ell_0$, and similarly for other length quantities.
The characteristic field magnitude is defined as $B_0 = A_0 / \ell_0$, which is used to normalize the magnetic field.
The normalized magnetic field is then
\begin{equation}\label{eq.2_normvecB}
\begin{aligned}
    \norm{\bvec{B}}(\norm{x}, \norm{y}) = \frac{\bvec{B}}{B_0}
    & = \norm{B}_x \uvec{x} + \norm{B}_y \uvec{y}
    \\
    & = \left( -\frac{\norm{y}_+}{\norm{r}_+^2}-\frac{\norm{y}_-}{\norm{r}_-^2} \right) \uvec{x}
    +\left( \frac{\norm{x}}{\norm{r}_+^2}+\frac{\norm{x}}{\norm{r}_-^2} \right) \uvec{y}
    \\
    & = \left( \frac{-2\norm{y}(\norm{r}^2-1)}{s^4} \right) \uvec{x}
    + \left( \frac{2\norm{x}(\norm{r}^2+1)}{s^4} \right) \uvec{y},
\end{aligned}
\end{equation}
where $\norm{r}^2 = \norm{x}^2 + \norm{y}^2$ is the normalized squared distance from the origin, and $s^2=\norm{r}_+ \norm{r}_-$ is the product of normalized distances from the two wires.
We use normalized expressions in most of this work, so that the investigation applies to a TWM of any physical scale.
The TWM characteristic scales and length quantities are summarized in Table~\ref{tab.2_1}.

The normalized field magnitude and the unit vector are
\begin{equation}\label{eq.2_normB}
    \norm{B}(\norm{x}, \norm{y}) = \frac{B}{B_0}
    = \frac{2 \norm{r}}{\norm{r}_+ \norm{r}_-} = \frac{2 \norm{r}}{s^2},
\end{equation}
\begin{equation}
    \uvec{b} = \left( \frac{-\norm{y}(\norm{r}^2-1)}{s^2 \norm{r}} \right)\uvec{x}
    + \left( \frac{\norm{x}(\norm{r}^2+1)}{s^2 \norm{r}} \right)\uvec{y}.
\end{equation}

\begin{table}[h]
\centering
\caption{\label{tab.2_1} Characteristic scales and length quantities of the TWM.}
\begin{tabular}{ll}
    \hline
    \multicolumn{2}{l}{\textbf{Characteristic scales}} \\
    \hline
    $\ell_0$  & Vertical distance from the origin to each wire \\
    $I_0$ & Current flowing in each wire \\
    $A_0 \equiv I_0 \mu_0 / 2\pi$ & Characteristic magnetic vector potential \\
    $B_0 \equiv A_0 / \ell_0$ & Characteristic magnetic field magnitude \\
    \hline
    \multicolumn{2}{l}{\textbf{Length quantities}} \\
    \hline
    $y_\pm \equiv y \pm \ell_0$ & Signed vertical distance from each wire \\
    $r^2 \equiv x^2 + y^2$ & Squared distance from the origin \\
    $r_\pm^2 \equiv x^2 + y_\pm^2$ & Squared distance from each wire \\
    \hline
    \multicolumn{2}{l}{\textbf{Normalized length quantities}} \\
    \hline
    $\norm{x} \equiv x / \ell_0$, $\norm{y} \equiv y / \ell_0$ & Normalized Cartesian coordinates \\
    $\norm{y}_\pm \equiv \norm{y} \pm 1$ & Normalized signed vertical distance from each wire \\
    $\norm{r}^2 \equiv \norm{x}^2 + \norm{y}^2$ & Normalized squared distance from the origin \\
    $\norm{r}_\pm^2 \equiv \norm{x}^2 + \norm{y}_\pm^2$ & Normalized squared distance from each wire\\
    $s^2 \equiv \norm{r}_+ \norm{r}_-$ & Product of normalized distances from the two wires \\
    \hline
\end{tabular}
\end{table}

Figure~\ref{fig.2_1}(a) shows the normalized TWM field magnitude in the $\norm{x}-\norm{y}$ plane, with the unit vector field shown as black arrows, the separatrix shown in gray, and the contour $\norm{B}=1$ in cyan.
The locations $\norm{r}_+ = 0$ and $\norm{r}_- = 0$ correspond to the positions of the bottom and top wires, respectively, and are indicated by yellow circles.
These points are singularities of the TWM field, where the field magnitude becomes infinite.
The magnetic X-point is located at the origin, $(\norm{x}, \norm{y}) = (0, 0)$, or equivalently $\norm{r}=0$, where the field magnitude vanishes.



\section{TWM coordinate system}\label{sec.2.2}

The TWM coordinate system $(\zeta, \eta)$ is an orthogonal curvilinear coordinate system adapted to the TWM field.
The coordinate $\zeta$ labels the closed field lines of the TWM, while $\eta$ is a periodic angle-like coordinate.

The coordinates are defined as follows:
\begin{subequations}\label{eq.2_zetaeta}
\begin{align}
    \zeta & = \ln s^2, 
    \mspace{185mu} \zeta \in (-\infty, +\infty),
    \\
    \eta & = \text{atan2} (2 \norm{x} \norm{y}, \norm{x}^2 - \norm{y}^2 + 1),
    \qquad \eta \in (-\pi, +\pi].
\end{align}
\end{subequations}

With these definitions, the magnetic field satisfies the following compact relationships:
\begin{subequations}
\begin{align}
    \label{eq.2_B_with_zetaeta}
    \bvec{B}
    & = \uvec{z} \times (A_0 \nabla \zeta) = A_0 \nabla \eta,
    \\
    \label{eq.2_normB_with_zetaeta}
    \norm{\bvec{B}}
    & = \uvec{z} \times \norm{\nabla} \zeta = \norm{\nabla} \eta,
\end{align}
\end{subequations}
where $\norm{\nabla} = \ell_0 \nabla = \uvec{x} \partial / \partial \norm{x} + \uvec{y} \partial / \partial \norm{y} + \uvec{z} \partial / \partial \norm{z}$ is the normalized gradient operator.

\begin{figure}
    \centering
    \includegraphics[width=15cm]{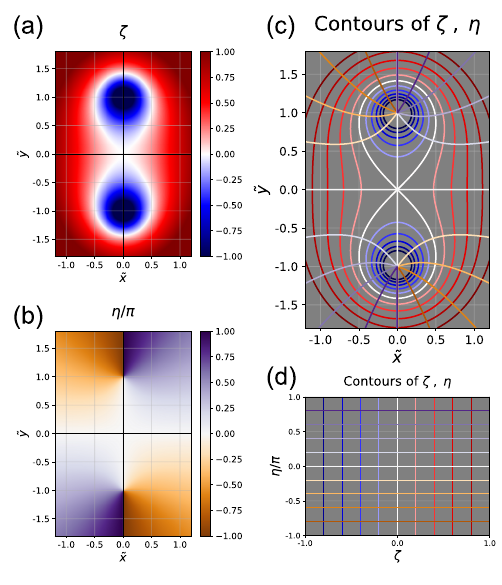}
    \caption[
    Spatial distribution and contours of the TWM coordinates $\zeta$ and $\eta$
    ]{
    Spatial distribution of (a) $\zeta$ and (b) $\eta/\pi$ in the $\norm{x}-\norm{y}$ plane.
    (c) Selected contours of $\zeta$ and $\eta$ in the $\norm{x}-\norm{y}$ plane.
    (d) Corresponding contours in the $\zeta-\eta$ plane.
    In both (c) and (d), the contour ranges are $\zeta \in [-1, +1]$ with $\Delta \zeta = 0.2$, and $\eta \in [-\pi, +\pi]$ with $\Delta \eta = 0.2 \pi$.
    }\label{fig.2_2}
\end{figure}

From Eq.~\eqref{eq.2_B_with_zetaeta}, it follows that $\bvec{B} \cdot \nabla \zeta = 0$, which shows that $\zeta$, and equivalently $s=\exp{(\zeta/2)}$, is constant along field lines.
Thus, $\zeta$ labels the field line contours.
These contours are geometrical shapes called Cassini ovals,\cite{Lawrence1972} whose foci are the wire positions.
The contour $\zeta=0$ ($s=1$) is the separatrix, which forms the shape known as the lemniscate of Bernoulli.
Inside the separatrix $(\zeta < 0)$, a given contour consists of two disconnected branches symmetric about the origin; outside the separatrix $(\zeta > 0)$, a given contour is a single connected oval.
Equation~\eqref{eq.2_B_with_zetaeta} also implies $\nabla \zeta \cdot \nabla \eta = 0$, indicating that the $\eta$ contours are orthogonal to the $\zeta$ contours, and thus to the magnetic field lines.

Therefore, the $\zeta$-coordinate is a radial-like coordinate that increases outward from the wires, while the $\eta$-coordinate is a periodic angle-like coordinate that parametrizes position along the field lines.
We refer to $\zeta$ as the field line label and $\eta$ as the orthogonal line label.
The magnetic X-point is located at $(\zeta,\eta) = (0,0)$ (see Fig.~\ref{fig.2_2}).

For a magnetic field with translational symmetry along the $z$-direction and no $z$-component, $\bvec{B}(x,y) = \nabla \times (A_z \uvec{z}) = \nabla A_z \times \uvec{z}$, where $A_z$ is the $z$-component of the magnetic vector potential.
Comparing this expression with Eq.~\eqref{eq.2_B_with_zetaeta}, we obtain $A_z = - A_0 \zeta$.

We can visualize any physical quantity in a field-aligned manner in the $\zeta-\eta$ plane, either by deriving its closed-form expression in $(\zeta, \eta)$, or by numerically applying the coordinate transformation using Eq.~\eqref{eq.2_zetaeta}.
For instance, Figure~\ref{fig.2_1}(b) shows the normalized TWM field magnitude in the $\zeta-\eta$ plane, which can be drawn from either approach.
The closed-form expression is
\begin{equation}\label{eq.2_normB_zetaeta_exact}
    \norm{B}(\zeta, \eta) = 32^{1/4} \exp \left( -\frac{3}{4}\zeta \right) \left( \cosh\zeta - \cos\eta \right)^{1/4},
\end{equation}
which is derived using the coordinate identities in Sec.~\ref{sec.2.5}.
Figure~\ref{fig.2_1}(b) also shows the unit vector field, the separatrix, and the contour $\norm{B}=1$.
This plot presents the separatrix as a vertical straight line at $\zeta = 0$, and all magnetic field unit vectors point in the positive $\eta$-direction.
Note that the transformation $(\norm{x}, \norm{y}) \mapsto (\zeta, \eta)$ is two-to-one: each point in $(\zeta, \eta)$ represents two points in the $\norm{x}-\norm{y}$ plane that are symmetric with respect to the origin.
This ambiguity may be removed by distinguishing the upper and lower half-planes, or by setting up the problem to be symmetric about the origin.



\section{Differential operators in the TWM coordinate system}\label{sec.2.3}

The preceding sections focused on the in-plane coordinates $(\zeta, \eta)$, but here we include the $z$-coordinate to form the three-dimensional system $(\zeta, \eta, z)$ for the formulation of vector quantities and differential operators.
The $z$-coordinate and its unit vector remain unchanged from the Cartesian system, but including them is useful for the complete formulation that follows.

The unit vectors of the TWM coordinate system are defined as $\uvecg{\zeta} = \nabla \zeta / |\nabla \zeta|$ and $\uvecg{\eta} = \nabla \eta / |\nabla \eta| = \uvec{b}$, together with the Cartesian unit vector $\uvec{z}$.
Physically, $\uvecg{\eta}$ is the direction parallel to the magnetic field, $\uvecg{\zeta}$ is the in-plane perpendicular direction, and $\uvec{z}$ is the out-of-plane perpendicular direction.
These unit vectors form a right-handed orthonormal basis satisfying
\begin{equation}
    \uvecg{\zeta} \times \uvecg{\eta} = \uvec{z}, \qquad
    \uvecg{\eta} \times \uvec{z} = \uvecg{\zeta}, \qquad
    \uvec{z} \times \uvecg{\zeta} = \uvecg{\eta} = \uvec{b}.
\end{equation}

To perform differential operations in the TWM coordinate system, it is necessary to determine the scale factor, $h$.
In a general curvilinear coordinate system $(q_1, q_2, q_3)$, the metric tensor has components
\begin{equation}
    g_{ij} = \frac{\partial \bvec{r}}{\partial q_i} \cdot \frac{\partial \bvec{r}}{\partial q_j},
\end{equation}
where $\bvec{r} = (x,y,z)$.\cite{Arfken2005}
The metric tensor encodes how distances and angles are measured in the coordinate system, and its components directly determine the differential elements and operators that follow.
For an orthogonal coordinate system, the off-diagonal components of the metric tensor vanish, i.e., $g_{ij}=0$ for $i \ne j$, and the metric is fully characterized by the scale factors, also known as the Lam\'e coefficients,
\begin{equation}
    h_i = \sqrt{g_{ii}} = \left| \frac{\partial \bvec{r}}{\partial q_i} \right| = |\nabla q_i|^{-1}.
\end{equation}

Since the TWM coordinate system is orthogonal, the scale factors are $h_\zeta = |\nabla \zeta|^{-1}$, $h_\eta = |\nabla \eta|^{-1}$, and $h_z = |\nabla z|^{-1} = 1$.
From Eq.~\eqref{eq.2_B_with_zetaeta}, $|\nabla \zeta| = |\nabla \eta|$, and therefore $h_\zeta = h_\eta$.
We thus define the common scale factor for $\zeta$ and $\eta$ as
\begin{equation}\label{eq.2_h}
    h = h_\zeta = h_\eta = \frac{1}{|\nabla \zeta|} = \frac{1}{|\nabla \eta|} = \frac{A_0}{B} = \frac{\ell_0}{\norm{B}},
\end{equation}
and its normalized form is
\begin{equation}\label{eq.2_normh}
    \norm{h} = \frac{h}{\ell_0} = \frac{1}{\norm{B}}.
\end{equation}
Since $h=A_0/B$, the coordinate system is singular at the magnetic X-point, where $B=0$ and $h \rightarrow \infty$, and at the wire locations, where $B \rightarrow \infty$ and $h \rightarrow 0$.
The differential operators given below are therefore defined only away from these singular points.

The equality $h_\zeta = h_\eta$ indicates that the coordinate transformation between the $\norm{x}-\norm{y}$ and $\zeta-\eta$ planes is conformal, i.e., angle-preserving.
This property simplifies the mathematical formulation, particularly when performing differential operations.
Note that the conformality applies only to the in-plane coordinates $(\zeta, \eta)$; the out-of-plane $z$-direction is unchanged from the Cartesian system, with $h_z = 1$.

The gradient, Laplacian, divergence, and curl operations are given as follows:
\begin{equation}\label{eq.2_formula_grad}
    \nabla a
    = \frac{1}{h} \frac{\partial a}{\partial \zeta} \uvecg{\zeta}
    + \frac{1}{h} \frac{\partial a}{\partial \eta} \uvecg{\eta}
    + \frac{\partial a}{\partial z} \uvec{z},
\end{equation}
\begin{equation}\label{eq.2_formula_lap}
    \nabla^2 a
    = \frac{1}{h^2} \frac{\partial^2 a}{\partial \zeta^2}
    + \frac{1}{h^2} \frac{\partial^2 a}{\partial \eta^2}
    +               \frac{\partial^2 a}{\partial z^2},
\end{equation}
\begin{equation}\label{eq.2_formula_div}
    \nabla \cdot \bvec{u}
    = \frac{1}{h^2} \frac{\partial (h u_\zeta)}{\partial \zeta}
    + \frac{1}{h^2} \frac{\partial (h u_\eta)}{\partial \eta} 
    +               \frac{\partial u_z}{\partial z},
\end{equation}
\begin{equation}\label{eq.2_formula_curl}
\begin{aligned}
    \nabla \times \bvec{u}
    = \left( \frac{1}{h} \frac{\partial u_z}{\partial \eta} - \frac{\partial u_\eta}{\partial z} \right) \uvecg{\zeta}
    + \left( \frac{\partial u_\zeta}{\partial z} - \frac{1}{h} \frac{\partial u_z}{\partial \zeta} \right) \uvecg{\eta}
    + \frac{1}{h^2} \left( \frac{\partial (h u_\eta)}{\partial \zeta} - \frac{\partial (h u_\zeta)}{\partial \eta} \right) \uvec{z},
\end{aligned}
\end{equation}
where $a$ and $\bvec{u}$ are arbitrary scalar and vector fields, respectively.

The corresponding normalized operators are obtained by replacing $\nabla$ with $\norm{\nabla} = \ell_0 \nabla$ and $h = \ell_0/\norm{B}$ with $\norm{h} = 1/\norm{B}$.
For instance, the normalized gradient is
\begin{equation}\label{eq.2_formula_normgrad}
    \norm{\nabla} a
    = \frac{1}{\norm{h}} \frac{\partial a}{\partial \zeta} \uvecg{\zeta}
    + \frac{1}{\norm{h}} \frac{\partial a}{\partial \eta} \uvecg{\eta}
    + \frac{\partial a}{\partial \norm{z}} \uvec{z},
\end{equation}
where $\norm{z} = z/\ell_0$.

For notational compactness, abbreviated notation is used throughout for partial derivatives, e.g., $\partial_\zeta a = \partial a / \partial \zeta$.
We also write $(\nabla a)_k = \uvec{k} \cdot \nabla a$, which denotes the $k$-component of the gradient of a scalar field $a$, so that $(\nabla a)_\zeta = h^{-1} \partial_\zeta a$, $(\nabla a)_\eta = h^{-1} \partial_\eta a$, and $(\nabla a)_z = \partial_z a$.

The material up to this point provides most of the information needed to follow the subsequent chapters on collisionless (Chapter~\ref{chap.3}) and collisional (Chapter~\ref{chap.4}) dynamics.
The remainder of this chapter, Secs.~\ref{sec.2.4}--\ref{sec.2.7}, presents additional properties of the TWM field and coordinate system that provide useful background and reference material.



\section{Conformal structure of the TWM coordinate system}\label{sec.2.4}

We provide two complementary demonstrations of the conformality of the TWM coordinate system.
The first uses complex analysis, and the second uses the Jacobian of the coordinate transformation.

Complex analysis provides a concise mathematical interpretation of the TWM coordinate transformation.
The normalized complex coordinate is defined as
\begin{equation}
    \norm{Z} = \norm{x} + i \norm{y}.
\end{equation}
We then define
\begin{equation}
    W = \zeta + i \eta = \ln ( \norm{Z}^2 + 1 )
      = \ln \left[ (\norm{x}^2-\norm{y}^2+1) + i(2\norm{x}\norm{y}) \right].
\end{equation}
Therefore, the real and imaginary parts of $W$ are
\begin{subequations}
\begin{align}
    \zeta & = \ln | \norm{Z}^2 + 1 | = \ln s^2,
    \\
    \eta & = \arg (\norm{Z}^2 + 1 ) = \text{atan2} (2 \norm{x} \norm{y}, \norm{x}^2 - \norm{y}^2 + 1),
\end{align}
\end{subequations}
which recover the coordinate definitions in Eq.~\eqref{eq.2_zetaeta}.

The complex function $W = \ln ( \norm{Z}^2 + 1 )$ is locally analytic on any simply connected domain that excludes $\norm{Z}^2 + 1 = 0$ and does not cross the chosen branch cut of the logarithm.
Therefore, the Cauchy-Riemann relations\cite{Arfken2005} are satisfied:
\begin{equation}
    \partial_{\norm{x}} \zeta = \partial_{\norm{y}} \eta,
    \qquad
    \partial_{\norm{y}} \zeta = -\partial_{\norm{x}} \eta.
\end{equation}
This implies
\begin{subequations}
\begin{align}
    \norm{\nabla} \zeta \cdot \norm{\nabla} \eta
    & = (\partial_{\norm{x}} \zeta) (\partial_{\norm{x}} \eta) + (\partial_{\norm{y}} \zeta) (\partial_{\norm{y}} \eta) = 0,
    \\
    | \norm{\nabla} \zeta |^2 & = (\partial_{\norm{x}} \zeta)^2 + (\partial_{\norm{y}} \zeta)^2
    \notag
    \\
    & = (\partial_{\norm{x}} \eta)^2 + (\partial_{\norm{y}} \eta)^2 = | \norm{\nabla} \eta |^2,
\end{align}
\end{subequations}
which shows the orthogonality and conformality of the TWM coordinate system.

The same conformal structure can be shown using the Jacobian of the coordinate transformation.
For the in-plane transformation $(\norm{x}, \norm{y}) \mapsto (\zeta, \eta)$, the forward Jacobian matrix is
\begin{equation}
    \norm{\overline{\overline{\mathbf{J}}}}_{\mathrm{f}}
    = \frac{\partial(\zeta, \eta)}{\partial(\norm{x}, \norm{y})}
    = \begin{bmatrix} \partial_{\norm{x}}\zeta & \partial_{\norm{y}}\zeta \\ \partial_{\norm{x}}\eta & \partial_{\norm{y}}\eta \end{bmatrix}
    = \norm{B} \begin{bmatrix} b_y & -b_x \\ b_x & b_y \end{bmatrix}
    = \frac{1}{\norm{h}} \begin{bmatrix} b_y & -b_x \\ b_x & b_y \end{bmatrix},
\end{equation}
where $b_x$ and $b_y$ are the Cartesian components of the unit vector $\uvec{b}$.
The Jacobian determinant is
\begin{equation}
    \left| \norm{\overline{\overline{\mathbf{J}}}}_{\mathrm{f}} \right|
    = \norm{B}^2 (b_x^2 + b_y^2) = \norm{B}^2 = \frac{1}{\norm{h}^2}.
\end{equation}
The matrix is the product of an isotropic scaling $\norm{B}$ and a rotation in $SO(2)$, which is precisely the statement that the transformation is conformal: it preserves angles and scales all lengths locally by the same factor $\norm{B} = 1/\norm{h}$.

The backward transformation $(\zeta, \eta) \mapsto (\norm{x}, \norm{y})$ has the backward Jacobian
\begin{equation}
    \norm{\overline{\overline{\mathbf{J}}}}_{\mathrm{b}}
    = \frac{\partial(\norm{x}, \norm{y})}{\partial(\zeta, \eta)}
    = \frac{1}{\norm{B}} \begin{bmatrix} b_y & b_x \\ -b_x & b_y \end{bmatrix}
    = \norm{h} \begin{bmatrix} b_y & b_x \\ -b_x & b_y \end{bmatrix}
    = \norm{\overline{\overline{\mathbf{J}}}}_{\mathrm{f}}^{-1},
    \qquad
    \left| \norm{\overline{\overline{\mathbf{J}}}}_{\mathrm{b}} \right| = \frac{1}{\norm{B}^2} = \norm{h}^2.
\end{equation}
Under the backward transformation, all coordinate-space lengths are scaled locally by the factor $\norm{h} = 1/\norm{B}$.

This scaling by $1/\norm{B}$ is illustrated in Fig.~\ref{fig.2_2}(c, d).
In the $\zeta-\eta$ plane (Fig.~\ref{fig.2_2}(d)), the $\zeta$ and $\eta$ contours form a uniform rectangular grid of equal size cells, as the contour spacings $\Delta \zeta$ and $\Delta \eta$ are chosen to be constant.
When the same contours are drawn in the $\norm{x}-\norm{y}$ plane (Fig.~\ref{fig.2_2}(c)), the cell sizes vary strongly with position through the local scaling factor $\norm{h} = 1/\norm{B}$.
Figure~\ref{fig.2_2}(c) shows that, near the X-point, where $\norm{B}$ is small and $\norm{h}$ is large, the contours are widely spaced; near the wires, where $\norm{B}$ is large and $\norm{h}$ is small, the contours are closely spaced.

The differentials transform as
\begin{equation}
\begin{aligned}
    \begin{bmatrix} d\zeta \\ d\eta \end{bmatrix}
    &= \norm{\overline{\overline{\mathbf{J}}}}_{\mathrm{f}} \begin{bmatrix} d\norm{x} \\ d\norm{y} \end{bmatrix}
    = \frac{1}{\norm{h}} \begin{bmatrix} b_y & -b_x \\ b_x & b_y \end{bmatrix} \begin{bmatrix} d\norm{x} \\ d\norm{y} \end{bmatrix},
    \\[6pt]
    \begin{bmatrix} d\norm{x} \\ d\norm{y} \end{bmatrix}
    &= \norm{\overline{\overline{\mathbf{J}}}}_{\mathrm{b}} \begin{bmatrix} d\zeta \\ d\eta \end{bmatrix}
    = \norm{h} \begin{bmatrix} b_y & b_x \\ -b_x & b_y \end{bmatrix} \begin{bmatrix} d\zeta \\ d\eta \end{bmatrix}.
\end{aligned}
\end{equation}
Both Jacobians are valid everywhere except at the singular points $\norm{r}_\pm = 0$ (the wires) and $\norm{r} = 0$ (the X-point), consistent with the singularities of the scale factor $h$ discussed in Sec.~\ref{sec.2.3}.
The unnormalized Jacobian follows from the normalized one by restoring the length scale $\ell_0$; for the transformation $(x, y) \mapsto (\zeta, \eta)$, the forward Jacobian is $\overline{\overline{\mathbf{J}}}_{\mathrm{f}} = \ell_0^{-1} \norm{\overline{\overline{\mathbf{J}}}}_{\mathrm{f}}$.

Finally, the scale factor yields the line and area elements directly, because the off-diagonal metric components are zero.
For in-plane displacements ($dz = 0$), the line element is
\begin{subequations}\label{eq.2_line_element}
\begin{align}
    (dl)^2 & = d\bvec{r} \cdot d\bvec{r}
    = h^2 \left[ (d\zeta)^2 + (d\eta)^2 \right]
    = \frac{A_0^2}{B^2} \left[ (d\zeta)^2 + (d\eta)^2 \right],
    \\
    (d\norm{l})^2 & = d\norm{\bvec{r}} \cdot d\norm{\bvec{r}}
    = \norm{h}^2 \left[ (d\zeta)^2 + (d\eta)^2 \right]
    = \frac{1}{\norm{B}^2} \left[ (d\zeta)^2 + (d\eta)^2 \right],
\end{align}
\end{subequations}
and the area element is
\begin{subequations}\label{eq.2_area_element}
\begin{align}
    dA & = dx\,dy = h^2\,d\zeta\,d\eta = \frac{A_0^2}{B^2}\,d\zeta\,d\eta,
    \\
    d\norm{A} & = d\norm{x}\,d\norm{y} = \norm{h}^2\,d\zeta\,d\eta = \frac{1}{\norm{B}^2}\,d\zeta\,d\eta.
\end{align}
\end{subequations}



\section{Coordinate identities \& Backward transformation}\label{sec.2.5}

This section collects several identities of the TWM coordinate system.
We first present equivalent forms of $s^4$.
\begin{equation}\label{eq.2_s4}
\begin{aligned}
s^4 = \norm{r}_+^2 \norm{r}_-^2 = \exp{(2\zeta)}
&
= (\norm{x}^2 + \norm{y}_+^2)(\norm{x}^2 + \norm{y}_-^2)
= \norm{x}^4 + \norm{y}^4 + 2 \norm{x}^2 \norm{y}^2 + 2 \norm{x}^2 - 2 \norm{y}^2 + 1
\\[6pt]
& 
= \norm{x}^4 + 2 (\norm{y}^2 + 1) \norm{x}^2 + (\norm{y}^2 - 1)^2
= \norm{y}^4 + 2 (\norm{x}^2 - 1) \norm{y}^2 + (\norm{x}^2 + 1)^2
\\[6pt]
&
= (\norm{x}^2 - \norm{y}^2 + 1)^2 + (2 \norm{x} \norm{y})^2 = |  (\norm{x}+i\norm{y})^2 + 1 |^2
\\[6pt]
&
= \norm{r}^4 + 2 \norm{r}^2 \cos{(2\theta)} + 1
= \left|  \norm{r}^2 e^{i2\theta} + 1 \right|^2
\\[6pt]
&
= \norm{r}^4 + 2 (\norm{x}^2 - \norm{y}^2) + 1
= (\norm{r}^2 + 1)^2 - 4\norm{y}^2
= (\norm{r}^2 - 1)^2 + 4\norm{x}^2
\\[6pt]
&
= \norm{r}_+^2 (\norm{r}_+^2 - 4\norm{y})
= \norm{r}_-^2 (\norm{r}_-^2 + 4\norm{y}),
\end{aligned}
\end{equation}
where $\theta = \text{atan2}(\norm{y}, \norm{x})$ is the polar angle in the $\norm{x}-\norm{y}$ plane.

Recall from Eq.~\eqref{eq.2_zetaeta} that $\eta = \text{atan2} (2 \norm{x} \norm{y}, \norm{x}^2 - \norm{y}^2 + 1)$.
Since the two arguments form a vector of magnitude $s^2$, as shown above, the trigonometric functions of $\eta$ are
\begin{subequations}\label{eq.2_trig_eta}
\begin{align}
    \cos\eta & = \frac{\norm{x}^2 - \norm{y}^2 + 1}{s^2},
    \\
    \sin\eta & = \frac{2 \norm{x} \norm{y}}{s^2},
    \\
    \tan\eta & = \frac{2 \norm{x} \norm{y}}{\norm{x}^2 - \norm{y}^2 + 1}.
\end{align}
\end{subequations}

We now express the combinations $\norm{x}^2 \pm \norm{y}^2$ in terms of the coordinates:
\begin{subequations}\label{eq.2_x2my2_r2}
\begin{align}
    \norm{x}^2 - \norm{y}^2 & = s^2 \cos\eta - 1 = \exp{(\zeta)} \cos\eta - 1,
    \\
    \norm{r}^2 = \norm{x}^2 + \norm{y}^2 & = \sqrt{s^4 - 2 s^2 \cos\eta + 1} = \sqrt{2}\,\exp{(\zeta/2)} \sqrt{\cosh\zeta - \cos\eta}.
\end{align}
\end{subequations}

\begin{figure}
    \centering
    \includegraphics[width=15cm]{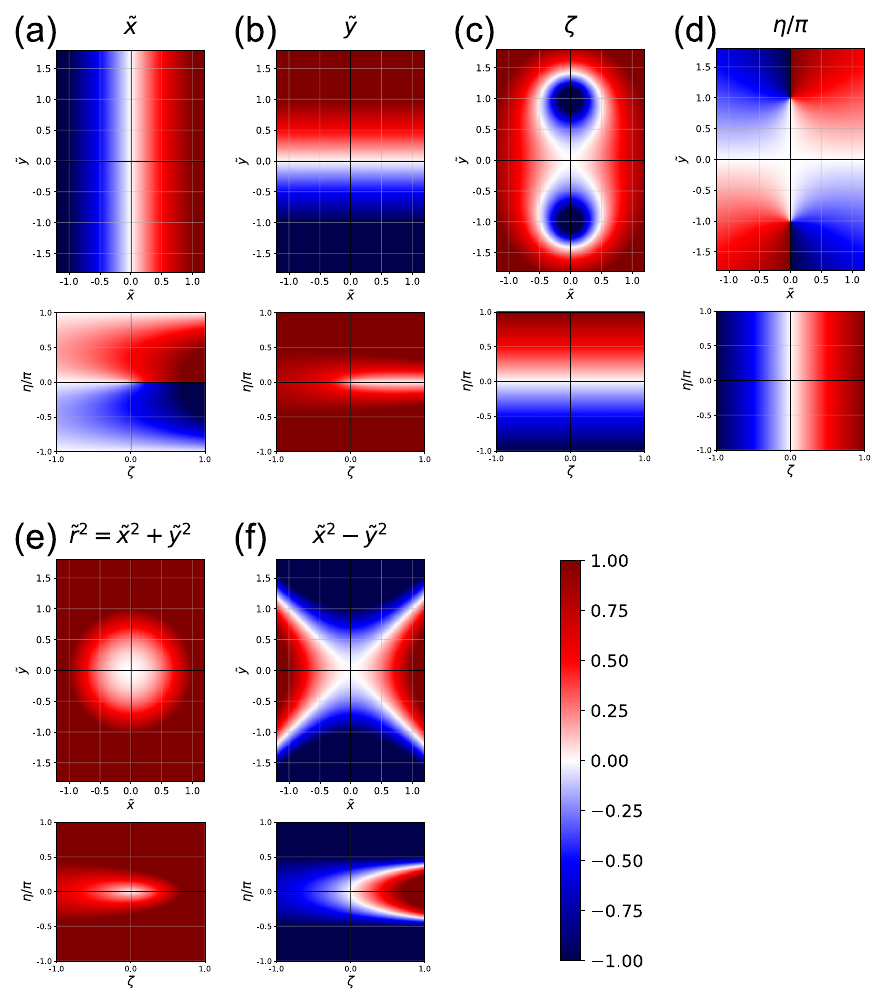}
    \caption[
    TWM coordinate quantities in the $\norm{x}-\norm{y}$ and $\zeta-\eta$ planes
    ]{
    Spatial distribution of (a) $\norm{x}$, (b) $\norm{y}$, (c) $\zeta$, (d) $\eta/\pi$, (e) $\norm{r}^2 = \norm{x}^2 + \norm{y}^2$, and (f) $\norm{x}^2 - \norm{y}^2$.
    For each quantity, the upper panel shows the distribution in the $\norm{x}-\norm{y}$ plane, and the lower panel shows the same distribution in the $\zeta-\eta$ plane.
    }\label{fig.2_3}
\end{figure}

The backward transformation $(\zeta, \eta) \mapsto (\norm{x}, \norm{y})$ then follows from Eq.~\eqref{eq.2_x2my2_r2} as
\begin{subequations}\label{eq.2_backward}
\begin{align}
    \norm{x}(\zeta, \eta) & = \sqrt{ \tfrac{1}{2}\left[ \norm{r}^2 + (\norm{x}^2-\norm{y}^2) \right] } \times \chi \times \text{sgn}(\eta),
    \\
    \norm{y}(\zeta, \eta) & = \sqrt{ \tfrac{1}{2}\left[ \norm{r}^2 - (\norm{x}^2-\norm{y}^2) \right] } \times \chi,
\end{align}
\end{subequations}
where $\chi = \pm 1$ selects the upper ($+$) or lower ($-$) half-plane.
The factor $\text{sgn}(\eta)$ fixes the sign of $\norm{x}$ relative to $\norm{y}$, since $2\norm{x}\norm{y} = s^2 \sin\eta$ implies $\text{sgn}(\norm{x}\norm{y}) = \text{sgn}(\eta)$ whenever $\norm{x}\norm{y} \neq 0$.
For the special case $\eta = 0$ with $\zeta > 0$, the factor $\text{sgn}(\eta)$ vanishes and Eq.~\eqref{eq.2_backward} fails to recover the two corresponding points, $(\norm{x}, \norm{y}) = (\pm\sqrt{s^2 - 1},\, 0)$ on the $\norm{x}$-axis, which are obtained as the limits $\eta \to 0^\pm$.
Figure~\ref{fig.2_3} visualizes several quantities in both the $\norm{x}-\norm{y}$ and $\zeta-\eta$ planes, illustrating the forward and backward transformations.

As an application of these identities, we derive the closed-form expression for the field magnitude, $\norm{B}(\zeta, \eta)$, shown earlier in Eq.~\eqref{eq.2_normB_zetaeta_exact}.
Starting from $\norm{B} = 2\norm{r}/s^2$ in Eq.~\eqref{eq.2_normB} and using $s^2 = \exp{(\zeta)}$ together with the expression for $\norm{r}^2$ in Eq.~\eqref{eq.2_x2my2_r2}, we obtain
\begin{equation}
\begin{aligned}
    \norm{B}(\zeta, \eta)
    = \frac{2 \norm{r}}{s^2}
    & = \frac{2 \left[ \sqrt{2}\,\exp{(\zeta/2)} \left( \cosh\zeta - \cos\eta \right)^{1/2} \right]^{1/2}}{\exp{(\zeta)}}
    \\
    & = 32^{1/4} \exp \left( -\frac{3}{4}\zeta \right) \left( \cosh\zeta - \cos\eta \right)^{1/4}.
\end{aligned}
\end{equation}



\section{Connection coefficients \& Gradient of the TWM field}\label{sec.2.6}

The connection coefficients $\Omega_\zeta$ and $\Omega_\eta$, also known as Ricci rotation coefficients,\cite{Arfken2005} arise when the unit vectors are differentiated:
\begin{equation}\label{eq.2_unitvec_deriv}
\begin{aligned}
    \partial_\zeta \uvecg{\zeta} = \Omega_\zeta \uvecg{\eta},
    \qquad
    \partial_\zeta \uvecg{\eta} = -\Omega_\zeta \uvecg{\zeta},
    \\
    \partial_\eta \uvecg{\zeta} = \Omega_\eta \uvecg{\eta},
    \qquad 
    \partial_\eta \uvecg{\eta} = -\Omega_\eta \uvecg{\zeta}.
\end{aligned}
\end{equation}
Because the TWM coordinate system is curvilinear, its unit vectors are not fixed but rotate from point to point.
The coefficients $\Omega_\zeta$ and $\Omega_\eta$ quantify how rapidly the basis rotates as one moves along the $\zeta$ and $\eta$ directions.

The connection coefficients can be written in several equivalent forms,
\begin{align}
    \Omega_\zeta
    = -\partial_\eta \ln h
    & = \frac{1}{4} \frac{\sin\eta}{\cosh\zeta - \cos\eta}
    = \frac{\norm{x} \norm{y}}{\norm{r}^4}
    \notag \\
    & = \frac{\uvec{b} \cdot \norm{\nabla} \norm{B}}{\norm{B}^2}
    = \frac{(\norm{\nabla} \norm{B})_\eta}{\norm{B}^2},
    \\
    \Omega_\eta
    = \partial_\zeta \ln h
    & = \frac{3}{4} - \frac{1}{4} \frac{\sinh\zeta}{\cosh\zeta - \cos\eta}
    = \frac{1}{2} \left( 1 - \frac{\norm{x}^2 - \norm{y}^2}{\norm{r}^4} \right)
    \notag \\
    & = \frac{(\uvec{b} \times \norm{\nabla} \norm{B})_z}{\norm{B}^2}
    = -\frac{(\norm{\nabla} \norm{B})_\zeta}{\norm{B}^2},
\end{align}
where $\norm{\nabla} = \ell_0 \nabla$ is the normalized gradient operator given in Eq.~\eqref{eq.2_formula_normgrad}.

The connection coefficients can be used to express the divergence and curl in alternative forms.
Using $\partial_\zeta \ln h = \Omega_\eta$ and $\partial_\eta \ln h = -\Omega_\zeta$, the divergence (Eq.~\eqref{eq.2_formula_div}) becomes
\begin{equation}\label{eq.2_div_omega}
\begin{aligned}
    \nabla \cdot \bvec{u}
    & = \frac{1}{h^2} \left[ \partial_\zeta (h u_\zeta) + \partial_\eta (h u_\eta) \right] + \partial_z u_z
    \\
    & = \frac{1}{h} \left( \partial_\zeta u_\zeta + \partial_\eta u_\eta + \Omega_\eta u_\zeta - \Omega_\zeta u_\eta \right) + \partial_z u_z,
\end{aligned}
\end{equation}
and the out-of-plane component of the curl (Eq.~\eqref{eq.2_formula_curl}) becomes
\begin{equation}\label{eq.2_curl_omega}
\begin{aligned}
    (\nabla \times \bvec{u})_z
    & = \frac{1}{h^2} \left[ \partial_\zeta (h u_\eta) - \partial_\eta (h u_\zeta) \right]
    \\
    & = \frac{1}{h} \left( \partial_\zeta u_\eta - \partial_\eta u_\zeta + \Omega_\eta u_\eta + \Omega_\zeta u_\zeta \right).
\end{aligned}
\end{equation}
The in-plane components of the curl contain no scale factor derivatives and are therefore unchanged.

\begin{figure}
    \centering
    \includegraphics[width=15cm]{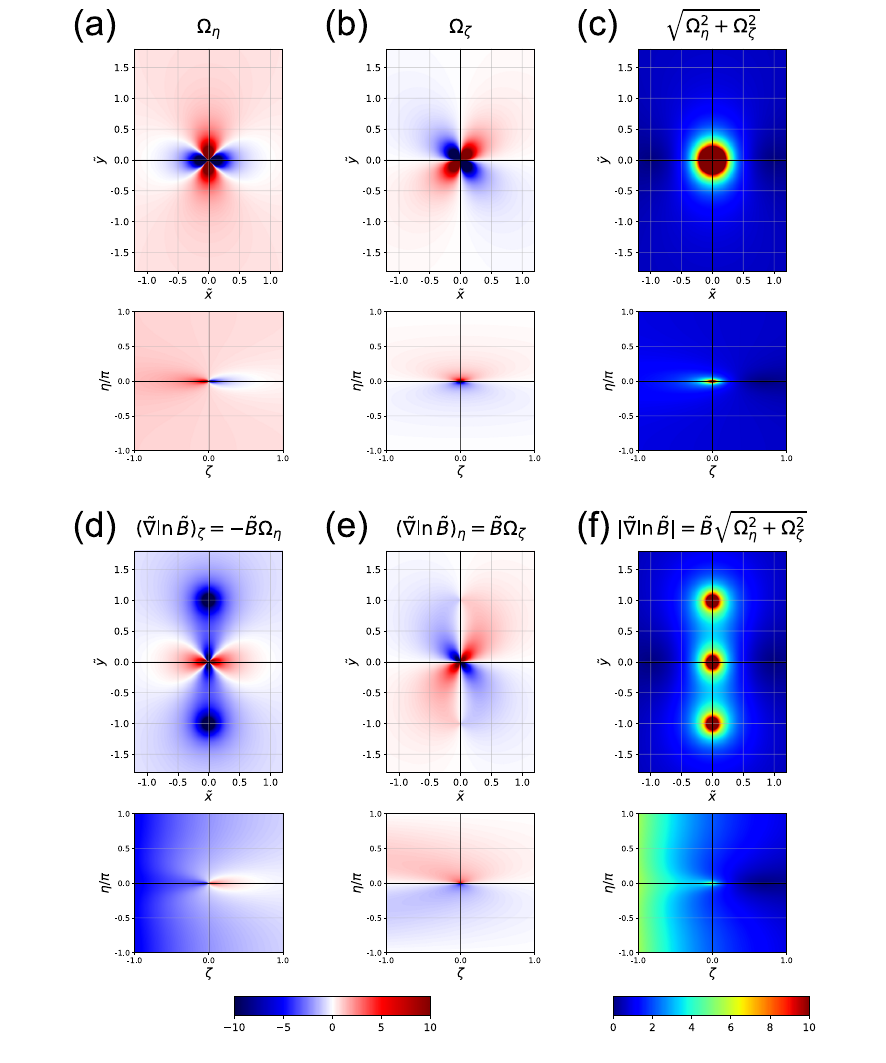}
    \caption[
    Connection coefficients and field magnitude gradient in the $\norm{x}-\norm{y}$ and $\zeta-\eta$ planes
    ]{
    Spatial distribution of (a) $\Omega_\eta$, (b) $\Omega_\zeta$, (c) $\sqrt{\Omega_\eta^2 + \Omega_\zeta^2}$, (d) $(\norm{\nabla} \ln \norm{B})_\zeta = -\norm{B} \Omega_\eta$, (e) $(\norm{\nabla} \ln \norm{B})_\eta = \norm{B} \Omega_\zeta$, and (f) $|\norm{\nabla} \ln \norm{B}| = \norm{B} \sqrt{\Omega_\eta^2 + \Omega_\zeta^2}$.
    For each quantity, the upper panel shows the distribution in the $\norm{x}-\norm{y}$ plane, and the lower panel shows the same distribution in the $\zeta-\eta$ plane.
    Panels (a), (b), (d), and (e) use the diverging color scale, and panels (c) and (f) use the sequential color scale.
    }\label{fig.2_4}
\end{figure}

Using the connection coefficients, the normalized gradient of the normalized field magnitude is
\begin{equation}\label{eq.2_gradB}
    \norm{\nabla} \norm{B}
    = (\norm{\nabla} \norm{B})_\zeta \uvecg{\zeta} + (\norm{\nabla} \norm{B})_\eta \uvecg{\eta}
    = \norm{B}^2 \left[ - \Omega_\eta \uvecg{\zeta} + \Omega_\zeta \uvecg{\eta} \right],
\end{equation}
and its magnitude is given by
\begin{equation}\label{eq.2_gradB_mag}
    \left| \norm{\nabla} \norm{B} \right|
    = \norm{B}^2 \sqrt{\Omega_\eta^2 + \Omega_\zeta^2}
    = \frac{2}{s^2} \sqrt{1 - \frac{4(\norm{x}^2 - \norm{y}^2)}{s^4}}.
\end{equation}
Dividing by $\norm{B}$ and $\norm{B}^2$ gives
\begin{equation}\label{eq.2_gradlnB_mag}
    \left| \frac{\norm{\nabla} \norm{B}}{\norm{B}} \right|
    = \left| \norm{\nabla} \ln \norm{B} \right|
    = \norm{B} \sqrt{\Omega_\eta^2 + \Omega_\zeta^2}
    = \frac{1}{\norm{r}} \sqrt{1 - \frac{4(\norm{x}^2 - \norm{y}^2)}{s^4}},
\end{equation}
\begin{equation}\label{eq.2_gradB_over_B2_mag}
    \left| \frac{\norm{\nabla} \norm{B}}{\norm{B}^2} \right|
    = \left| \frac{\norm{\nabla} \ln \norm{B}}{\norm{B}} \right|
    = \sqrt{\Omega_\eta^2 + \Omega_\zeta^2}
    = \frac{1}{2 \norm{r}^2} \sqrt{s^4 - 4(\norm{x}^2 - \norm{y}^2)}.
\end{equation}
The connection coefficients and $\norm{\nabla} \ln \norm{B}$ are shown in Fig.~\ref{fig.2_4}.



\section{Geometry of the TWM field lines (Cassini ovals)}\label{sec.2.7}

This section examines the geometric properties of the TWM field lines.
The TWM field lines are Cassini ovals labeled by $\zeta = \ln s^2$, as explained in Sec.~\ref{sec.2.2}.
In this section, we mostly refer to the field lines as ovals and use $s^2$ to label them, for convenience.
The oval $s^2=1$ is the separatrix.

\begin{figure}
    \centering
    \includegraphics[width=14cm]{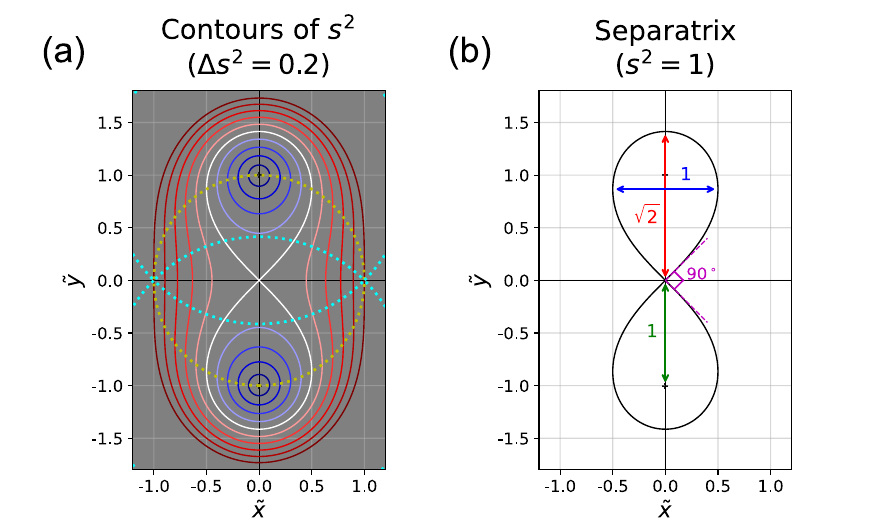}
    \caption[
    Geometric properties of the Cassini ovals
    ]{
    Geometric properties of the Cassini ovals, with the wires (foci) marked by black $+$ symbols at $(\norm{x}, \norm{y}) = (0, \pm 1)$.
    (a) Contours of $s^2 = 0.2, 0.4, \ldots, 1.8, 2.0$ (spacing $\Delta s^2 = 0.2$).
    The yellow dotted curve is the unit circle $\norm{r} = 1$, on which the leftmost and rightmost points of every oval with $s^2 \le 2$ lie.
    The cyan dotted curves are the two circles of radius $\sqrt{2}$ centered at the wires, which together form the contour $\norm{B} = 1$.
    (b) The separatrix ($s^2=1$), the lemniscate of Bernoulli.
    The horizontal width equals $1$ at $\norm{y} = \pm\sqrt{3}/2$ (blue), the vertical maximum is $|\norm{y}|_{\max} = \sqrt{2}$ (red), and the wires lie a distance $1$ from the origin (green).
    At the X-point, the separatrix forms a right angle ($90^\circ$) (magenta).
    }\label{fig.2_5}
\end{figure}

We first determine the spatial extent of an oval in the $\norm{x}$ and $\norm{y}$ directions.
From the identity in Eq.~\eqref{eq.2_s4}, an oval is the level set
\begin{equation}\label{eq.2_FL_levelset}
    (\norm{x}^2 + \norm{y}^2 - 1)^2 + 4\norm{x}^2 = s^4.
\end{equation}
The horizontal extent is found where the tangent is vertical, i.e., where the $\norm{y}$-derivative of Eq.~\eqref{eq.2_FL_levelset} vanishes, which gives $\norm{y}(\norm{r}^2 - 1) = 0$.
This yields two candidate loci: the axis $\norm{y} = 0$ and the unit circle $\norm{r} = 1$.
Substituting $\norm{y} = 0$ gives $|\norm{x}| = \sqrt{s^2 - 1}$, which is real only for $s^2 \ge 1$;
substituting $\norm{r} = 1$ gives $|\norm{x}| = s^2/2$, with $\norm{y}^2 = 1 - s^4/4$, which is real only for $s^2 \le 2$.
Since $s^2/2 \ge \sqrt{s^2 - 1}$ for all $s^2$, with equality at $s^2 = 2$, the maximum horizontal extent is
\begin{equation}\label{eq.2_xmax}
    |\norm{x}|_{\max} =
    \begin{cases}
        s^2/2, & s^2 \le 2, \quad \text{on the unit circle } \norm{r} = 1,
        \\[8pt]
        \sqrt{s^2 - 1}, & s^2 \ge 2, \quad \text{on the $\norm{x}$-axis}.
    \end{cases}
\end{equation}
A notable consequence is that the leftmost and rightmost points of all ovals with $s^2 \le 2$ lie exactly on the unit circle $\norm{r} = 1$, shown as the yellow dotted curve in Fig.~\ref{fig.2_5}(a).
As $s^2$ increases past $2$, the point of maximum $|\norm{x}|$ leaves the unit circle and moves outward along the $\norm{x}$-axis.

The vertical extent is found similarly where the tangent is horizontal, i.e., where the $\norm{x}$-derivative of Eq.~\eqref{eq.2_FL_levelset} vanishes, which gives $\norm{x}(\norm{r}^2 + 1) = 0$.
Since $\norm{r}^2 + 1 > 0$, the only solution is $\norm{x} = 0$, so all vertical extrema lie on the $\norm{y}$-axis.
Setting $\norm{x} = 0$ in Eq.~\eqref{eq.2_FL_levelset} gives $(\norm{y}^2 - 1)^2 = s^4$, hence $\norm{y}^2 = 1 \pm s^2$, so
\begin{equation}\label{eq.2_ymax}
    |\norm{y}|_{\max} = \sqrt{1 + s^2},
    \qquad
    |\norm{y}|_{\min} = \sqrt{1 - s^2} \quad (s^2 < 1).
\end{equation}
The minimum applies only to the inner ovals ($s^2 < 1$, inside the separatrix): each encloses a wire, and the two roots $\sqrt{1 \pm s^2}$ are its two crossings of the $\norm{y}$-axis, one at $|\norm{y}| < 1$ and the other at $|\norm{y}| > 1$.
For the outer ovals ($s^2 > 1$, outside the separatrix), only $\sqrt{1 + s^2}$ is real, and the oval crosses the $\norm{x}$-axis so that its minimum $|\norm{y}|$ is zero.

In addition, we note a special case of the field magnitude contours.
Setting $\norm{B} = 2\norm{r}/s^2 = 1$ gives $s^4 = 4\norm{r}^2$, which combined with Eq.~\eqref{eq.2_FL_levelset} reduces to
\begin{equation}\label{eq.2_Beq1_circles}
    \norm{x}^2 + \norm{y}_\pm^2 = 2.
\end{equation}
The contour $\norm{B} = 1$ therefore consists of two circles of radius $\sqrt{2}$ centered at the wires $(0, \pm 1)$, shown as the cyan dotted curves in Fig.~\ref{fig.2_5}(a) and the cyan solid curves in Fig.~\ref{fig.2_1}(a).

Figure~\ref{fig.2_5}(b) shows the separatrix ($s^2 = 1$), the lemniscate of Bernoulli, with its key dimensions.
Evaluating the extent formulas, Eqs.~\eqref{eq.2_xmax} and \eqref{eq.2_ymax}, at $s^2 = 1$ gives the horizontal extent $|\norm{x}|_{\max} = 1/2$ at $\norm{y} = \pm\sqrt{3}/2$, so each oval spans a full width of $1$.
The vertical maximum is $|\norm{y}|_{\max} = \sqrt{2}$, at the tips on the $\norm{y}$-axis.

The upper and lower ovals of the separatrix meet at the X-point, where the separatrix crosses itself at a right angle ($90^\circ$).
Using the polar form $(\norm{x}, \norm{y}) = \norm{r}(\cos\theta, \sin\theta)$ and $s^2 = 1$, Eq.~\eqref{eq.2_FL_levelset} becomes $\norm{r}^2 = -2\cos 2\theta$.
Therefore, the separatrix approaches the X-point where $\cos 2\theta = 0$, giving $\theta = \pm 45^\circ$ and $\pm 135^\circ$.
These are the diagonal lines $(\norm{y} = \pm\norm{x})$, which are mutually perpendicular.

We next turn to the total arclength and enclosed area of the ovals.
Here, we provide the exact expressions for the total normalized arclength $\norm{L}_{\text{FL}}$ and total normalized enclosed area $\norm{A}_{\text{FL}}$ (subscript FL for ``field line'').
Note again that, for $0<s^2<1$ (equivalently, $\zeta < 0$), a contour consists of two disconnected ovals symmetric about the origin; in this case, $\norm{L}_{\text{FL}}$ and $\norm{A}_{\text{FL}}$ denote the sums over both ovals.

The integrands are the line and area elements derived in Sec.~\ref{sec.2.4}: along an oval ($s^2$ fixed) the arclength element is $d\norm{l} = \norm{h}\,d\eta$, and the area element is $d\norm{A} = \norm{h}^2\,d\zeta\,d\eta$.
The total normalized arclength and enclosed area are then
\begin{subequations}
\begin{align}
    \norm{L}_{\text{FL}}(\zeta_0) = \frac{ L_{\text{FL}}(\zeta_0) }{ \ell_0 }
    & = 2 \int_{-\pi}^{+\pi} \norm{h}(\zeta_0, \eta) \, d\eta
    = 2 \int_{-\pi}^{\pi} \frac{1}{\norm{B}(\zeta_0,\eta)} \, d\eta,
    \label{eq.2_arclength_integral}
    \\[4pt]
    \norm{A}_{\text{FL}}(\zeta_0) = \frac{ A_{\text{FL}}(\zeta_0) }{ \ell_0^2 }
    & = 2 \int_{-\pi}^{\pi} \int_{-\infty}^{\zeta_0} \norm{h}^{2}(\zeta,\eta) \, d\zeta\,d\eta
    = 2 \int_{-\pi}^{\pi} \int_{-\infty}^{\zeta_0} \frac{1}{\norm{B}^{2}(\zeta,\eta)} \, d\zeta\,d\eta,
    \label{eq.2_area_integral}
\end{align}
\end{subequations}
where $\zeta_0 = \ln s^2$ labels the chosen oval, and the factor of $2$ arises because sweeping $\eta$ over its full range covers only one half of the physical plane, as discussed in Sec.~\ref{sec.2.2}.

For compactness, the exact expressions are written as functions of $s^2$, in terms of the complete elliptic integrals, $K$ and $E$.
The expression for the total normalized arclength was given by Tamiozzo:\cite{Tamiozzo2019}
\begin{equation}
\norm{L}_{\text{FL}}(s^2) = 
\begin{cases}
    4 s^2 K\left( \sqrt{ \frac{1-\sqrt{1-s^4}}{2} } \right)
    ,& 0 < s^2 < 1,
    \\[6pt]
    4 \sqrt{s^2} K\left( \sqrt{ \frac{1-\sqrt{1-s^{-4}}}{2} } \right)
    ,& s^2 \ge 1.
\end{cases}
\end{equation}
The total normalized enclosed area is derived analytically as
\begin{equation}
\norm{A}_{\text{FL}}(s^2) = 
\begin{cases}
    2 \left[ E(s^2) + (s^4 - 1) K(s^2) \right]
    ,& 0 < s^2 < 1,
    \\[6pt]
    2s^2 E(s^{-2})
    ,& s^2 \ge 1.
\end{cases}
\end{equation}
Differentiating the enclosed area with respect to $s^2$ gives
\begin{equation}\label{eq.2_dArea}
    \frac{d\norm{A}_{\text{FL}}}{d(s^2)} =
    \begin{cases}
        2 s^2 K(s^2), & 0 < s^2 < 1, \\[4pt]
        2 K(s^{-2}), & s^2 > 1,
    \end{cases}
\end{equation}
which diverges logarithmically at the separatrix, since $K(k) \to \infty$ as $k \to 1$.
This divergence marks the separatrix as a critical contour, across which the inner ovals merge into a single outer oval.
The complete elliptic integrals used above are defined by\cite{Whittaker2021}
\begin{subequations}
\begin{align}
    K(k) & = \int_0^{\pi/2} \frac{1}{\sqrt{1-k^2\sin^2{\theta}}} \, d\theta,
    \\
    E(k) & = \int_0^{\pi/2} \sqrt{1-k^2 \sin^2{\theta}} \, d\theta.
\end{align}
\end{subequations}
For the separatrix, $\norm{L}_{\text{FL}}(s^2=1) = 4 K( 1/\sqrt{2} ) = \sqrt{8} \varpi \approx 7.416$ and $\norm{A}_{\text{FL}}(s^2=1) = 2$, where $\varpi$ is the lemniscate constant.\cite{Whittaker2021}

\begin{figure}
    \centering
    \includegraphics[width=14cm]{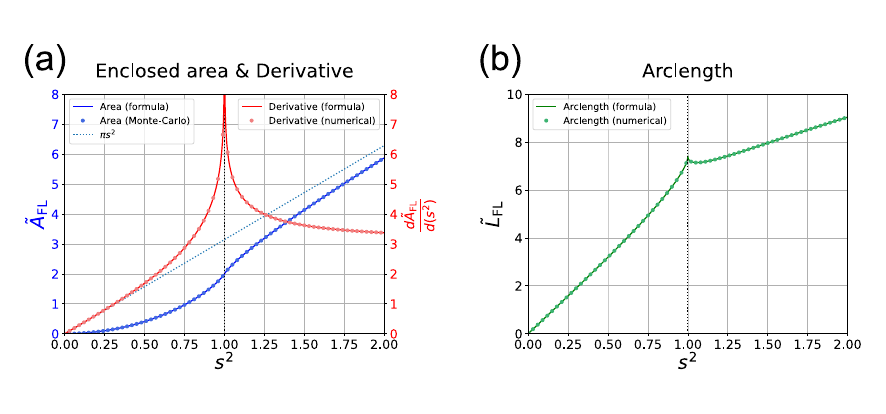}
    \caption[
    Total arclength and enclosed area of the Cassini ovals
    ]{
    (a) Enclosed area $\norm{A}_{\text{FL}}$ (left axis, blue) and its derivative $d\norm{A}_{\text{FL}}/d(s^2)$ (right axis, red), and (b) arclength $\norm{L}_{\text{FL}}$ (green) of the TWM field lines (Cassini ovals) versus $s^2$: analytic expressions (solid) validated against independent numerical results (dots).
    The vertical dotted line marks the separatrix ($s^2 = 1$); the dotted line in (a) is the large-$s^2$ asymptote $\pi s^2$.
    }\label{fig.2_6}
\end{figure}

Figure~\ref{fig.2_6} shows the total normalized enclosed area $\norm{A}_{\text{FL}}$ and arclength $\norm{L}_{\text{FL}}$ as functions of $s^2$, with the analytic expressions validated against independent numerical results.
In both panels, the separatrix $(s^2 = 1)$ is marked by the vertical dotted line, separating the disconnected ovals $(s^2 < 1)$ from the single connected oval $(s^2 > 1)$.
Panel (a) shows the enclosed area $\norm{A}_{\text{FL}}$ (left axis), checked against a Monte-Carlo estimate, in which random points are sampled uniformly over a rectangular region and the enclosed area is the fraction with smaller $s^2$ times the area of that sampling region.
The dotted line $\pi s^2$ is the large $s^2$ asymptote, which $\norm{A}_{\text{FL}}$ approaches as the oval becomes circular.
Panel (a) also shows the derivative of the enclosed area $d\norm{A}_{\text{FL}}/d(s^2)$ (right axis), checked against a finite difference (numerical) derivative of the exact area formula.
The derivative diverges at the separatrix, consistent with Eq.~\eqref{eq.2_dArea}.
Panel (b) shows the arclength $\norm{L}_{\text{FL}}$, checked against direct numerical integration of Eq.~\eqref{eq.2_arclength_integral}, which peaks at the separatrix value $\norm{L}_{\text{FL}}(s^2=1) \approx 7.416$ before decreasing slightly and rising again.



\section{Summary}\label{sec.2.8}

In this chapter, we have presented the TWM magnetic field, which is described by a closed-form analytic expression and offers a tractable setting for studying a true magnetic null.
The first three sections provided the essential material needed for the studies of collisionless (Chapter~\ref{chap.3}) and collisional (Chapter~\ref{chap.4}) dynamics: the TWM field and its normalization (Sec.~\ref{sec.2.1}), the field-aligned coordinate system (Sec.~\ref{sec.2.2}), and the scale factor and differential operators (Sec.~\ref{sec.2.3}).
The remaining sections collected additional information for reference: demonstrations of conformality through complex analysis and the Jacobian (Sec.~\ref{sec.2.4}), a set of coordinate identities and the backward transformation (Sec.~\ref{sec.2.5}), the connection coefficients and the gradient of the field magnitude (Sec.~\ref{sec.2.6}), and the geometric properties of the Cassini ovals (Sec.~\ref{sec.2.7}).


\chapter{Collisionless Dynamics of Charged Particles}\label{chap.3}

The two-wire model (TWM) presented in Chapter~\ref{chap.2} provides a clean setting for studying plasma dynamics near a true magnetic null.
In this chapter, we investigate the collisionless dynamics of charged particles in the TWM, in which the motion of each particle is determined solely by the externally applied TWM field, with no influence from collisions or self-consistent electromagnetic fields induced by other particles.
This idealized regime isolates the role of the X-point geometry, and provides a baseline understanding of charged particle motion.

The central physical question is how the breakdown of magnetic moment conservation near the X-point shapes single particle trajectories.
Far from the X-point, particles are strongly magnetized and gyrate tightly about closed field lines, with the magnetic moment acting as an adiabatic invariant.
Near the X-point, however, the field magnitude weakens and the spatial gradient of the field becomes large, so the magnetic moment is no longer conserved.
Particles passing through this region experience abrupt changes in their gyration, which can lead to a qualitatively different trajectory.
This chapter investigates both the statistical character of the magnetic moment shifts and the cross-separatrix migration that they enable.

The chapter is organized as follows.
Section~\ref{sec.3.1} formulates the single particle equation of motion in the TWM field, identifies the two constants of motion governing the dynamics, and presents the numerical scheme used to integrate particle orbits.
Section~\ref{sec.3.2} characterizes the chaotic shifts of the magnetic moment that occur whenever a particle traverses the X-point region, and shows that the long time statistics of these shifts are determined entirely by the two invariants.
Section~\ref{sec.3.3} investigates the cross-separatrix migration enabled by the magnetic moment shifts, derives a threshold energy for migration from an effective potential analysis, and formulates an empirical expression for the migration confinement time.
Section~\ref{sec.3.4} summarizes the chapter.

The work presented in this chapter has been published as Ahn~\textit{et al.}\cite{Ahn2024}
The notation in this chapter differs from that publication for consistency with the rest of this work.
The field line label was written as $\norm{\psi}$ in the publication, but it is denoted $\zeta$ throughout this work (Chapter~\ref{chap.2}); the two are the same quantity.
The cyclotron frequency is signed, $\omega_{c}=qB/m$, in the publication, but it is taken as unsigned here, $\omega_{c}=|q|B/m$, with the sign of the charge carried separately by $\sigma=q/|q|$.



\section{Single particle motion \& Test particle simulation}\label{sec.3.1}

We begin by formulating the equation of motion for a single particle in the two-wire model (TWM) field, identifying the constants of motion, and describing the numerical scheme used in the rest of the chapter.
The equation of motion is
\begin{equation}\label{eq.3_eqom}
    \frac{d}{dt}\bvec{v} = \frac{q}{m} \left( \bvec{v} \times \bvec{B} \right).
\end{equation}
Here, we introduce additional characteristic scales for a specific particle with charge, $q$, and mass, $m$.
The characteristic cyclotron frequency is $\omega_{c0}=|q|B_0/m$, and the characteristic gyro-period is $\tau_0=2\pi/\omega_{c0}$.
The characteristic speed and energy are $v_0=|q|A_0/m=\ell_0\omega_{c0}$ and $W_0=mv_0^2/2$.
These scales are used to normalize corresponding quantities with the same units.
Using $\omega_{c0}$, the equation of motion becomes
\begin{equation}\label{eq.3_eqom_2}
    \frac{d}{dt}\bvec{v} = \sigma \omega_{c0} \left( \bvec{v} \times \norm{\bvec{B}} \right),
\end{equation}
where $\sigma=q/|q|$ is the sign of the charge.

There are two constants of motion in this system: the total kinetic energy, $W_\text{total}$, and the base field line value, $\zeta_\text{base}$.

The Lorentz force $\bvec{F} = q\bvec{v}\times\bvec{B}$ is perpendicular to $\bvec{v}$, so the magnetic field alone does no work on particles, i.e.,
\begin{equation}
    dW = \bvec{F} \cdot d\bvec{x} = \bvec{F} \cdot \bvec{v}\,dt = 0.
\end{equation}
Therefore, the total kinetic energy, $W_\text{total} = mv^2/2$, is conserved, and so is the normalized total kinetic energy:
\begin{equation}
    \norm{W}_\text{total} = \frac{W_\text{total}}{W_0} = \norm{v}^2.
\end{equation}

The Lagrangian $\mathcal{L}$ of the system is
\begin{equation}
    \mathcal{L} = \frac{m v^2}{2} + q \bvec{v}\cdot\bvec{A} - q\phi = \frac{m v^2}{2} - q v_z A_0 \zeta(x,y),
\end{equation}
where $\bvec{A} = -A_0\zeta(x,y)\uvec{z}$ from Chapter~\ref{chap.2} and $\phi = 0$ are used.
The axial canonical momentum $p_{z,\text{can}}$ is then
\begin{equation}
    p_{z,\text{can}} = \frac{\partial \mathcal{L}}{\partial v_z} = m v_z - q A_0 \zeta(x,y).
\end{equation}
By the Euler-Lagrange equation, i.e.,
\begin{equation}
    \frac{ dp_{z,\text{can}} }{ dt } =
    \frac{ d }{ dt } \left( \frac{\partial \mathcal{L}}{\partial v_z} \right) =
    \frac{\partial \mathcal{L}}{\partial z} = 0,
\end{equation}
we see that $p_{z,\text{can}}$ is a constant of motion in this system.\cite{Kim1983}
We now define the base field line value $\zeta_\text{base}$ using $p_{z,\text{can}}$ as
\begin{equation}
    \zeta_\text{base} = \frac{p_{z,\text{can}}}{-qA_0} =
    \zeta(x,y) - \frac{m}{qA_0} v_z =
    \zeta(x,y) - \frac{v_z}{\sigma v_0}  =
    \zeta(x,y) - \sigma \norm{v}_z,
\end{equation}
which is therefore conserved as well.
Here, we used the fact that $\sigma = \sigma^{-1}$, since $\sigma = q/|q| = \pm 1$.

The two invariants have the following physical interpretations.
The base field line can be treated as the guiding field line that a particle follows and gyrates about, which is distinct from the gyro-center.
The normalized total kinetic energy, $\norm{W}_\text{total}$, governs the travel speed and the Larmor radius.
Hence, these two invariants are used to distinguish particles in the TWM system.

We have built a test particle simulation code in C for the long time integration of charged particle orbits in the TWM field.
The code uses the Boris method,\cite{Qin2013, Wei2015} a second-order explicit leapfrog scheme that splits each time step into three substeps: a half-push by the electric field, a rotation of the velocity vector by the magnetic field, and a final half-push by the electric field.
The full update over one time step is
\begin{subequations}\label{eq.3_boris}
\begin{align}
    \bvec{v}^- & = \bvec{v}^{k-1/2} + \frac{q}{m} \bvec{E}^k \frac{\Delta t}{2},
    \\
    \bvec{v}' & = \bvec{v}^- + \bvec{v}^- \times \uvec{b}^k \tan\left(\theta_c^k/2\right),
    \\
    \bvec{v}^+ & = \bvec{v}^- + \bvec{v}' \times \uvec{b}^k \sin\left(\theta_c^k\right),
    \\
    \bvec{v}^{k+1/2} & = \bvec{v}^+ + \frac{q}{m} \bvec{E}^k \frac{\Delta t}{2},
    \\
    \bvec{x}^{k+1} & = \bvec{x}^k + \bvec{v}^{k+1/2} \Delta t,
\end{align}
\end{subequations}
where $\theta_c^k = \sigma \omega_c^k \Delta t = (q/m) B^k \Delta t$ with $\omega_c^k = |q| B^k/m$, $\uvec{b}^k = \bvec{B}^k/|\bvec{B}^k|$, and the superscript $k$ denotes the time step index.
The Boris method is a volume-preserving integrator,\cite{Qin2013} a property that bounds the energy error globally over arbitrarily many time steps and makes it well suited for the long time integrations performed in this chapter.
In the present collisionless TWM problem, the electric field is absent, so the algorithm conserves the kinetic energy of each particle exactly at every time step, up to floating-point roundoff.

Unless stated otherwise, the test particle simulations in this chapter use the TWM field with $I_0=\SI{1}{kA}$ and $\ell_0=\SI{0.1}{m}$, resulting in $A_0 = \SI{200}{\mu\tesla\meter}$ and $B_0=\SI{2000}{\mu\tesla}$.
Particles are electrons, which give $q/m=\SI{-1.76e11}{C/kg}$, $\tau_0=\SI{17.86}{\nano\second}$, and $W_0=\SI{3517.64}{eV}$.
Note that our results can be readily applied to any charged particles, such as ions, since we use normalized quantities.
The time step of the simulation is set to $\Delta t=\SI{1}{\pico\second}$, which is much smaller than the system gyro-period $\tau_0$, in order to minimize numerical errors.

\begin{figure}
    \centering
    \includegraphics[width=15cm]{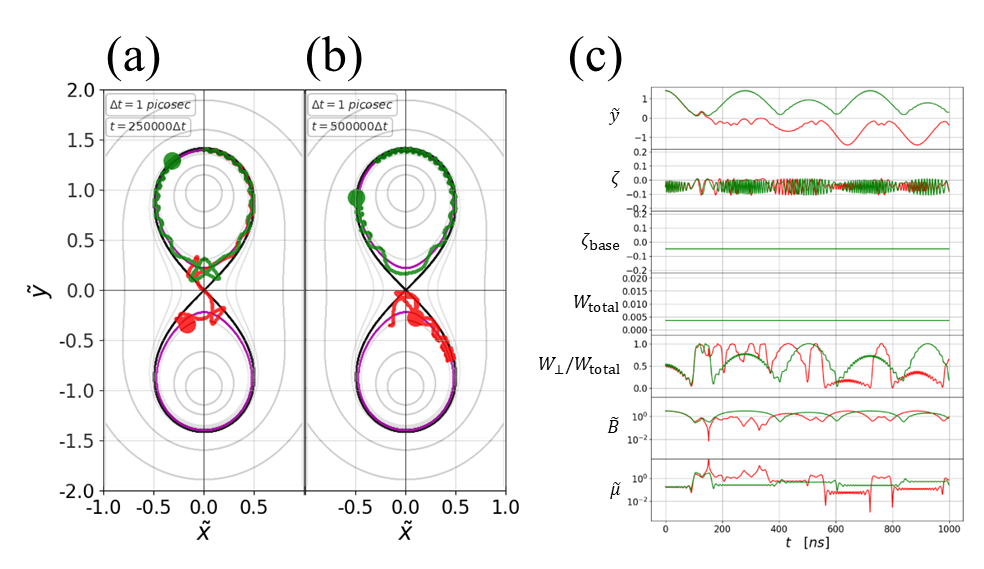}
    \caption[
    Two test particles with marginally different injection angles in the TWM
    ]{
    Two particles with the same invariants ($\zeta_\text{base}=-0.05, \norm{W}_\text{total}=0.0035$) are simulated in the TWM configuration.
    These particles start from the top of $\zeta=-0.05$ with $\tan^{-1}{(v_y/v_x)}=45^\circ$ (red) and $46^\circ$ (green).
    The magenta field line indicates the location of $\zeta_\text{base}=-0.05$.
    Traces of the particles are shown in (a) for $t=0\text{--}250\:\si{ns}$ and in (b) for $t=250\text{--}500\:\si{ns}$.
    Temporal traces of various physical quantities of interest for the red and the green particles are shown in (c) from $t=0$ to $\SI{1000}{ns}$ as red and green lines, respectively.
    }\label{fig.3_1}
\end{figure}

Simulation results are shown in Fig.~\ref{fig.3_1} as an example.
For this case, we have simulated two independent particles that follow the base field line of $\zeta_\text{base}=-0.05$ (the magenta field line) with the total energy of $\norm{W}_\text{total}=0.0035$, whose corresponding dimensional values are $A_0\zeta_\text{base}=\SI{-10}{\mu\tesla\meter}$ and $W_\text{total}=\SI{12.31}{eV}$.
The two particles start their motions from the top of $\zeta=-0.05$ with $\tan^{-1}{(v_y/v_x)}=45^\circ$ (red) and $46^\circ$ (green), respectively.
Figure~\ref{fig.3_1}(a,b) show the traces of the particle motions in the time range of $t=0\text{--}250\:\si{ns}$ and $t=250\text{--}500\:\si{ns}$, respectively.
Figure~\ref{fig.3_1}(c) shows temporal evolutions of various parameters which are, from the top, $\norm y$, $\zeta$, $\zeta_\text{base}$, $\norm{W}_\text{total}$, $W_\perp/W_\text{total}$, $\norm{B}$, and $\norm{\mu}$, where the red and green lines correspond to the red and green particles from Fig.~\ref{fig.3_1}(a,b), respectively.
Here, the normalized magnetic moment $\norm\mu$ is defined as $\norm{\mu}=(W_\perp/W_\text{total})/\norm{B}$, where $W_\perp$ is the kinetic energy perpendicular to the magnetic field.

There are two interesting phenomena to observe in this example, which are the main topics of this chapter, i.e., the magnetic moment shifts and the associated migrations.

First, as can be seen from the last plot of Fig.~\ref{fig.3_1}(c), the particles' normalized magnetic moments $\norm{\mu}$ make seemingly random shifts to various values.
The $\norm{\mu}$ shifts occur whenever the particle travels near the X-point, where the field gradient is large.
This observation, including its sensitivity to the initial injection angle, i.e., the difference between the green and red particles, is further detailed in Sec.~\ref{sec.3.2}.

Second, a particle may jump to follow other closed field lines.
For $\zeta_\text{base}<0$, the base field line is two closed loops separated by the separatrix, as indicated by the magenta lines in Fig.~\ref{fig.3_1}(a,b).
A particle typically gyrates about and travels along the closed field line on one side.
However, if the Larmor radius becomes large enough, due to occasional magnetic moment shifts, the particle can transport to the other closed field line in the TWM configuration, as can be seen with the red particle in Fig.~\ref{fig.3_1}(a).
During this transport, due to the reversal of the magnetic field direction, the particle follows one cycle of the betatron orbit.\cite{Steinhauer2011}
This transport phenomenon is termed \textit{migration} in this work.
The first temporal plot of Fig.~\ref{fig.3_1}(c) shows that the red particle migrates at $t\approx\SI{100}{ns}$, that is, $\norm{y}$ goes from positive to negative, and stays on the negative side.
Note that the green particle does not migrate at least up to $t=\SI{1000}{ns}$ in this example; however, it can eventually migrate to the other side as well, since it satisfies the necessary condition for migration, which is discussed in Sec.~\ref{sec.3.3}.



\section{Magnetic moment shift}\label{sec.3.2}

We have observed that the behaviors of the $\norm\mu$ shift are different for different initial injection angles with the same two invariants (see the $\norm\mu$ traces in Fig.~\ref{fig.3_1}(c)).
This can be explained by an inhomogeneous magnetic field causing chaotic particle orbits,\cite{Numata2002, Numata2003, Buchner1989} which result in different realizations of $\norm{\mu}$.
Because the $\norm{\mu}$ shifts are chaotic, we investigate the shifts statistically.

First, we consider overall consequences of the initial injection angles on the $\norm{\mu}$ shift.
This is done with seven particles whose invariants are the same, i.e., $\zeta_\text{base}=-0.05$ and $\norm{W}_\text{total}=0.002$.
All of them start from the top of $\zeta=-0.05$, but with different initial injection angles from $0^\circ$ to $30^\circ$ with an interval of $5^\circ$.

\begin{figure}
    \centering
    \includegraphics[width=7.5cm]{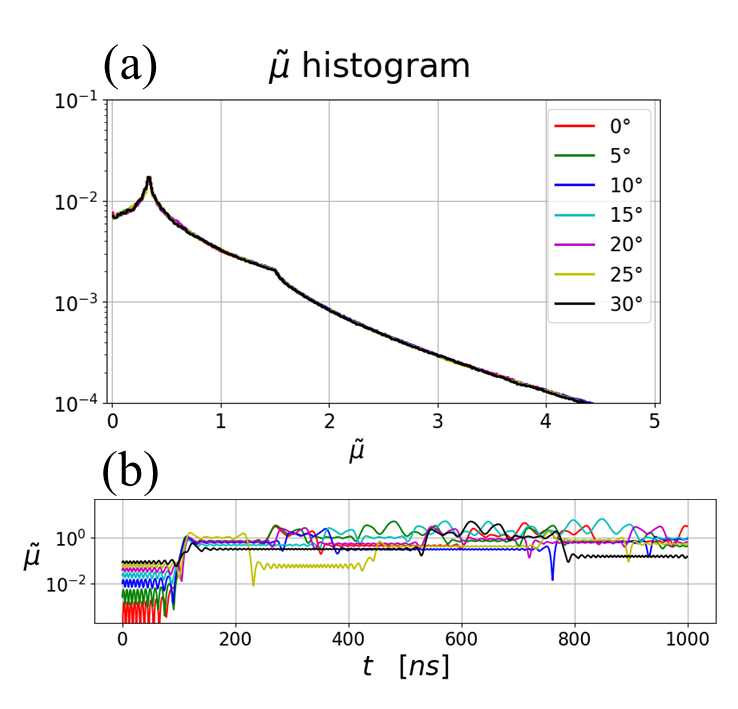}
    \caption[
    Histograms of $\norm{\mu}$ for seven particles with the same invariants and different injection angles
    ]{
    (a) Histograms of the $\norm{\mu}$ for seven particles obtained by simulating them up to $t=\SI{10}{ms}$.
    Invariant pairs of the seven particles are identical, i.e., $\zeta_\text{base}=-0.05$ and $\norm{W}_\text{total}=0.002$.
    They start from the top of $\zeta=-0.05$ having different initial injection angles from $0^\circ$ to $30^\circ$ with a $5^\circ$ interval.
    (b) Each realization of the $\norm\mu$ for the seven particles from $t=0$ to $1000\:\si{ns}$ showing that they are all different while the histograms are the same.
    }\label{fig.3_2}
\end{figure}

Figure~\ref{fig.3_2}(a) shows histograms of the $\norm\mu$ for the seven particles obtained by simulating them up to $t=\SI{10}{ms}$.
It clearly shows that the different initial injection angles all lead to the same histogram, while their individual realizations are different as shown in Fig.~\ref{fig.3_2}(b).
Although not shown here, many more cases with random positions and injection angles ($0$ to $2\pi$) for the same invariant pair are simulated, and the resultant histograms are found to be statistically indistinguishable.
The same independence from initial conditions holds for any invariant pair; the invariant pair alone determines the histogram shape.
Therefore, we can infer that the path, which a particle covers over long times, is independent of the initial condition and distinguished by the invariant pair of the particle.

\begin{figure}
    \centering
    \includegraphics[width=14cm]{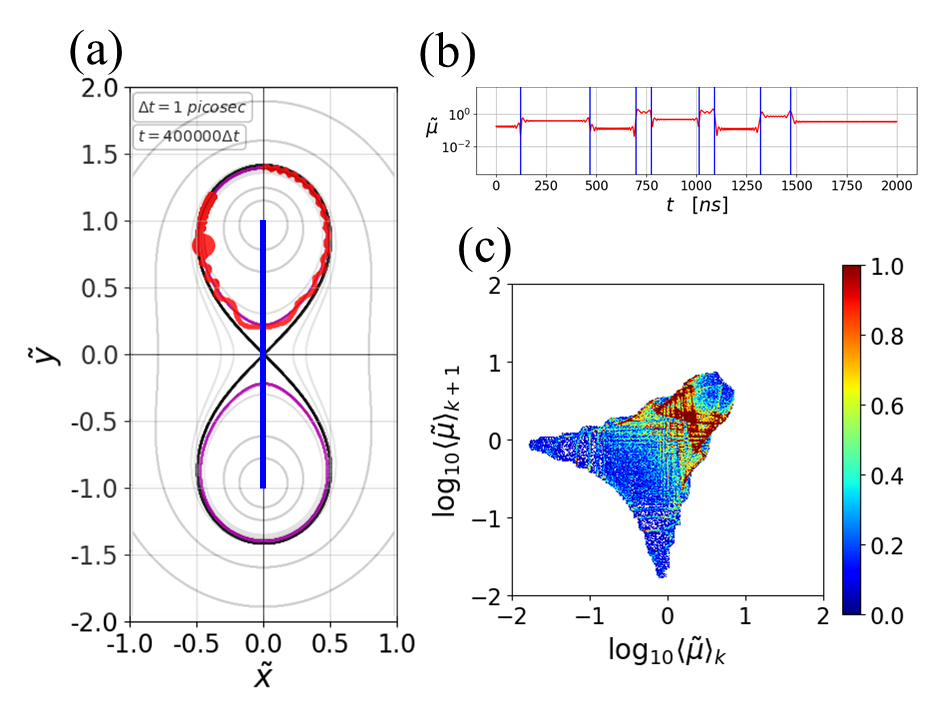}
    \caption[
    Correlation between consecutive $\langle\norm\mu\rangle$ values across the X-point reference line
    ]{
    (a) Trajectory (red) of a single particle starting from the top with $\zeta_\text{base}=-0.05$ (magenta), $\norm{W}_\text{total}=0.00225$ and $\tan^{-1}{(v_y/v_x)}=45^\circ$ from $t=0$ to $400\:\si{ns}$.
    The vertical reference line (blue) connects the positions of the two wires where the spatial gradient of the magnetic field is large.
    (b) Temporal evolution of the $\norm{\mu}$ from $t=0$ to $2000\:\si{ns}$, where the blue vertical lines indicate when the particle crosses the reference line.
    (c) Normalized 2D histogram for $\langle \norm{\mu} \rangle_k$ (abscissa) and $\langle \norm{\mu} \rangle_{k+1}$ (ordinate) indicating the correlation between the two consecutive values of $\langle \norm\mu \rangle$.
    }\label{fig.3_3}
\end{figure}

\begin{figure}
    \centering
    \includegraphics[width=15cm]{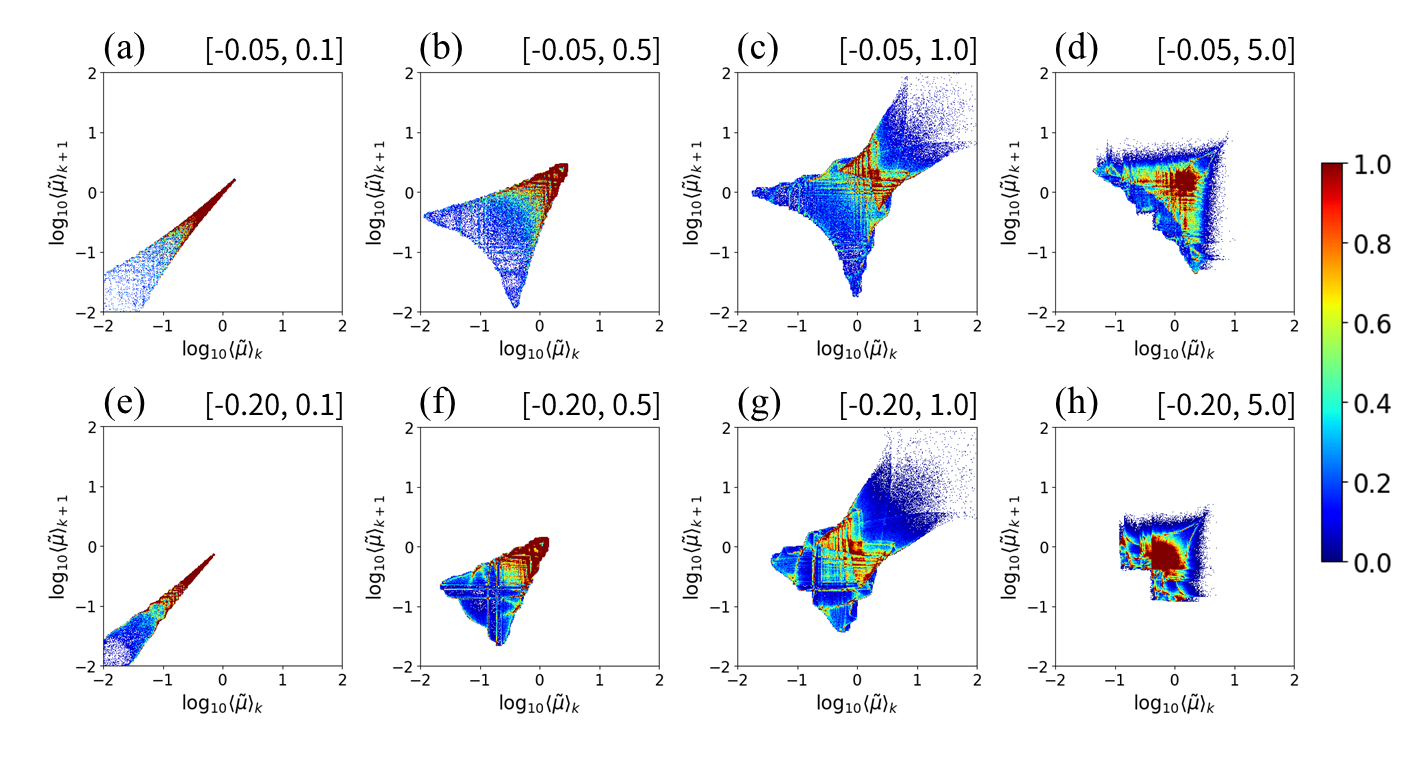}
    \caption[
    Correlation histograms of $\langle\norm\mu\rangle$ for different invariant pairs
    ]{
    Normalized 2D histograms for $\langle \norm{\mu} \rangle_k$ (abscissa) and $\langle \norm{\mu} \rangle_{k+1}$ (ordinate) indicating the correlation between the two consecutive values of $\langle \norm\mu \rangle$ with different pairs of the two invariants denoted as $[\zeta_\text{base}, \norm{W}_\text{total}/\norm{W}_\text{base}]$ on each plot.
    }\label{fig.3_4}
\end{figure}

Next, we statistically investigate how the current value of $\norm\mu$ is correlated with the next one.
Figure~\ref{fig.3_3}(a) shows the trajectory (red) of a single particle having the invariant values of $\zeta_\text{base}=-0.05$ (magenta) and $\norm{W}_\text{total}=0.00225$ with the initial injection angle of $\tan^{-1}{(v_y/v_x)}=45^\circ$.
We find that the $\norm{\mu}$ shifts abruptly whenever the particle passes near the large gradient region around the X-point indicated by the blue line in Fig.~\ref{fig.3_3}(a), which is the reference line connecting the positions of the two wires (see Fig.~\ref{fig.2_4} and Eq.~\eqref{eq.2_gradlnB_mag}).
It can be seen clearly from Fig.~\ref{fig.3_3}(b) that the abrupt shifts of the $\norm\mu$ coincide with the blue vertical lines that indicate the times when the particle passes the reference line.
Thus, we define an expectation value of the $\norm\mu$, denoted as $\langle \norm{\mu} \rangle_k$, to be the arithmetic average of $\norm\mu$ values in-between two consecutive ($k^\text{th}$ and $(k+1)^\text{th}$) crossings of the reference line.

This single particle is simulated up to $\SI{30}{ms}$ to obtain the correlation between $\langle \norm{\mu} \rangle_k$ and $\langle \norm{\mu} \rangle_{k+1}$.
This is visualized as a 2D normalized histogram with $400\times 400$ square bins in Fig.~\ref{fig.3_3}(c).
As attested by the histogram, the behavior of the $\langle \norm\mu \rangle$ shifts is not completely random, indicating that the next $\langle \norm\mu \rangle$ can be probabilistically predicted given the current $\langle \norm\mu \rangle$ despite the stochasticity.
In addition, multiple simulations with the same pair of the invariants are performed to confirm that the resultant histograms are independent of the initial injection positions or angles.

We have also investigated how different sets of the invariants affect the correlation between the current and next $\langle \norm\mu \rangle$.
For this purpose, we introduce a new quantity, the base energy, defined as $\norm{W}_\text{base}=\zeta_\text{base}^2$.
This quantity is the normalized threshold energy for a particle to cross the separatrix, i.e., a particle can cross the separatrix only if $\norm{W}_\text{total}/\norm{W}_\text{base} > 1$.
Its role as the threshold and the physical meaning of $\norm{W}_\text{base}$ are derived and discussed in Sec.~\ref{sec.3.3}.
From here on, we use $\zeta_\text{base}$ and $\norm{W}_\text{total}/\norm{W}_\text{base}$ as a pair of the two invariants to distinguish a particle.

Figure~\ref{fig.3_4} shows the 2D histograms illustrating correlations between the current and next $\langle \norm\mu \rangle$ for different sets of the invariants.
Figure~\ref{fig.3_4}(a)--(d) are of $\zeta_\text{base}=-0.05$, while Fig.~\ref{fig.3_4}(e)--(h) are of $\zeta_\text{base}=-0.20$ with different values of the $\norm{W}_\text{total}/\norm{W}_\text{base}$.
The particles with a lower energy ($\norm{W}_\text{total}/\norm{W}_\text{base}=0.1$) have the sharper histograms (Fig.~\ref{fig.3_4}(a,e)), meaning that the amounts of $\langle \norm\mu \rangle$ shifts are relatively small and the shifts are more predictable.
This is because the particle travels more slowly, so the magnetic field varies more slowly along its path.
The particles with $\norm{W}_\text{total}/\norm{W}_\text{base}=1.0$ move faster and can reach the X-point.
Hence, they can feel more rapid changes of the magnetic field, and so the histograms are wider (Fig.~\ref{fig.3_4}(c,g)).
Therefore, we conclude that in the TWM configuration the magnetic moment shift is more stochastic for a higher energy since such particles travel faster and go closer to the large gradient region.



\section{Migration}\label{sec.3.3}

The field line label $\zeta < 0$ corresponds to two separate closed field lines located above and below the magnetic X-point inside the separatrix, as explained earlier (see magenta lines in Fig.~\ref{fig.3_1}(a,b)).
Migration is defined, in this work, as the phenomenon where a particle jumps to follow the other field line, i.e., the sign of the $\norm{y}$-position of a particle flips, with the same $\zeta$ in the TWM configuration.

We find that the probability of migration depends on the two invariants, and there is a minimum required energy.
These findings can be explained by the trajectory regions that the effective potential renders accessible.\cite{Kim1983}
Since there is no $z$-component of the magnetic field in the TWM configuration, we divide the equation of motion \eqref{eq.3_eqom_2} into $xy$- and $z$-components to derive the effective potential as
\begin{equation}
    \frac{d}{dt}\bvec{v} = \sigma \omega_{c0} \left( \bvec{v}\times\norm{\bvec{B}} \right)
    = \sigma v_0\bvec{v}\times\left( \uvec{z}\times\nabla\zeta \right)
    = \sigma v_0\left( \bvec{v}\cdot\nabla\zeta \right)\uvec{z} - \sigma v_0v_z\nabla\zeta,
\end{equation}
where $\norm{\bvec{B}} = \uvec{z} \times \norm{\nabla} \zeta$ from Eq.~\eqref{eq.2_normB_with_zetaeta} and $\omega_{c0} \ell_0 = v_0$ are used.
Then, taking the $xy$-component, we have
\begin{equation}\label{eq.3_effpotderive}
\begin{aligned}
    \frac{d}{dt}\bvec{v}_{xy}
    & = - \sigma v_0 v_z \nabla\zeta
    = - v_0^2 \left( \zeta-\zeta_\text{base} \right) \nabla\zeta
    \\
    & = - \frac{1}{m}\nabla \left[ \frac{1}{2} m v_0^2 \left( \zeta(x,y)-\zeta_\text{base} \right)^2 \right]
    \\
    & = - \frac{1}{m}\nabla \left[ W_0 \left( \zeta(x,y)-\zeta_\text{base} \right)^2 \right]
    \equiv - \frac{1}{m}\nabla U_\text{eff}(x,y),
\end{aligned}
\end{equation}
where $v_z = \sigma v_0 \left( \zeta-\zeta_\text{base} \right)$ follows from the definition of $\zeta_\text{base}$, and
\begin{equation}\label{eq.3_effpot}
    U_\text{eff}(x,y)
    = W_0 \left( \zeta(x,y)-\zeta_\text{base} \right)^2
    = W_\text{base} \left( \frac{\zeta(x,y)}{\zeta_\text{base}}-1 \right)^2.
\end{equation}
The base energy was defined earlier as $\norm{W}_\text{base}=\zeta^2_\text{base}=W_\text{base}/W_0$.

Note that $U_\text{eff}$ has a direct physical interpretation: since the conservation of $\zeta_\text{base}$ ties $v_z$ to position through $v_z = \sigma v_0 \left( \zeta-\zeta_\text{base} \right)$, the effective potential is simply the axial kinetic energy expressed as a function of position, $U_\text{eff}(x,y) = W_z(x,y) = m v_z^2 / 2$.
Then, the total energy of a particle is
\begin{equation}
\begin{aligned}
    W_\text{total}
    & = W_{xy}(v_x, v_y) + W_z(v_z) 
    \\
    & = W_{xy}(v_x, v_y) + U_\text{eff}(x,y),
\end{aligned}
\end{equation}
which is both the sum of in-plane and axial kinetic energy, and the sum of the in-plane kinetic energy and the effective potential.
Consequently, $U_\text{eff}$ cannot exceed $W_\text{total}$, and a particle's trajectory is restricted to a range of $\zeta$ around $\zeta_\text{base}$, with maximum and minimum bounds, $\zeta_\text{max}$ and $\zeta_\text{min}$, given by
\begin{subequations}\label{eq.3_zeta_maxmin}
\begin{align}
    \zeta_\text{max,min} & = \zeta_\text{base} \pm \sqrt{\norm{W}_\text{total}},
    \\[4pt]
    \frac{\zeta_\text{max,min}}{\zeta_\text{base}} & = 1 \pm \sqrt{\frac{\norm{W}_\text{total}}{\norm{W}_\text{base}}},
\end{align}
\end{subequations}
where, in the second form, the $(+, -)$ signs correspond to $(\zeta_\text{max}, \zeta_\text{min})$ for $\zeta_\text{base} > 0$ and to $(\zeta_\text{min}, \zeta_\text{max})$ for $\zeta_\text{base} < 0$, since dividing by a negative $\zeta_\text{base}$ reverses the ordering.
The equations tell us that a particle with $\zeta_\text{base}<0$ can have a trajectory outside the separatrix only if $\norm{W}_\text{total} > \norm{W}_\text{base}$.
Only in this case can the particle cross the separatrix and the X-point.
Therefore, this is precisely the necessary condition for migration; in other words, $\norm{W}_\text{base}$ is the threshold energy.

\begin{figure}
    \centering
    \includegraphics[width=14cm]{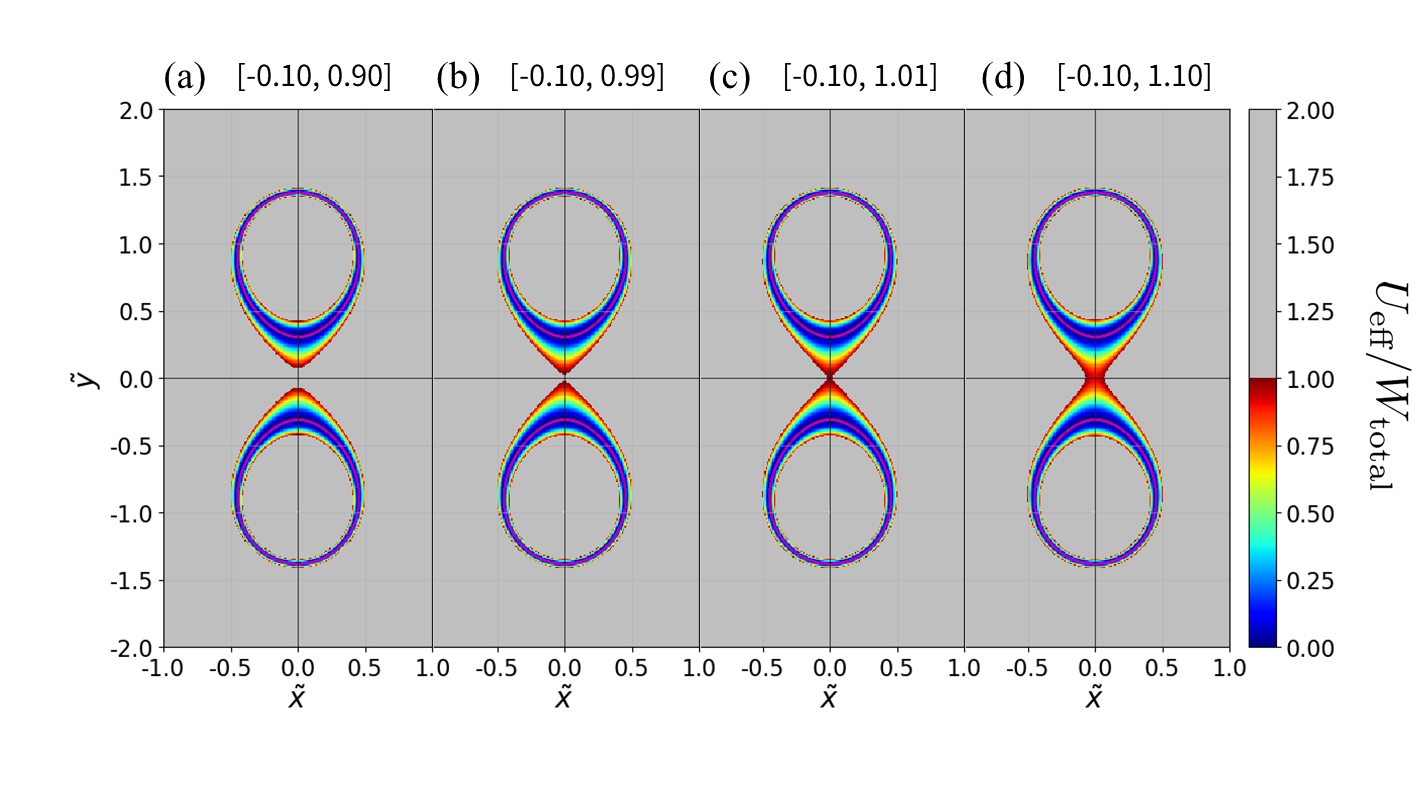}
    \caption[
    Effective potential contours for particles with different kinetic energies
    ]{
        (a)--(d) Contour plots of $U_\text{eff}/W_\text{total}$ with various pairs of invariants denoted as $[\zeta_\text{base}, \norm{W}_\text{total}/\norm{W}_\text{base}]$ on each plot, showing that a particle can migrate to the opposite side of the closed field lines only if $\norm{W}_\text{total}/\norm{W}_\text{base} > 1$.
        Gray regions indicate inaccessible areas for the particle.
        Magenta lines correspond to $\zeta_\text{base}=-0.10$.
    }\label{fig.3_5}
\end{figure}

Figure~\ref{fig.3_5}(a)--(d) are contour plots of $U_\text{eff}/W_\text{total}$ for different energies with the same $\zeta_\text{base}=-0.10$ (magenta lines).
This shows the two bounds, i.e., $\zeta_\text{max}$ and $\zeta_\text{min}$, for the particle motion.
It is impossible for a particle to travel into the region of $U_\text{eff}/W_\text{total} > 1$ depicted as the gray area.
It clearly shows that the accessible area is divided into upper and lower sides, and the two sides become connected when $\norm{W}_\text{total}/\norm{W}_\text{base} > 1$.
This means that migration is only possible when this condition is satisfied, which is the same conclusion discussed above.

A particle with $\norm{W}_\text{total}/\norm{W}_\text{base} > 1$ travels on one side of the TWM configuration, experiences occasional magnetic moment shifts (discussed in Sec.~\ref{sec.3.2}), and can eventually have a Larmor radius large enough to cross the $\norm{y}=0$ axis to migrate to the other side.
This is how the migration occurs, and it is significant for it represents a possibility of collisionless transport loss of particles that reside inside the separatrix.

\begin{figure}
    \centering
    \includegraphics[width=14cm]{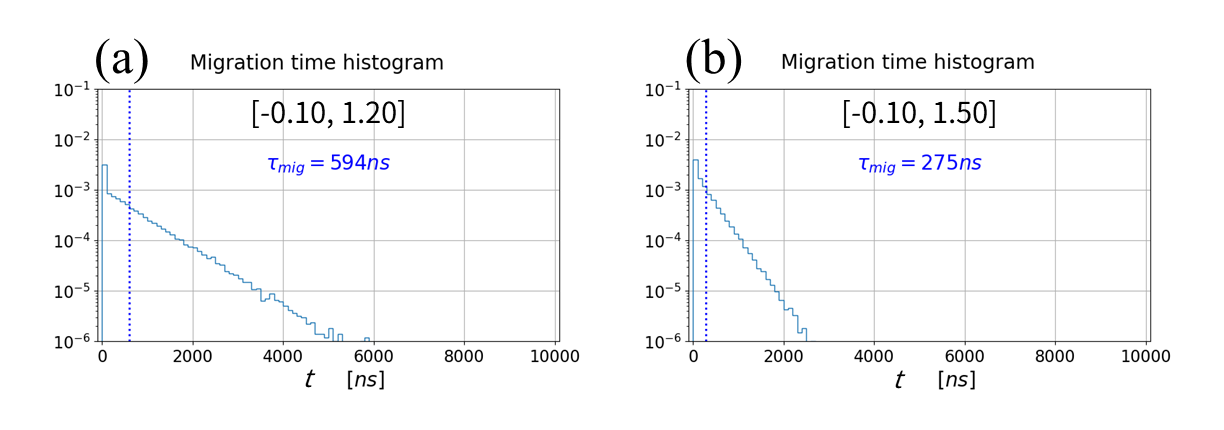}
    \caption[
    Histograms of migration times for two invariant pairs
    ]{
        Histograms of the times a particle spent in-between consecutive migrations for the invariant pair $[\zeta_\text{base}, \norm{W}_\text{total}/\norm{W}_\text{base}]$ of (a) $[-0.10, 1.20]$ and (b) $[-0.10, 1.50]$.
        The migration confinement time, $\tau_\text{mig}$, which is the average value from the histogram, is depicted in (a) and (b) with blue vertical dotted lines.
    }\label{fig.3_6}
\end{figure}

We have recorded the times a particle spent in-between consecutive migrations, and the resulting histograms are shown in Fig.~\ref{fig.3_6} for $[\zeta_\text{base}, \norm{W}_\text{total}/\norm{W}_\text{base}]$ equal to (a) $[-0.10, 1.20]$ and (b) $[-0.10, 1.50]$.
As attested by the figure, the times are nearly exponentially distributed, indicating that migration is an almost memoryless phenomenon.
There exists a sharp outlier at very short times, which may be caused by the gyrating orbits of a particle very close to the X-point.
The average time is termed the migration confinement time, $\tau_\text{mig}$, and we have $\tau_\text{mig}=594\:\si{ns}$ and $275\:\si{ns}$ for Fig.~\ref{fig.3_6}(a,b), respectively.
This quantity represents how probable migration is for a particle with a certain pair of the invariants.
We see that the migration confinement time is shorter for a particle with a higher energy.

We have seen that collisionless transport is possible through the magnetic moment shift, when a particle with sufficient energy travels into the region of large magnetic field gradient around the X-point.
Thus, we have estimated the normalized migration confinement time $\norm{\tau}_\text{mig}=\tau_\text{mig}/\tau_0$ for 600 different pairs of the invariants so that the overall quantitative behavior of $\tau_\text{mig}$ can be empirically formulated.
We have simulated a particle with six values of $\zeta_\text{base}$, which are $-0.005, -0.010, -0.020, -0.050, -0.100,$ and $-0.200$.
For each $\zeta_\text{base}$, we sample 100 different values of the excess energy $\norm{W}_\text{total}-\norm{W}_\text{base}$, from $0.01\norm{W}_\text{base}$ to unity.

\begin{figure}
    \centering
    \includegraphics[width=15cm]{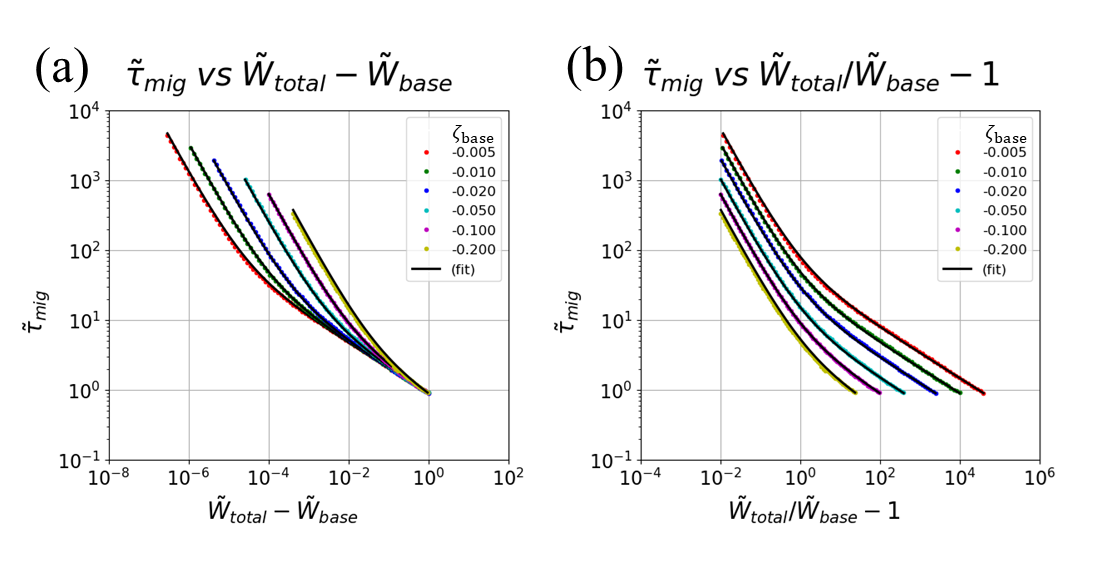}
    \caption[
    Estimated migration confinement time from 600 invariant pairs
    ]{
    Estimated normalized migration confinement time $\norm{\tau}_\text{mig}$ from 600 different pairs of $\zeta_\text{base}$ and $\norm{W}_\text{total}$ as a function of (a) $\norm{W}_\text{total}-\norm{W}_\text{base}$ and (b) $\norm{W}_\text{total}/\norm{W}_\text{base}-1$.
    We have simulated a particle with six different values of $\zeta_\text{base}$, which are $-0.005, -0.010, -0.020, -0.050, -0.100,$ and $-0.200$.
    For each $\zeta_\text{base}$, we have 100 different values of $\norm{W}_\text{total}$.
    Black lines are the fitted lines obtained by the empirical expression \eqref{eq.3_empexp}.
    }\label{fig.3_7}
\end{figure}

Figure~\ref{fig.3_7} shows the estimated $\norm{\tau}_\text{mig}$ as a function of (a) $\norm{W}_\text{total}-\norm{W}_\text{base}$, i.e., how much excess energy a particle has, and (b) $\norm{W}_\text{total}/\norm{W}_\text{base}-1$, i.e., how much fractional excess energy with respect to the base energy $\norm{W}_\text{base}$ (or the threshold energy) a particle has.
We now construct an empirical expression for $\norm{\tau}_\text{mig}$ based on the numerical results.
As can be seen from Fig.~\ref{fig.3_7}, the trend follows two different power laws.
First of all, there seems to be, from Fig.~\ref{fig.3_7}(a), a converging power law at large $\norm{W}_\text{total}-\norm{W}_\text{base}$ with different values of $\zeta_\text{base}$; thus, we introduce a $\kappa_0 (\norm{W}_\text{total}-\norm{W}_\text{base})^{\kappa_1}$ factor in the formula.
Second, from Fig.~\ref{fig.3_7}(b) we observe a similar slope for different values of $\zeta_\text{base}$ for small $\norm{W}_\text{total}/\norm{W}_\text{base}-1$ with different offsets.
Therefore, we include a $[({\norm{W}_\text{total}}/{\norm{W}_\text{base}}-1)^{\kappa_2} - \norm{W}_\text{base}^{\kappa_3} + 1]$ factor in the formula as well.
Together, we have
\begin{equation}\label{eq.3_empexp}
\begin{aligned}
    & \norm{\tau}_\text{mig} \left( \zeta_\text{base}, \norm{W}_\text{total} \right)=
    \kappa_0 \left(\norm{W}_\text{total}-\norm{W}_\text{base}\right)^{\kappa_1}
    \times \left[ \left(\frac{\norm{W}_\text{total}}{\norm{W}_\text{base}}-1\right)^{\kappa_2} - \norm{W}_\text{base}^{\kappa_3} + 1\right],
    \\
    & \qquad\qquad \kappa_0 = 0.90, \qquad \kappa_1=-0.36,
    \qquad  \kappa_2=-0.70, \qquad \kappa_3=0.70,
\end{aligned}
\end{equation}
where the fitting parameters $\kappa_0, \kappa_1, \kappa_2$, and $\kappa_3$ are determined by fitting to the numerically obtained $\norm{\tau}_\text{mig}$.
The black lines in Fig.~\ref{fig.3_7}(a,b) are the fitted lines, which show good agreement with the data.

\begin{figure}
    \centering
    \includegraphics[width=9cm]{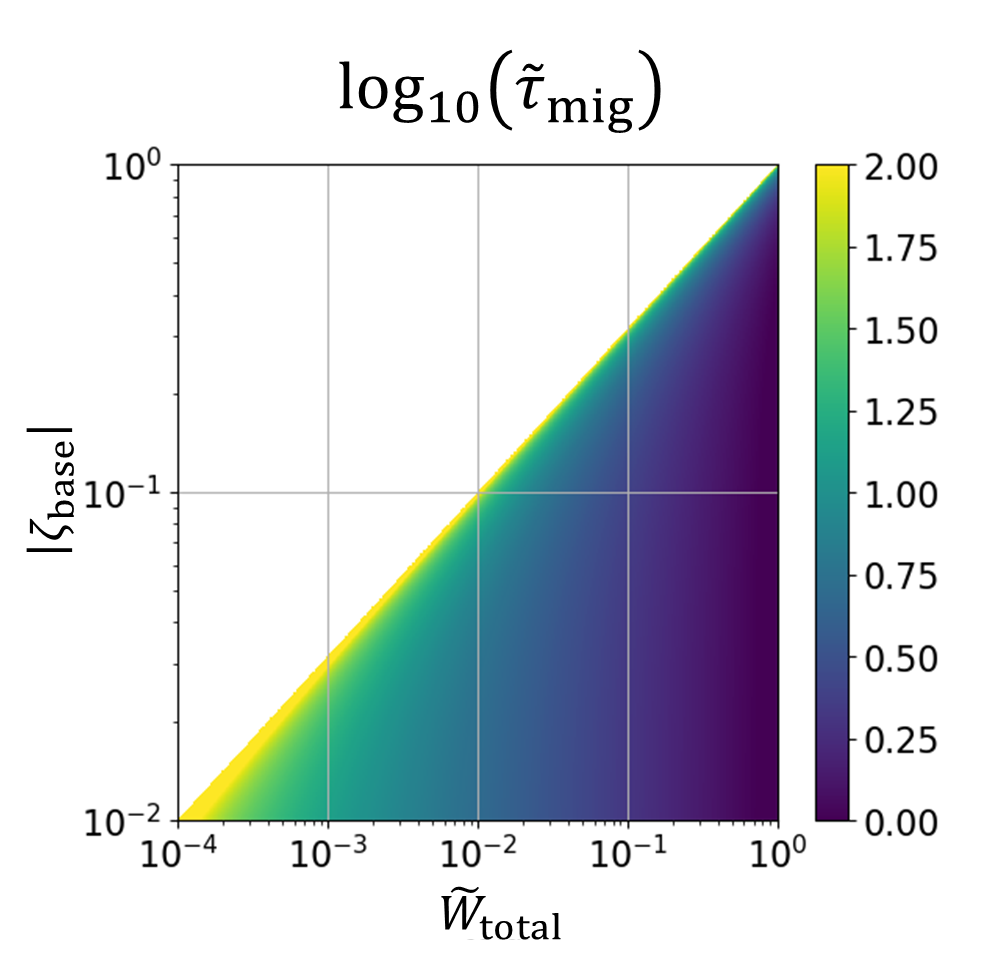}
    \caption[
    Migration confinement time as a function of the two invariants
    ]{
    $\norm{\tau}_\text{mig}$ as a function of $|\zeta_\text{base}|$ and $\norm{W}_\text{total}$ obtained by the empirical expression \eqref{eq.3_empexp}, showing the quantitative trend of $\norm{\tau}_\text{mig}$ with the two invariants of a particle.
    }\label{fig.3_8}
\end{figure}

Equipped with the empirical expression \eqref{eq.3_empexp}, we show the quantitative trend of $\norm{\tau}_\text{mig}$ as a function of $\zeta_\text{base}$ and $\norm{W}_\text{total}$ in Fig.~\ref{fig.3_8}.
Note that we use the absolute value of $\zeta_\text{base}$ since the confined particles have $\zeta_\text{base}<0$.
It is clear that $\norm{\tau}_\text{mig}$ is finite only when $\norm{W}_\text{total} > \norm{W}_\text{base} = \zeta_\text{base}^2$, as a particle cannot migrate otherwise.
The migration confinement time, $\norm{\tau}_\text{mig}$, decreases with increasing $\norm{W}_\text{total}$ and decreasing $|\zeta_\text{base}|$; both trends are consistent with our intuition.
We emphasize that our results can be readily applied to any charged particle species since we have used normalized quantities.

\begin{figure}
    \centering
    \includegraphics[width=15cm]{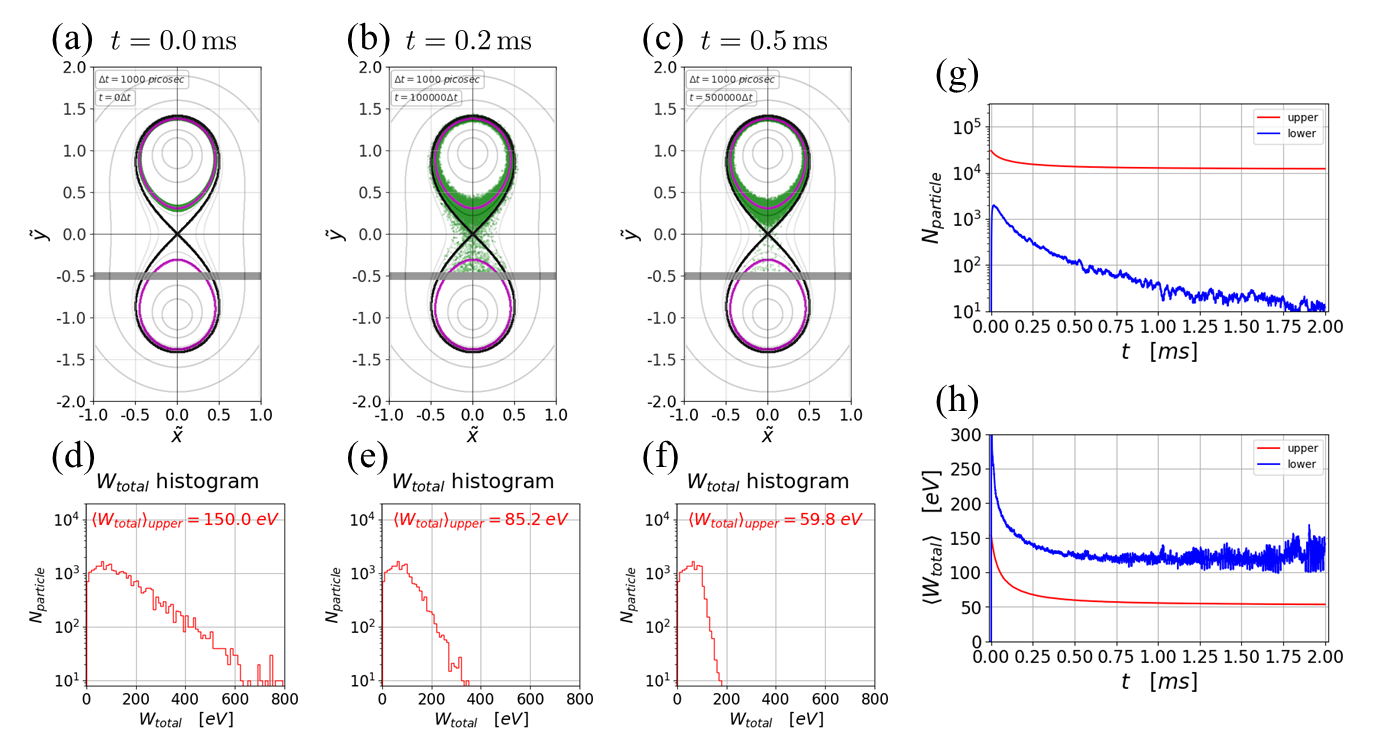}
    \caption[
    Collisionless migration losses of deuterons across a virtual divertor
    ]{
    An example of particle losses due to the collisionless migration phenomenon with the loss boundary, acting as a divertor in a tokamak, located at $\norm{y}=-0.5$ (thick gray horizontal line).
    Initially, 30,000 Maxwellian deuterons (green dots) are randomly placed on $\zeta = -0.1$ (magenta line) with $v_z = 0$ on the upper side of the TWM configuration, so that they share the same invariant, $\zeta_\text{base} = -0.1$.
    (a), (b), and (c) show the spatial distribution of deuterons at times $t=0.0, 0.2,$ and $0.5\:\si{ms}$, respectively.
    (d), (e), and (f) are the histograms of $W_\text{total}$ at these times.
    (g) and (h) show temporal evolutions of the number of particles and the average energy $\langle W_\text{total} \rangle$, respectively, with red for the upper side and blue for the lower side.
    }\label{fig.3_9}
\end{figure}

This migration phenomenon causes particles to become completely lost from the system if there is a loss boundary on the other side of the TWM configuration such as a divertor in a tokamak.
To investigate how relevant this collisionless transport can be, we create an environment similar to that of a tokamak, although crude since there are no axial (or toroidal) magnetic fields.
The characteristic parameters are $I_0=\SI{100}{kA}$ and $\ell_0=\SI{1.0}{m}$, with a deuteron species ($q/m=\SI{4.79e7}{C/kg}$).
We then have $\tau_0=\SI{6555.33}{\nano\second}$ and $W_0=\SI{9584.85}{eV}$, and so $\Delta t=\SI{1}{\nano\second}$ is used for this simulation.
We randomly place 30,000 deuterons on the upper side of $\zeta_\text{base}=-0.1$ with in-plane isotropically distributed initial injection angles.
These ions are initialized with a Maxwellian distribution having $T=(2/3)\langle W_\text{total} \rangle = \SI{100}{eV}$ (see Fig.~\ref{fig.3_9}(d)).
Since all the ions have $\zeta_\text{base}=-0.1$, we have $W_\text{base}=\SI{95.85}{eV}$ which means only some of the deuterons with high enough energy may migrate to the lower side and eventually be lost to the loss boundary (virtual divertor) located at $\norm{y}=-0.5$.

Figure~\ref{fig.3_9}(a)--(c) show the spatial distribution of deuterons (green dots) at times $t=0.0, 0.2,$ and $0.5\:\si{ms}$, respectively.
Figure~\ref{fig.3_9}(d)--(f) show the $W_\text{total}$ histograms at these times.
It is clear that as time progresses, the high energy tail of the distribution is quickly lost to the virtual divertor, since high energy particles have shorter migration confinement times.
This can also be seen from the decreasing average energy of the particles denoted as $\langle W_\text{total} \rangle$.
By the time $t=\SI{0.5}{ms}$, most of the deuterons with $W_\text{total} > W_\text{base}$ have migrated to the lower side, and only the low energy deuterons remain confined on the upper side.
Figure~\ref{fig.3_9}(g,h) show temporal evolutions of the number of particles and $\langle W_\text{total} \rangle$, respectively, for the upper (red) and lower (blue) sides of the TWM configuration.
We note that even at around $t=\SI{2.0}{ms}$, there exist particles migrating to the lower side due to the random nature of the phenomenon.
Through this example, we can see that even if particles follow the base field line inside the separatrix, they may migrate to the other side and become lost to a boundary.

As mentioned earlier, this result may not be directly applied to tokamak edge physics as the system lacks not only the axial (toroidal) magnetic fields but also collisions which restore a Maxwellian distribution.
Nevertheless, as a typical transport time and an ion collision time in a tokamak are of the order of a few to tens of $\si{ms}$,\cite{Wesson2011} we see that the collisionless transport time scale associated with the migration is much faster.
If axial magnetic fields had been included, they would have slowed migration transport, while collisions would have enhanced it.
Thus, the migration may play a non-negligible role in determining an edge transport time scale as well as a radial profile in the scrape-off-layer region of a tokamak.



\section{Summary}\label{sec.3.4}

In this chapter, we have numerically investigated collisionless single charged particle motion in the TWM configuration, focusing on the magnetic moment shift and the associated migration transport near the X-point, where the conservation of the magnetic moment breaks down due to the large spatial gradient of the magnetic field magnitude.

From the Lagrangian analysis, two system invariants are identified: the total kinetic energy, $W_\text{total}$, and the base field line value, $\zeta_\text{base}$, derived from the conserved axial canonical momentum.
We have found that although particles with the same invariant pair and different initial conditions produce different realizations of chaotic $\norm{\mu}$ shifts, their long time trajectories converge to the same statistical distribution.
The current and next $\langle\norm{\mu}\rangle$ values are statistically correlated, and the joint distribution is completely determined by the invariant pair.

When a $\norm{\mu}$ shift produces a Larmor radius large enough for a particle to cross the separatrix, the particle migrates to the corresponding branch of its base field line on the other side of the X-point.
Using the effective potential analysis, we have derived a necessary condition for migration: the total energy of a particle must exceed the threshold energy, $W_\text{base} = W_0 \zeta_\text{base}^2$.
The inter-migration time is exponentially distributed, indicating that migration is a memoryless random process, and the migration confinement time $\tau_\text{mig}$ depends only on the invariant pair.
An empirical expression for $\norm{\tau}_\text{mig}(\zeta_\text{base}, \norm{W}_\text{total})$ has been formulated from simulations of 600 different invariant pairs.
A tokamak-like demonstration shows that migration preferentially loses high energy particles from inside the separatrix, on a time scale much faster than typical tokamak transport and collision times.
The migration phenomenon investigated here is therefore a candidate mechanism contributing to edge transport, though direct application to tokamak edge physics requires an axial guide field and finite collisions.

Beyond the TWM itself, the framework developed in this chapter is relevant to any plasma system in which a true magnetic null governs particle dynamics, including magnetic confinement experiments and magnetic reconnection sites.
The identification of the controlling invariants, the breakdown of magnetic moment conservation near the null, and the threshold energy and exponential migration statistics provide a minimal geometric and statistical baseline against which the effects present in real systems, such as collisions, guide fields, and self-consistent fields, can be assessed.


\chapter{Collisional Dynamics of Magnetized Low Temperature Plasmas}\label{chap.4}

Chapter~\ref{chap.3} investigated the collisionless dynamics of charged particles in the two-wire model (TWM), where each particle moves under the externally applied field alone.
In this chapter, we turn to the collisional regime, in which a plasma fills the TWM configuration, collisions and self-consistent electric fields shape the dynamics, and the description shifts from individual orbits to macroscopic transport.
The physical question is how the presence of a true magnetic null affects the cross-field transport and the steady-state density profile of a magnetized plasma confined by the TWM field.

Collisional cross-field transport in a magnetized plasma is strongly suppressed compared to parallel transport, with the degree of suppression set by the local magnetization.
At the X-point, however, the field magnitude vanishes and particles are locally unmagnetized, so the cross-field transport suppression is locally relaxed.
As we will show, this localized relaxation can affect the density profile over a much wider region due to rapid parallel equilibration along closed field lines.

To analyze the collisional dynamics in a physically intuitive way, we adopt a classical drift-diffusion model (DD model),\cite{Hagelaar2007,Chen2016} which captures the essential anisotropy between parallel and cross-field transport in a magnetized plasma without invoking the kinetic detail of full distribution functions.
The DD model expresses the cross-field particle flux as a product of the unmagnetized flux and a magnetization reduction factor that depends on the local field magnitude, which makes the role of the magnetic field geometry transparent.
We derive the DD model and reduce it to one-dimensional models by applying $\eta$-averaging along the field lines and a fast parallel equilibration closure.
The reduced models are used to analyze the cross-field density profile, and they predict density plateau formation near the separatrix in the strongly magnetized regime.
The predictions are then compared with particle-in-cell (PIC) simulation results, and the dependence of the plateau on the magnetization is confirmed by a wire current scan.

The chapter is organized as follows.
Section~\ref{sec.4.1} formulates the DD model in the TWM coordinate system and identifies a sharp localized relaxation of the cross-field transport suppression at the X-point.
Section~\ref{sec.4.2} derives the reduced DD models by applying $\eta$-averaging along the field lines and a fast parallel equilibration closure.
Section~\ref{sec.4.3} uses the reduced models to analyze the cross-field density profile and predicts density plateau formation near the separatrix in the strongly magnetized regime.
Section~\ref{sec.4.4} presents two-dimensional electrostatic PIC simulations of the TWM configuration performed with the EDIPIC-2D code and identifies the density plateau formation in the steady-state profiles.
Section~\ref{sec.4.5} reduces the PIC data to $\eta$-averaged profiles for a quantitative comparison with the reduced models, and examines the dependence of the plateau on the magnetization through a wire current scan.
Section~\ref{sec.4.6} summarizes the chapter.

The work presented in this chapter has been submitted as Ahn~\textit{et al.}\cite{Ahn2026}
The presentation here is expanded with additional detail and derivations for completeness.



\section{Drift-diffusion model \& Magnetization structure}\label{sec.4.1}

We begin with the momentum equation for a plasma species,
\begin{equation}\label{eq.4_mom_eq}
    m n \left( \partial_t \bvec{u} + \bvec{u} \cdot \nabla \bvec{u} \right)
    = - \nabla p + n q \left( \bvec{E} + \bvec{u} \times \bvec{B} \right)
    - m n \nu \bvec{u},
\end{equation}
where $m$, $q$, and $n$ are the particle mass, charge, and density, respectively, $\bvec{u}$ is the flow velocity, $\bvec{E}$ and $\bvec{B}$ are electric and magnetic fields, $p$ is the scalar pressure, and $\nu$ is the momentum transfer collision frequency with neutrals.

Under the standard drift-diffusion approximation for collisional magnetized plasmas, a steady-state is assumed and the inertial terms on the left-hand side of Eq.~\eqref{eq.4_mom_eq} are neglected:\cite{Hagelaar2007, Chen2016}
\begin{equation}\label{eq.4_mom_eq_2}
    0
    = - \nabla p + n q \left( \bvec{E} + \bvec{u} \times \bvec{B} \right)
    - m n \nu \bvec{u}.
\end{equation}

In this work, we make the following additional assumptions:
all physical quantities are uniform along the $z$-direction, so that $\partial_z = 0$;
the magnetic field is prescribed by the externally applied static two-wire model (TWM) field;
the electric field is given by $\bvec{E}(x,y) = -\nabla \phi$, where $\phi(x,y)$ is the electrostatic potential determined self-consistently by the charge density and boundary conditions;
and the temperature, $T$, and the collision frequency, $\nu$, are spatially uniform, with the scalar pressure given by $p = nT$.
Applying these assumptions to Eq.~\eqref{eq.4_mom_eq_2} yields
\begin{equation}\label{eq.4_mom_eq_3}
    0 = -T \nabla n - n q \nabla \phi + n q \bvec{u} \times \bvec{B} - m n \nu \bvec{u}.
\end{equation}

We further restrict the drift-diffusion model (DD model) to plasma states assumed to be symmetric about the origin.
Therefore, all quantities considered in the DD model are taken to respect this symmetry, and the upper and lower half-planes need not be distinguished, avoiding the complication from the two-to-one TWM coordinate map $(\norm{x}, \norm{y}) \mapsto (\zeta, \eta)$ (see Chapter~\ref{chap.2}).

We now decompose the momentum equation, Eq.~\eqref{eq.4_mom_eq_3}, into the $\zeta$, $\eta$, and $z$ components of the TWM coordinate system.
Since $\bvec{B} = B \uvecg{\eta}$, the orthonormal basis relation $(\uvecg{\zeta} \times \uvecg{\eta} = \uvec{z})$ gives the components of $\bvec{u} \times \bvec{B}$ as
\begin{subequations}
\begin{align}
    (\bvec{u} \times \bvec{B})_\zeta &= -u_z B,
    \\
    (\bvec{u} \times \bvec{B})_\eta &= 0,
    \\
    (\bvec{u} \times \bvec{B})_z &= u_\zeta B.
\end{align}
\end{subequations}
Together with $\partial_z = 0$, which gives $(\nabla n)_z = (\nabla \phi)_z = 0$, the three components of Eq.~\eqref{eq.4_mom_eq_3} are
\begin{subequations}
\begin{align}
    m n \nu u_\zeta + n q B u_z &= - T (\nabla n)_\zeta - n q (\nabla \phi)_\zeta,
    \label{eq.4_mom_eq_zeta}
    \\
    m n \nu u_\eta &= - T (\nabla n)_\eta - n q (\nabla \phi)_\eta,
    \label{eq.4_mom_eq_eta}
    \\
    m n \nu u_z - n q B u_\zeta &= 0.
    \label{eq.4_mom_eq_z}
\end{align}
\end{subequations}
The $\eta$-component is decoupled from the magnetic force, while the $\zeta$-component and $z$-component are coupled through $\bvec{u} \times \bvec{B}$.
This $\zeta-z$ coupling is a direct consequence of $\bvec{B}$ being purely in the plane, i.e., the in-plane cross-field flow $u_\zeta$ produces a Lorentz force in the $z$-direction, and the out-of-plane perpendicular flow $u_z$ produces a Lorentz force in the $\zeta$-direction.

We now solve for the three velocity components.
The $\eta$-component, Eq.~\eqref{eq.4_mom_eq_eta}, yields
\begin{equation}
    u_\eta = -\frac{D}{n} (\nabla n)_\eta - \sigma \mu (\nabla \phi)_\eta,
\end{equation}
where we have introduced the diffusion coefficient $D = T / (m\nu)$, the mobility $\mu = |q| / (m\nu)$, and the sign of the charge $\sigma = q / |q|$.
The $z$-component, Eq.~\eqref{eq.4_mom_eq_z}, gives the coupling between $u_z$ and $u_\zeta$,
\begin{equation}\label{eq.4_uz}
    u_z = \frac{q B}{m \nu} u_\zeta = \sigma \beta u_\zeta,
\end{equation}
where $\beta = \omega_c / \nu = |q| B / (m\nu)$ is the Hall parameter and $\omega_c = |q| B / m$ is the cyclotron frequency.
The sign of the charge is carried separately by $\sigma = q/|q|$, following the same convention as Chapter~\ref{chap.3}, which is convenient when treating multiple species.

Substituting Eq.~\eqref{eq.4_uz} into Eq.~\eqref{eq.4_mom_eq_zeta} eliminates $u_z$,
\begin{equation}
    m n \nu (1 + \beta^2) u_\zeta = - T (\nabla n)_\zeta - n q (\nabla \phi)_\zeta.
\end{equation}
Solving for $u_\zeta$ gives
\begin{equation}
    u_\zeta = - f \frac{D}{n} (\nabla n)_\zeta - f \sigma \mu (\nabla \phi)_\zeta.
\end{equation}
In this work, we define $f = 1/(1+\beta^2)$ as the magnetization reduction factor, since it quantifies the reduction of cross-field flux due to magnetization.\cite{Boeuf2023, Watanabe2023, Chen2016}

Finally, the particle flux in the TWM coordinate system can be written as
\begin{equation}\label{eq.4_flux_vec}
    \bvecg{\Gamma} = n \bvec{u} = \Gamma_\zeta \uvecg{\zeta} + \Gamma_\eta \uvecg{\eta} + \Gamma_z \uvec{z}
\end{equation}
with components
\begin{subequations}\label{eq.4_flux_comp}
\begin{align}
    \Gamma_\zeta & = n u_\zeta = - f D (\nabla n)_\zeta - f \sigma \mu n (\nabla \phi)_\zeta,
    \label{eq.4_flux_comp_zeta} \\
    \Gamma_\eta & = n u_\eta = - D (\nabla n)_\eta - \sigma \mu n (\nabla \phi)_\eta,
    \label{eq.4_flux_comp_eta} \\
    \Gamma_z & = n u_z = \sigma \beta \Gamma_\zeta.
    \label{eq.4_flux_comp_z}
\end{align}
\end{subequations}

Because $\uvecg{\eta} \parallel \bvec{B}$ and $\uvecg{\zeta} \perp \bvec{B}$, $\Gamma_\eta$ is the parallel flux, whereas $\Gamma_\zeta$ is the in-plane cross-field flux reduced by the factor $f$.
Note that, to avoid ambiguity, we do not use the generic subscript $\perp$, because both $\uvecg{\zeta}$ and $\uvec{z}$ are perpendicular to $\bvec{B}$ (Chapter~\ref{chap.2}).
We instead label vector components explicitly with subscripts $\zeta$, $\eta$, and $z$.
Also, throughout this work, the term cross-field refers specifically to the in-plane perpendicular $\zeta$-direction, since perpendicular transport in the $z$-direction is not the main focus of the work.

As discussed in Chapter~\ref{chap.2}, the TWM field is strongly nonuniform and contains an X-point.
Since the magnetic field vanishes at the X-point, the local magnetization necessarily becomes weak in its vicinity, even when the plasma is strongly magnetized over most of the domain.
In this work, the degree of magnetization is characterized by the Hall parameter, $\beta$, defined as the ratio of the cyclotron frequency to the collision frequency.
To express $\beta$ in terms of the normalized field magnitude, $\norm{B}=B/B_0$, we adopt the characteristic cyclotron frequency $\omega_{c0} = |q|B_0/m$, with an optional species subscript (e.g., $\omega_{ce0}$).
For reference, the parameter set used for the PIC simulation presented in Sec.~\ref{sec.4.4} is $(\ell_0, I_0) = (\SI{0.036}{m}, \SI{500}{A})$, which gives the TWM characteristic field magnitude $B_0 = \SI{2.778}{\milli\tesla}$, and for electrons, $\omega_{ce0} = \SI{4.886e8}{\per\second}$.
Since $\omega_c = |q|B/m$ and $B = B_0 \norm{B}$, the Hall parameter can be expressed as
\begin{equation}\label{eq.4_beta}
    \beta = \frac{\omega_c}{\nu} = \frac{\omega_{c0}}{\nu}\norm{B}.
\end{equation}

\begin{figure}
    \centering
	\includegraphics[width=11cm]{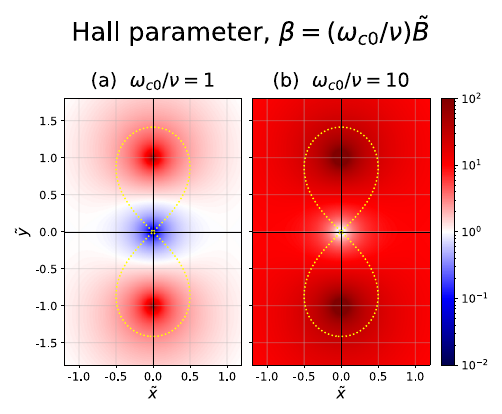}
    \caption[
    Spatial distribution of the Hall parameter for two values of $\omega_{c0}/\nu$
    ]{
    The spatial distribution of the Hall parameter, $\beta$, in the $\norm{x}-\norm{y}$ plane, for (a) $\omega_{c0}/\nu = 1$ and (b) $\omega_{c0}/\nu = 10$.
    The separatrix is indicated by the yellow dotted curve.
    }\label{fig.4_1}
\end{figure}

The ratio $\omega_{c0} / \nu$ provides a reference level of magnetization, but the actual degree of magnetization remains position dependent, even with a spatially uniform $\nu$.
Figure~\ref{fig.4_1} shows the spatial distribution of the Hall parameter in the $\norm{x}-\norm{y}$ plane, for (a) $\omega_{c0}/\nu = 1$ and (b) $\omega_{c0}/\nu = 10$.
Even for the larger ratio, the small region near the X-point is effectively unmagnetized, whereas most of the domain is strongly magnetized.
Throughout this work, the strongly magnetized regime refers to $\omega_{c0}/\nu \gg 1$, where the plasma is strongly magnetized over most of the domain even though unmagnetized locally at the X-point.

\begin{figure}
    \centering
	\includegraphics[width=15cm]{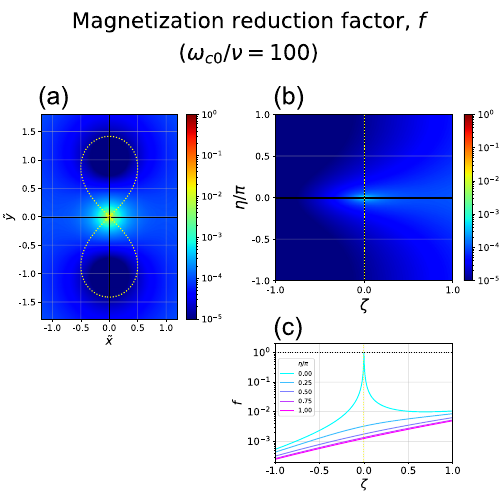}
    \caption[
    Spatial distribution and field-line profiles of the magnetization reduction factor
    ]{
    The spatial distribution of the magnetization reduction factor, $f=1/(1+\beta^2)$, for $\omega_{c0}/\nu = 100$, in (a) the $\norm{x}-\norm{y}$ plane and (b) the $\zeta-\eta$ plane.
    (c) Profiles of $f$ as functions of $\zeta$ for selected values of $\eta$, with $\eta/\pi = 0.00,\, 0.25,\, 0.50,\, 0.75$, and $1.00$.
    The separatrix is indicated by the yellow dotted curve in all panels, and the black dotted line in (c) marks $f=1$, the maximum possible value.
    }\label{fig.4_2}
\end{figure}

The effect of this nonuniform magnetization on cross-field transport is captured by the factor
\begin{equation}\label{eq.4_f}
    f = \frac{1}{1+\beta^2} = \frac{1}{1+(\omega_{c0}/\nu)^2\norm{B}^2},
\end{equation}
which serves as the reduction factor in the cross-field flux in Eq.~\eqref{eq.4_flux_comp_zeta}.
Since $0 < f \le 1$, it provides a direct measure of the suppression of cross-field transport by magnetization.
When $f \approx 1$, the plasma is weakly magnetized, so the cross-field flux approaches its unmagnetized drift-diffusion limit.
Figure~\ref{fig.4_2} shows the spatial distribution of $f$ for $\omega_{c0}/\nu = 100$, in (a) the $\norm{x}-\norm{y}$ plane and (b) the $\zeta-\eta$ plane.
The plots show that $f$ becomes close to $1$ in a narrow region near the X-point.
Figure~\ref{fig.4_2}(c) shows that, along $\eta/\pi = 0$, which corresponds to the orthogonal path through the X-point, $f$ rises sharply to unity at $\zeta = 0$.
Therefore, the cross-field transport is expected to be least suppressed near the X-point and strongly suppressed elsewhere.

\section{Formulation of reduced models}\label{sec.4.2}

We now formulate reduced DD models based on $\eta$-averaging.
These models capture how the localized enhancement of cross-field transport around the X-point shapes the density structure near the separatrix.
Because $\eta$ is a periodic coordinate along each closed field line, $\eta$-averaging provides a natural reduction that removes variation along the field line while retaining the dependence on $\zeta$ across field lines.

For a scalar quantity $a(\zeta, \eta)$, we define the $\eta$-average as
\begin{equation}\label{eq.4_eta_avg}
    \langle a \rangle_\eta (\zeta) = \frac{1}{2\pi} \int_{-\pi}^{\pi} a(\zeta, \eta)\, d\eta.
\end{equation}
Any scalar quantity can then be decomposed into its $\eta$-average and fluctuation:
\begin{equation}\label{eq.4_eta_avg_fluc}
    a(\zeta, \eta) = \langle a \rangle_\eta (\zeta) +  a'(\zeta, \eta),
\end{equation}
where $\langle a' \rangle_\eta = 0$ by definition.

Note that the $\eta$-average is a coordinate average, rather than a physical arclength average along a field line; the latter would include an additional factor $h$ in the integrand.
This choice is particularly useful because $\eta$-averaging satisfies the following properties:
\begin{subequations}\label{eq.4_eta_avg_partial_zetaeta}
\begin{align}
    \langle \partial_\zeta a \rangle_\eta & = \partial_\zeta \langle a \rangle_\eta,
    \label{eq.4_eta_avg_partial_zeta}
    \\
    \langle \partial_\eta  a \rangle_\eta & = \frac{1}{2\pi} [ a ]_{\eta = -\pi}^{\eta = \pi} = 0.
    \label{eq.4_eta_avg_partial_eta}
\end{align}
\end{subequations}
These identities show that $\eta$-averaging commutes with $\partial_\zeta$, and that the $\eta$-average of an $\eta$-derivative vanishes by periodicity.

We now apply $\eta$-averaging to the particle continuity equation.
Assuming negligible source and sink terms, the steady-state particle continuity equation is
\begin{equation}\label{eq.4_continuity}
    \nabla \cdot \bvecg{\Gamma} = 
    h^{-2} \left[ \partial_\zeta (h \Gamma_\zeta) + \partial_\eta (h \Gamma_\eta) \right] = 0
\end{equation}
where Eq.~\eqref{eq.2_formula_div} has been used.
Multiplying it by $h^2$ and taking the $\eta$-average gives
\begin{equation}\label{eq.4_continuity_etaavg}
    \left\langle \partial_\zeta (h \Gamma_\zeta) + \partial_\eta (h \Gamma_\eta) \right\rangle_\eta
    = \partial_\zeta \langle h \Gamma_\zeta \rangle_\eta = 0
\end{equation}
using the properties of Eqs.~\eqref{eq.4_eta_avg_partial_zeta} and~\eqref{eq.4_eta_avg_partial_eta}.
Therefore, we introduce the conserved quantity
\begin{equation}\label{eq.4_G}
    G = \langle h \Gamma_\zeta \rangle_\eta
    = \frac{1}{2\pi} \int_{-\pi}^{\pi} \Gamma_\zeta h\, d\eta
    = \text{const}.
\end{equation}

Because the scale factor $h$ converts the coordinate integral into the physical arclength integral, $G$ is proportional to the total cross-field particle flux through the closed field lines.
Its constancy follows directly from steady-state particle conservation: in the absence of source and sink terms, any region bounded by two closed field lines can have no net particle accumulation, so the total particle inflow must equal the total outflow.
As explained in Chapter~\ref{chap.2}, sweeping $\eta$ over its full range represents only one half of the physical plane.
Therefore, the total cross-field particle flux across the closed field lines, per unit length in the $z$-direction, is $4 \pi G$.

The conserved quantity $G$ is an exact consequence of steady-state particle continuity, but it does not by itself provide a closure for the reduced model.
To obtain such a closure, we consider strongly anisotropic transport, in which transport along the closed field lines is much faster than transport across them.
Under this ordering, fast parallel transport rapidly smooths parallel variation, and the system approaches parallel force balance to leading order, so that the parallel flux $\Gamma_\eta$ becomes small.
This motivates the closure introduced below.

Under the present assumptions of spatially uniform $T$ and $\nu$, the diffusion coefficient $D = T/(m\nu)$ is also uniform.
Using Eq.~\eqref{eq.4_flux_comp_eta}, together with the Einstein relation, $\mu/D = |q|/T$,\cite{Chen2016} the parallel flux can be written as
\begin{equation}\label{eq.4_Gamma_eta}
\begin{aligned}
    \Gamma_\eta
    & = - D (\nabla n)_\eta - \sigma \mu n (\nabla \phi)_\eta
    \\
    & = - D h^{-1} \left[ (\partial_\eta n) + \frac{q}{T} n (\partial_\eta \phi) \right]
    \\
    & = - D h^{-1} e^{-q\phi/T} \partial_\eta \left( n e^{q\phi/T} \right)
    \\
    & = - D h^{-1} e^{-q\phi/T} \partial_\eta \mathcal{N},
\end{aligned}
\end{equation}
where we define
\begin{equation}\label{eq.4_mathcal_N}
    \mathcal{N}(\zeta,\eta) = n (\zeta, \eta) \exp \left( \frac{q}{T}\phi(\zeta,\eta) \right).
\end{equation}
Then, the closure $\Gamma_\eta \approx 0$ implies $\partial_\eta \mathcal{N} \approx 0$ to leading order, so that $\mathcal{N}(\zeta, \eta) \approx \mathcal{N}(\zeta)$, i.e., $\mathcal{N}$ is constant along each field line.
Hence, the density approximately takes the Boltzmann response form
\begin{equation}\label{eq.4_Boltzmann_response}
    n(\zeta,\eta) \approx \mathcal{N}(\zeta) \exp \left( -\frac{q}{T}\phi(\zeta,\eta) \right).
\end{equation}
The cross-field flux, Eq.~\eqref{eq.4_flux_comp_zeta}, can also be rewritten in a similar form,
\begin{equation}\label{eq.4_Gamma_zeta}
\begin{aligned}
    \Gamma_\zeta
    & = - f D h^{-1} e^{-q\phi/T} \partial_\zeta \left( n e^{q\phi/T} \right)
    \\
    & = - f D h^{-1} e^{-q\phi/T} \partial_\zeta  \mathcal{N}.
\end{aligned}
\end{equation}
Multiplying by $h$ and using the definition of $G$, we obtain
\begin{equation}
\begin{aligned}
    G = \langle h \Gamma_\zeta \rangle_\eta
    & = \left\langle - f D e^{-q\phi/T} ( \partial_\zeta \mathcal{N} ) \right\rangle_\eta
    \\
    & = - D ( \partial_\zeta \mathcal{N} ) \left\langle f e^{-q\phi/T} \right\rangle_\eta,
\end{aligned}
\end{equation}
where $\partial_\zeta \mathcal{N}$ has been taken outside the $\eta$-average because it only depends on $\zeta$ under this closure.
The reduced model for the density profile is therefore
\begin{equation}\label{eq.4_reduced_model_gen}
    \frac{d \mathcal{N}}{d\zeta} = - \frac{G}{D}\frac{1}{\left\langle f e^{-q\phi/T} \right\rangle_\eta},
\end{equation}
with the density, $n$, given by Eq.~\eqref{eq.4_Boltzmann_response}.

We further simplify the model by adopting the diffusion-only approximation, in which the electric field is neglected and the potential, $\phi$, is spatially uniform.
The factor $e^{-q\phi/T}$ may then be absorbed into the definition of $\mathcal{N}(\zeta)$, so that Eq.~\eqref{eq.4_Boltzmann_response} reduces to $n(\zeta,\eta) \approx n(\zeta)$, and Eq.~\eqref{eq.4_reduced_model_gen} simplifies to
\begin{equation}\label{eq.4_reduced_model_diff}
    \frac{d n}{d\zeta} = - \frac{G}{D}\frac{1}{\langle f \rangle_\eta}.
\end{equation}

In this work, we refer to Eq.~\eqref{eq.4_reduced_model_gen} as the general reduced model and Eq.~\eqref{eq.4_reduced_model_diff} as the diffusion-only reduced model.
We use these models in the next section to illustrate the resulting density structure near the separatrix.



\section{Density profile analysis from the reduced models}\label{sec.4.3}

We now use the general reduced model, Eq.~\eqref{eq.4_reduced_model_gen}, and the diffusion-only reduced model, Eq.~\eqref{eq.4_reduced_model_diff}, to examine how the localized enhancement of cross-field transport near the X-point modifies the density profile in the TWM field.
We emphasize that the analyses in this section are intended to illustrate the density profile structure produced by the reduced models, rather than to test the validity of the underlying closure.
The reduced model predictions are compared against self-consistent kinetic simulations in the next section.

In the following analyses, the region of interest (ROI) is $\zeta \in [-0.2, +0.2]$, an interval centered on the separatrix (see Fig.~\ref{fig.2_2}(c,d)).
We consider a situation in which plasma undergoing net outward cross-field transport enters this interval from the inner side, crosses the separatrix, and exits toward larger $\zeta$.
The system is in steady-state, and the ROI is assumed to be source-free.
The reduced model analysis therefore focuses on the density profile within this interval under steady net outward transport.

\begin{figure}
    \centering
    \includegraphics[width=15cm]{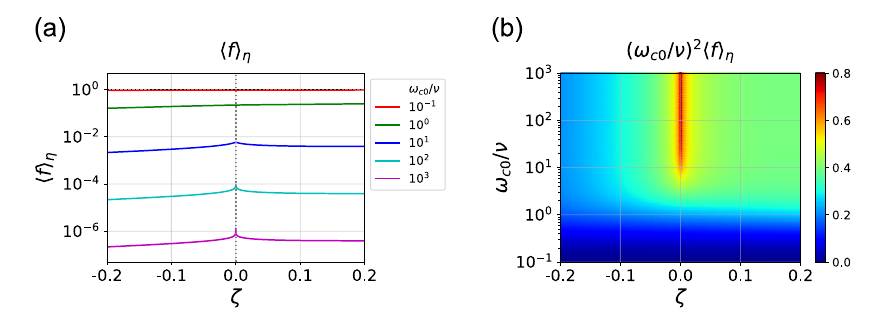}
    \caption[
    Profiles of the $\eta$-averaged magnetization reduction factor
    ]{
    (a) Profiles of the $\eta$-averaged reduction factor, $\langle f \rangle_\eta$, versus $\zeta$ for $\omega_{c0}/\nu = 10^{-1}, 10^{0}, 10^{1}, 10^{2},$ and $10^{3}$.
    The gray dotted line at $\zeta=0$ indicates the separatrix.
    The black dotted line marks the maximum possible value.
    (b) The rescaled quantity $(\omega_{c0}/\nu)^2\langle f\rangle_\eta$ as a function of $\zeta$ and $\omega_{c0}/\nu$, for $10^{-1} \le \omega_{c0}/\nu \le 10^{3}$.
    For both panels, the ROI, $\zeta \in [-0.2, +0.2]$, is shown.
    }\label{fig.4_3}
\end{figure}

In the diffusion-only reduced model, $\langle f \rangle_\eta$ directly determines the density gradient profile and is therefore the key quantity to examine.
Figure~\ref{fig.4_3}(a) shows the profiles of $\langle f \rangle_\eta$, for $\omega_{c0}/\nu = 10^{-1}, 10^{0}, 10^{1}, 10^{2},$ and $10^{3}$ in the ROI.
Except for the weakly magnetized cases, $\langle f \rangle_\eta$ shows a clear localized peak near the separatrix.
This peak is a direct geometric consequence of the X-point:
$\norm{B} \rightarrow 0$ at the null implies $f \rightarrow 1$ locally (see Fig.~\ref{fig.4_2}), and $\eta$-averaging picks up this enhancement most strongly on the field line that passes through the X-point itself, $\zeta = 0$.

The trend of $\langle f \rangle_\eta$ for the strongly magnetized regime $(\omega_{c0}/\nu \gg 1)$ is shown more clearly in Fig.~\ref{fig.4_3}(b), which plots the rescaled quantity $(\omega_{c0}/\nu)^2\langle f\rangle_\eta$ as a function of $\zeta$ and $\omega_{c0}/\nu$.
Once $\omega_{c0}/\nu$ becomes sufficiently large, the overall $\zeta$-dependent profile of the rescaled quantity is only weakly affected by further increases in $\omega_{c0}/\nu$.
This behavior is consistent with the large magnetization limit
\begin{equation}\label{eq.4_ea_f_rescaled}
    \lim_{\omega_{c0}/\nu \rightarrow \infty}
    \left( \frac{\omega_{c0}}{\nu} \right)^2 \langle f \rangle_\eta
    =
    \left\langle \norm{B}^{-2} \right\rangle_\eta = \left\langle \norm{h}^{2} \right\rangle_\eta.
\end{equation}
This indicates that, in the strongly magnetized regime, the shape of $\langle f \rangle_\eta$ is governed mainly by the TWM field geometry.
At the same time, a narrow localized enhancement remains evident near $\zeta=0$, showing that the near-separatrix transport feature associated with the X-point persists.

We next use the diffusion-only reduced model, Eq.~\eqref{eq.4_reduced_model_diff},
\begin{equation*}
    \frac{dn}{d\zeta} = -\frac{G}{D}\frac{1}{\langle f\rangle_\eta},
\end{equation*}
to illustrate the resulting density gradient and density profiles.
Since $G/D$ is constant, the $\zeta$-dependence of $dn/d\zeta$ is determined directly by $1/\langle f\rangle_\eta$. 
Here, we focus on how the density profile shape depends on $\omega_{c0}/\nu$, rather than on its overall magnitude.
To compare the profile shapes, we fix the values of both $dn/d\zeta$ and $n$ at the inner boundary, $\zeta=-0.2$. 
The fixed gradient determines the constant $G/D$ for each case, while the fixed density sets the integration constant.
This choice allows the profile shapes to be compared directly.

\begin{figure}
    \centering
    \includegraphics[width=15cm]{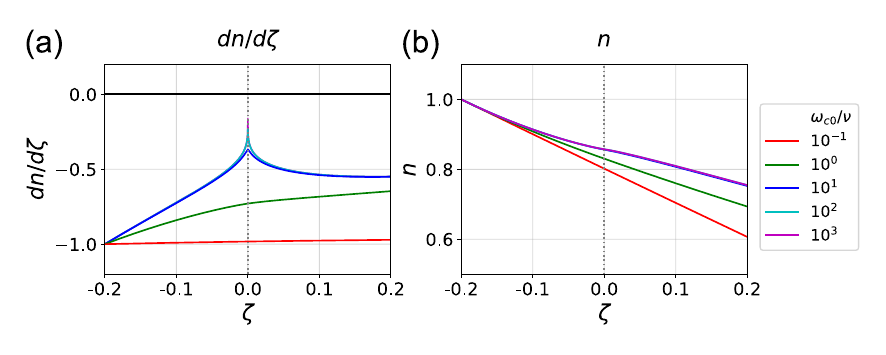}
    \caption[
    Density gradient and density profiles from the diffusion-only reduced model
    ]{
    (a) Density gradient and (b) density profiles from the diffusion-only reduced model, Eq.~\eqref{eq.4_reduced_model_diff}, for $\omega_{c0}/\nu = 10^{-1}, 10^{0}, 10^{1}, 10^{2},$ and $10^{3}$.
    For all cases, the inner boundary values at $\zeta = -0.2$ are fixed as $dn/d\zeta=-1$ and $n=1$, so that the comparison focuses only on profile shapes.
    The resulting values for $G/D$ are approximately $0.94$, $0.16$, $2.2\times10^{-3}$, $2.2\times10^{-5}$, and $2.2\times10^{-7}$, respectively.
    The gray dotted line indicates the separatrix.
    }\label{fig.4_4}
\end{figure}

Figure~\ref{fig.4_4} shows the density gradient and density profiles obtained numerically from the diffusion-only reduced model.
For $\omega_{c0}/\nu = 10^{1}$, $10^{2}$, and $10^{3}$, the density gradient magnitude $|dn/d\zeta|$ develops a pronounced local minimum at $\zeta = 0$, i.e., a localized relaxation near the separatrix.
After integration, this produces a plateau-like density profile in the vicinity of the separatrix.
In this work, the term density plateau refers to a localized region in which $|dn/d\zeta|$ has a local minimum, corresponding to a relative flattening of the density profile.
The profiles for $\omega_{c0}/\nu = 10^{2}$ and $10^{3}$ are nearly indistinguishable, consistent with the earlier result that the rescaled quantity $(\omega_{c0}/\nu)^2\langle f\rangle_\eta$ changes only weakly with further increases in $\omega_{c0}/\nu$ in the strongly magnetized regime.
Within the diffusion-only reduced model, density plateau formation is therefore a feature of the strongly magnetized regime.
By contrast, for $\omega_{c0}/\nu = 10^{-1}$ and $10^{0}$, $\langle f \rangle_\eta$ lacks a sharp localized peak (see Fig.~\ref{fig.4_3}(a)), so $|dn/d\zeta|$ has no local minimum, and no pronounced flattening appears in the corresponding density profiles.

\begin{figure}
    \centering
    \includegraphics[width=12cm]{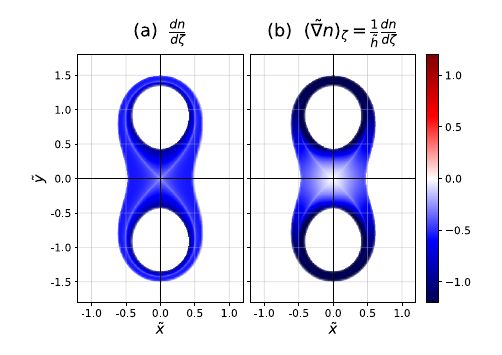}
    \caption[
    Physical space representation of the diffusion-only reduced model result
    ]{
    Physical space representation of the diffusion-only reduced model result for $\omega_{c0}/\nu = 10^{3}$, shown in the $\norm{x}-\norm{y}$ plane.
    (a) The coordinate gradient, $dn/d\zeta$, constant along each field line.
    (b) The normalized physical cross-field gradient, $(\norm{\nabla}n)_\zeta = \norm{h}^{-1}(dn/d\zeta)$, varying along each field line through $\norm{h}$.
    }\label{fig.4_5}
\end{figure}

Figure~\ref{fig.4_5} shows the physical space representation of the diffusion-only reduced model result in the $\norm{x}-\norm{y}$ plane for $\omega_{c0}/\nu = 10^{3}$.
Figure~\ref{fig.4_5}(a) shows the coordinate gradient, $dn/d\zeta$, which depends only on $\zeta$ and is therefore constant along each field line.
Figure~\ref{fig.4_5}(b) shows the normalized physical cross-field gradient, $(\norm{\nabla}n)_\zeta = \norm{h}^{-1}(dn/d\zeta)$, which acquires an additional $\eta$-dependence through the scale factor $\norm{h}$.
The density plateau identified above is a global feature shared by the closed field lines surrounding the separatrix.
In addition, Fig.~\ref{fig.4_5}(b) reveals a local relaxation of the physical cross-field gradient at the X-point itself, where $\norm{h} \rightarrow \infty$ further suppresses $(\norm{\nabla} n)_\zeta$.
This representation provides a reference for visual comparison with the PIC results in the next section.

We now extend the analysis to the general reduced model, Eq.~\eqref{eq.4_reduced_model_gen},
\begin{equation*}
    \frac{d \mathcal{N}}{d\zeta} = - \frac{G}{D}\frac{1}{\left\langle f e^{-q\phi/T} \right\rangle_\eta},
\end{equation*}
in which the electric field contribution is included, and the density, $n$, is given by the Boltzmann response $n = \mathcal{N} e^{-q\phi/T}$ from Eq.~\eqref{eq.4_Boltzmann_response}.
Compared to the diffusion-only model, the relevant $\eta$-averaged reduction factor becomes $\langle f e^{-q\phi/T} \rangle_\eta$ rather than $\langle f \rangle_\eta$.

In a fully self-consistent treatment, the potential $\phi$ would be determined by the charge density and boundary conditions.
Such a treatment is beyond the scope of the reduced model analysis and is addressed using PIC simulation in the next section.
Within the reduced model framework, we treat $\phi$ as a prescribed input and show how a given electrostatic structure modifies the resulting density profile.

For the present illustration, we prescribe a linear potential profile depending only on $\zeta$, characterized by a dimensionless constant $\alpha$,
\begin{equation}\label{eq.4_alpha}
    \alpha = -\frac{q}{T} \partial_\zeta \phi.
\end{equation}
Physically, this corresponds to applying an external in-plane perpendicular electric field, $E_\zeta = -h^{-1}\partial_\zeta\phi$, with the corresponding electric force on the species, $F_\zeta = qE_\zeta = \alpha T h^{-1}$.
With this convention, positive $\alpha$ produces an electric force in the positive $\zeta$-direction, pushing the species outward, and vice versa.
The general reduced model can produce a two-dimensional density profile, $n(\zeta,\eta)$, but the specific choice of potential profile here yields a one-dimensional density, $n(\zeta)$.

To illustrate the role of $\alpha$, we fix $\omega_{c0}/\nu = 10^{2}$, which places the system in the strongly magnetized regime.
Unlike the diffusion-only analysis, the value of $G/D$ is held fixed at $5 \times 10^{-5}$ for all cases, so that the resulting profiles differ only through the effect of the electric field.
The values of $\alpha$ are $-1.0, -0.5, 0.0, 0.5,$ and $1.0$ for different cases.
Finally, the boundary condition $n(\zeta = -0.2) = 1$ sets the integration constant for all cases.

\begin{figure}
    \centering
    \includegraphics[width=15cm]{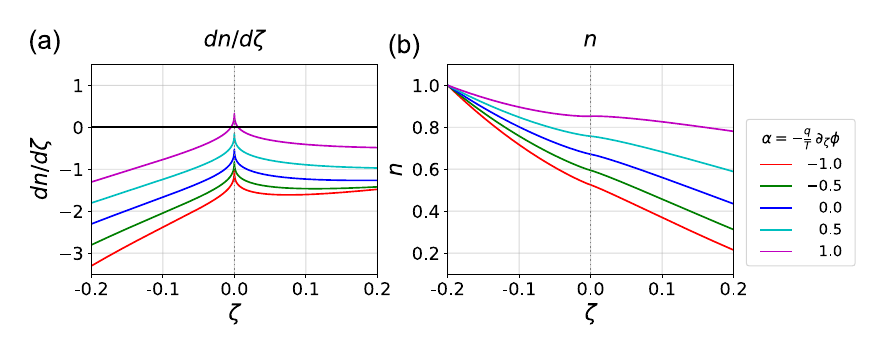}
    \caption[
    Density gradient and density profiles from the general reduced model
    ]{
    (a) Density gradient and (b) density profiles from the general reduced model, Eq.~\eqref{eq.4_reduced_model_gen}, for $\alpha = -(q/T)(\partial_\zeta\phi) = -1.0, -0.5, 0.0, 0.5,$ and $1.0$.
    For all cases, $\omega_{c0}/\nu = 10^{2}$, $G/D = 5 \times 10^{-5}$, and $n(\zeta=-0.2) = 1$.
    The gray dotted line indicates the separatrix.
    }\label{fig.4_6}
\end{figure}

In Fig.~\ref{fig.4_6}(a), the $\alpha = 0$ case is equivalent to the diffusion-only result shown earlier, and a finite $\alpha$ shifts the $dn/d\zeta$ profile while preserving the localized peak at $\zeta = 0$.
For $\alpha > 0$, the outward drift contributes to the overall flux, so diffusion requires a smaller density gradient; for $\alpha < 0$, the inward drift opposes the flux, so diffusion requires a steeper gradient.
The corresponding density profiles in Fig.~\ref{fig.4_6}(b) show the integrated effect of this shift, with the density plateau near the separatrix preserved in all cases.
This example uses a globally uniform $\alpha$ to demonstrate how an electric field can shift the overall density gradient.
More generally, a nonuniform electric field would reshape the gradient locally and produce a more complex density profile.

The reduced model analyses predict density plateau formation near the separatrix.
This arises from the localized cross-field transport enhancement at the X-point and the parallel equilibration along the closed field lines.
This prediction, however, rests on the closure assumption underlying the reduced models.
The closure assumes fast parallel transport along each closed field line, so that the quantity $\mathcal{N}$ defined in Eq.~\eqref{eq.4_mathcal_N} is weakly dependent on $\eta$.

Using the decomposition $\mathcal{N} = \langle \mathcal{N} \rangle_\eta + \mathcal{N}'$ from Eq.~\eqref{eq.4_eta_avg_fluc}, the closure can be stated precisely as
\begin{equation}\label{eq.4_closure_form}
    |\langle \mathcal{N} \rangle_\eta(\zeta)| \gg |\mathcal{N}'(\zeta, \eta)|,
\end{equation}
for a field line labeled by $\zeta$.
When plasma data are available from kinetic simulation or experimental measurement, the closure can be tested directly by measuring the relative fluctuation amplitude,
\begin{equation}\label{eq.4_epsilon}
    \epsilon_\mathcal{N}(\zeta)
    = \frac{\sqrt{\langle (\mathcal{N}'(\zeta,\eta))^2 \rangle_\eta}}{|\langle \mathcal{N} \rangle_\eta(\zeta)|},
\end{equation}
where $\epsilon_\mathcal{N}(\zeta) \ll 1$ indicates that the closure holds on the field line at $\zeta$.
The fluctuation diagnostic $\epsilon_\mathcal{N}$ is applied to the PIC results in the next section.
An additional analytic estimate of the closure inequality is presented in Appendix~\ref{sec.B.1}, based on order-of-magnitude scaling of the fluctuation equation using the characteristic gradient length scale of $\langle \mathcal{N} \rangle_\eta$.
The resulting criterion is sufficient but approximate, and identifies the regime in which the closure should hold rather than a sharp threshold.

The analyses in this section have treated the electric field as prescribed and the cross-field flux as an imposed parameter.
In a self-consistent plasma, the potential $\phi$ is determined by Poisson's equation with the appropriate boundary conditions, and the particle flux is governed by the overall plasma dynamics.
Whether the density plateau predicted by the reduced models persists under these self-consistent conditions, and whether the closure assumption holds, are the questions addressed in the next section using electrostatic PIC simulations.
Appendix~\ref{sec.B.2} additionally presents a brief analysis for the special case of a quasineutral ambipolar plasma.



\section{PIC simulation \& Steady-state profiles}\label{sec.4.4}

Particle-in-cell (PIC) simulation provides a self-consistent kinetic complement to the fluid-level reduced models of Secs.~\ref{sec.4.1}--\ref{sec.4.3}.
In this section, we present two-dimensional electrostatic simulations of a low temperature plasma in the TWM field, performed with the EDIPIC-2D code.\cite{EDIPIC2D}
We first describe the simulation setup, then present the steady-state two-dimensional profiles of the results and identify the density plateau formation predicted by the reduced models, along with the plasma conditions relevant to the model comparison.
The cross-field profiles of the simulation data, the empirical validation of the parallel equilibration closure, the quantitative comparison with the reduced model predictions, and the dependence of the plateau formation on the magnetization are presented in Sec.~\ref{sec.4.5}.

\begin{table}
\caption{\label{tab.4_simparams}PIC simulation configuration.}
\centering
\setlength{\extrarowheight}{0.05cm}
\begin{tabular}{lccc}
    \hline
    Parameter & Symbol & Value & Unit \\
    \hline
    \noalign{\vspace{0.1cm}}
    \multicolumn{4}{c}{\textbf{Simulation domain}} \\
    Cell size & $\Delta x = \Delta y$ & $1$ & mm \\
    Number of cells in $x$ & $N_{cx}$ & $73$ & \\
    Number of cells in $y$ & $N_{cy}$ & $129$ & \\
    Box length in $x$ & $L_x$ & $73$ & mm \\
    Box length in $y$ & $L_y$ & $129$ & mm \\[0.1cm]
    \hline
    \noalign{\vspace{0.1cm}}
    \multicolumn{4}{c}{\textbf{Computational parameters}} \\
    Time step (electrons) & $\Delta t$ & $0.5$ & ns \\
    Time step (ions, He$^+$) & $\Delta t_i = 5\Delta t$ & $2.5$ & ns \\
    Macroparticle weight & $\text{MW}$ & $1000$ & \\[0.1cm]
    \hline
    \noalign{\vspace{0.1cm}}
    \multicolumn{4}{c}{\textbf{Background neutrals}} \\
    Neutral species & --- & helium & \\
    Neutral temperature & $T_n$ & $300$ & K \\
    Neutral pressure & $p_n$ & $10$ & mTorr \\[0.1cm]
    \hline
    \noalign{\vspace{0.1cm}}
    \multicolumn{4}{c}{\textbf{Primary electron injection}} \\
    Injection rate per rod & $R_{\text{inject}}$ & $1000$ & MP/$\Delta t$\,$^{*}$ \\
    Primary electron energy & $W_{\text{prim}}$ & $50$ & eV \\[0.1cm]
    \hline
    \noalign{\vspace{0.1cm}}
    \multicolumn{4}{c}{\textbf{TWM magnetic field}} \\
    Wire vertical position & $\ell_0$ & $36$ & mm \\
    Wire current & $I_0$ & $500$ & A \\
    Characteristic field magnitude & $B_0 = A_0/\ell_0$ & $2.778$ & mT \\[0.1cm]
    \hline
\end{tabular}

\vspace{0.15cm}
{\footnotesize\centering $^{*}$\,Macroparticles per electron time step.\par}
\end{table}

We perform the simulations with EDIPIC-2D,\cite{EDIPIC2D} a two-dimensional electrostatic PIC code that has been verified through benchmarks and applied to a wide range of low temperature plasma studies.\cite{Charoy2019, Villafana2021, Sun2023, Rauf2023, Son2023, Villafana2024}
Charged particles are advanced with the explicit Boris algorithm,\cite{Qin2013, Birdsall1991} the same integration scheme used for the test particle simulations of Chapter~\ref{chap.3}.
The electrostatic potential is obtained by solving Poisson's equation on the Cartesian grid; a static external magnetic field can also be prescribed.
Electron-neutral and ion-neutral collisions are calculated by the Monte Carlo collision (MCC) method.\cite{Birdsall1991}

\begin{figure}
    \centering
    \includegraphics[width=12cm]{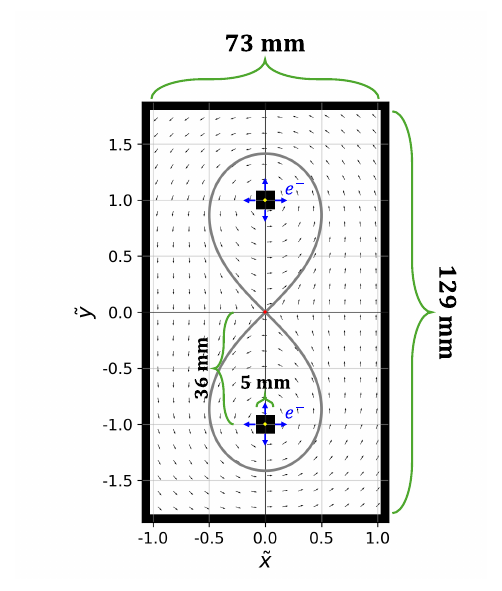}
    \caption[
    PIC simulation domain in the $\norm{x}-\norm{y}$ plane
    ]{
    Simulation domain, drawn to scale in the $\norm{x}-\norm{y}$ plane.
    The actual box dimensions are $L_x \times L_y = \SI{73}{mm} \times \SI{129}{mm}$, and $\ell_0 = \SI{36}{mm}$ is the distance from the X-point to the center of each rod.
    The side length of the rods is $\SI{5}{mm}$.
    The black regions mark the four outer walls and the two square rods, all grounded conductors held at $\phi = \SI{0}{V}$.
    The Cartesian coordinates are shifted and normalized by $\ell_0$ so that the axis labels are $\norm{x}$ and $\norm{y}$, as in Sec.~\ref{sec.2.1}.
    The unit vector of the prescribed TWM magnetic field is indicated by the black arrows, the separatrix is shown in gray, and the X-point is the red dot at the center, $(\norm{x},\norm{y}) = (0,0)$.
    Each rod emits monoenergetic \SI{50}{eV} primary electrons, indicated by the blue arrows.
    The domain is initially empty of plasma.
    }\label{fig.4_7}
\end{figure}

The simulation domain is shown in Fig.~\ref{fig.4_7}.
It is a rectangular box of dimensions $L_x \times L_y = \SI{73}{mm} \times \SI{129}{mm}$, with the lower-left corner at $(x, y) = (0, 0)$.
Two square objects of side length $\SI{5}{mm}$ are placed at $(x, y) = (\SI{36.5}{mm}, \SI{28.5}{mm})$ and $(\SI{36.5}{mm}, \SI{100.5}{mm})$.
These objects represent the two square rods that enclose the current sources of the TWM.
The currents themselves are not simulated; instead, the TWM magnetic field is prescribed as if a line current $I_0$ is concentrated at the center of each rod.
With this prescribed TWM field, the magnetic X-point lies at the center of the box, $(\SI{36.5}{mm}, \SI{64.5}{mm})$, and $\ell_0 = \SI{36}{mm}$ is the distance from the X-point to each line current, i.e., to the center of each rod.

In all Cartesian plane figures in this section, the spatial coordinates are shifted to place the X-point at the origin and normalized by $\ell_0$, so the physical space is represented in the $\norm{x}-\norm{y}$ plane, as in Chapter~\ref{chap.2}.
Consequently, the closed field lines in the simulation domain (i.e., not touching the boundary) span approximately $\zeta \in [-1.5, +0.7]$, from the rod side to the outer side, so the separatrix ($\zeta = 0$) lies well inside the box.

Electron (e$^-$) and singly-charged helium ion (He$^+$) macroparticles are evolved in 2D3V phase space, $(x, y, v_x, v_y, v_z)$, with each macroparticle representing $1000$ physical particles.
The domain is uniform along the $z$-direction, and only $(x, y)$ positions are tracked during the simulation.
For diagnostics and analysis, $(x, y)$ are shifted and normalized to $(\norm{x}, \norm{y})$, following the same convention as the figures.
The electron time step is $\Delta t = \SI{0.5}{ns}$, and ions are subcycled with $\Delta t_i = 5\Delta t = \SI{2.5}{ns}$ to reduce computational cost.
The cell size is $\Delta x = \Delta y = \SI{1}{mm}$, giving a total of $73 \times 129$ cells.
Particles are advanced by the self-consistent electric field and the prescribed TWM magnetic field.
The electric field is obtained from Poisson's equation, with the charge density on the grid as the source.
The four outer walls and the two rods are particle-absorbing boundaries held at $\phi = 0$.
The TWM field is set by $(\ell_0, I_0) = (\SI{36}{mm}, \SI{500}{A})$, which gives the characteristic field magnitude $B_0 = A_0/\ell_0 = \SI{2.778}{mT}$.

The background neutral gas is helium at temperature $T_n = \SI{300}{K}$ and pressure $p_n = \SI{10}{mTorr}$.
Electrons undergo elastic scattering, electronic excitation, and ionization, while He$^+$ ions undergo resonant charge exchange; these events are calculated by the null-collision MCC method.\cite{Birdsall1991}
The cross section data are obtained from various sources.\cite{Ralchenko2008, Maiorov2009, LXCat}

The simulation proceeds as follows.
The domain is initialized with no charged particles.
Each rod emits monoenergetic \SI{50}{eV} primary electrons at a constant rate of $1000$ macroparticles per $\Delta t$.
Since the magnetic field is strong near the rods, the primary electrons are well confined there and ionize the background helium to produce electron-ion pairs.
Most particles are produced near the rods and undergo cross-field transport across the separatrix and eventually to the outer walls, filling the whole domain.
The simulation is run until the total macroparticle count and the mean kinetic energy of all particles become stationary.
The configuration is symmetric with respect to the origin by construction: the magnetic field, the rod geometry and potential, and the primary electron emission all respect this symmetry.
The symmetry is not enforced by the simulation, so residual statistical noise remains.
The PIC simulation configuration is summarized in Table~\ref{tab.4_simparams}.

\begin{figure}
    \centering
    \includegraphics[width=15cm]{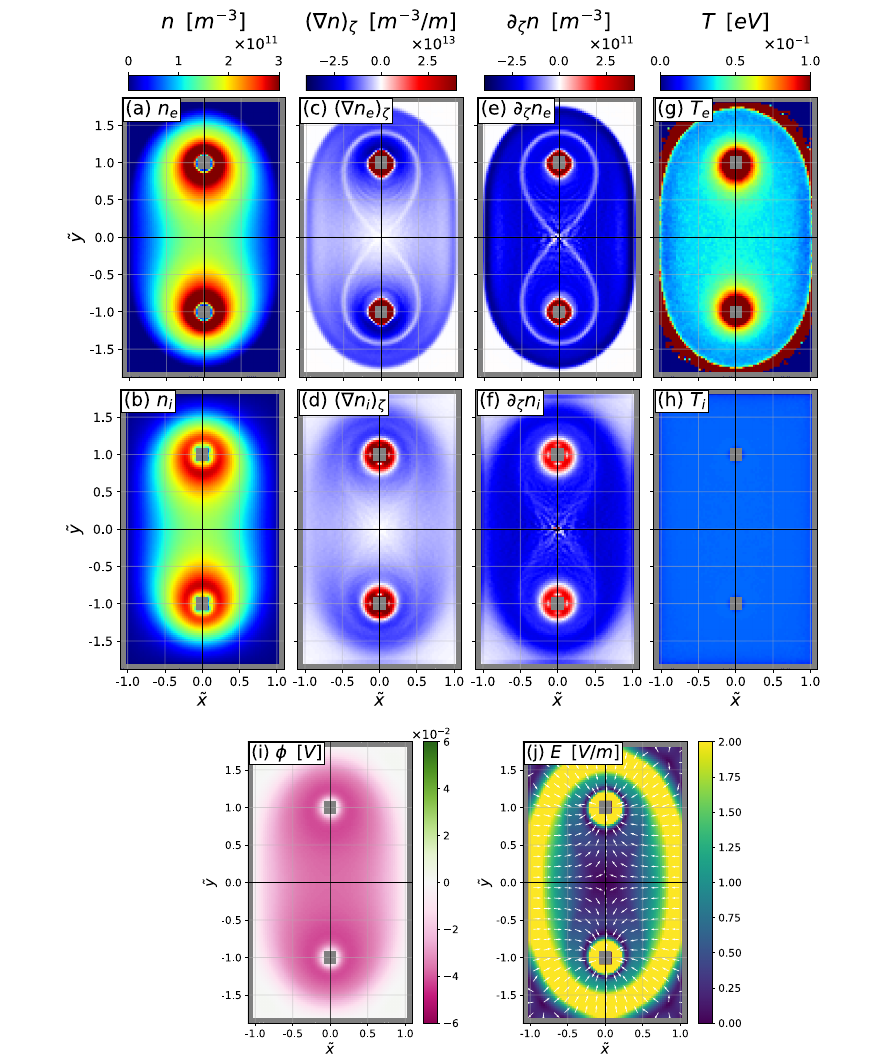}
    \caption[
    Steady-state PIC profiles in the $\norm{x}-\norm{y}$ plane
    ]{
    Time-averaged steady-state PIC profiles in the $\norm{x}-\norm{y}$ plane.
    Panels (a--h) show electrons (top row) and ions (bottom row) with a shared colorbar per column: (a,b) density, $n$, (c,d) physical and (e,f) coordinate cross-field density gradients, $(\nabla n)_\zeta$ and $\partial_\zeta n$, and (g,h) temperature, $T$.
    (i) Electrostatic potential, $\phi$, and (j) electric field magnitude, $E = |\bvec{E}|$, with the unit vector $\bvec{E}/E$ overlaid as white arrows.
    The gray squares are the rods, and the X-point is at the origin.
    }\label{fig.4_8}
\end{figure}

The simulation proceeds in successive runs of $\num{e7} \Delta t$ each, with every new run continuing from the final state of the previous one.
For each run, the total macroparticle count and the mean kinetic energy of all particles are monitored.
Steady-state is declared once both quantities vary by less than $0.1\%$ within a single run.
The subsequent analysis is performed on data from the final run after this criterion is satisfied.

The full 2D3V phase-space data of all macroparticles are saved at each data dump.
The plasma quantities are then computed in post-processing by depositing the relevant moments of the particle distribution onto the $\norm{x}-\norm{y}$ grid.
To reduce statistical noise, $500$ data dumps uniformly spaced over the entire final steady-state run are averaged, with a dump interval of $\Delta t_{\text{dump}} = \num{2e4} \Delta t = \SI{e4}{ns}$.

Figure~\ref{fig.4_8} presents the steady-state PIC profiles in the $\norm{x}-\norm{y}$ plane.
The profiles are symmetric about the origin, which is a consequence of the symmetric configuration of the simulation setup.
Figures~\ref{fig.4_8}(a) and (b) show the density profiles for electrons and ions, respectively.
The two profiles appear similar; both peak near the rods and decrease outward.
The densities remain on the order of $\SI{e11}{\per\cubic\meter}$ across much of the bulk before falling further toward the outer edges of the domain, and contours of constant density approximately follow the field line geometry of the TWM.
Note that, in the following analysis, we focus mostly on the near-separatrix region, which is the ROI, $\zeta \in [-0.2, +0.2]$, introduced in Sec.~\ref{sec.4.3}.

The important observation in Fig.~\ref{fig.4_8} is the density plateau formation, predicted by the reduced models in Sec.~\ref{sec.4.3}.
Figures~\ref{fig.4_8}(c) and (d) show the physical cross-field gradient of density, $(\nabla n)_\zeta$, for electrons and ions, respectively.
Figures~\ref{fig.4_8}(e) and (f) show the coordinate cross-field gradient, $\partial_\zeta n = h (\nabla n)_\zeta$, in which the scale factor cancels.
For electrons, $(\nabla n_e)_\zeta$ shows a clear band of relaxed gradient magnitude along the separatrix, with an additional localized relaxation in the region near the X-point, where the scale factor $h$ diverges; $\partial_\zeta n_e$ shows only the band along the separatrix.
These electron panels indicate the density plateau formation, and the structures resemble the predictions of Fig.~\ref{fig.4_5}, though the plateau appears visibly broader in the simulation.
The coordinate gradient of the ion density in Fig.~\ref{fig.4_8}(f) shows only faint structure along the separatrix.
Further discussion of the density plateaus and their comparison to the reduced models is presented in Sec.~\ref{sec.4.5}.

Figure~\ref{fig.4_8}(g) shows the electron temperature profile.
Near the rods, $T_e$ is significantly higher due to the injected primary electrons.
In the ROI, however, $T_e \lesssim \SI{0.1}{eV}$, which is well below the few-eV range typically reported for low pressure plasma discharges.
This unusually cold $T_e$ arises from a combination of physical and numerical effects.
Electrons are strongly magnetized and cannot easily diffuse across the field lines.
Over the long confinement times, the high energy tail of the electron distribution is continually depleted by inelastic collisions with neutrals: electrons lose energy in large discrete steps to excitation and ionization of helium, with thresholds at $\SI{19.8}{eV}$ and $\SI{24.6}{eV}$, respectively.\cite{Ralchenko2008}
The tail is partially repopulated by numerical thermalization, an artifact of the discrete macroparticle distribution that is analogous to Coulomb collisions in real plasmas.
This effect was quantified by Jubin~\textit{et al.}\cite{Jubin2024} using EDIPIC-2D; they showed that the numerical thermalization timescale can become comparable to or shorter than that of MCC-modeled electron-neutral collisions.
Each cycle of high energy tail depletion and repopulation continuously transfers energy from the electrons to the neutrals, steadily cooling the electrons.
This cooling further suppresses the cross-field transport of the bulk electrons, so the long confinement and the cooling reinforce each other.
In addition, the simulation does not include volume recombination, which would otherwise remove cold electron-ion pairs.
Nonetheless, this unusually low $T_e$ does not affect the central result of this work, since the density plateau formation discussed earlier is controlled by $\omega_{c0}/\nu$, which is quantified below.

Figure~\ref{fig.4_8}(h) shows that the ion temperature is approximately uniform at $T_i \approx \SI{0.023}{eV}$, slightly below the neutral temperature $T_n \approx \SI{0.026}{eV}$ ($\SI{300}{K}$).
This small deficit arises from the selective loss of high energy ions.
Ions are produced by ionization of the background neutrals and are thus born with $T_i \approx T_n$, but high energy ions are easily lost to the walls, truncating the high energy tail of the ion distribution.
Resonant charge exchange with cold neutrals acts to restore $T_i$ toward $T_n$, but does not fully complete this restoration.

Within the ROI, $T_e$ varies by approximately $5\%$ and $T_i$ by approximately $0.5\%$, so both may be treated as approximately constant in the analysis that follows.
Although not shown here, the MCC collision frequencies $\nu_e$ and $\nu_i$ are similarly uniform in the ROI, varying by approximately $3\%$ and $0.2\%$, respectively.
The corresponding magnetization ratios in the ROI, computed using these collision frequencies, are $\omega_{ce0}/\nu_e \approx 403$ and $\omega_{ci0}/\nu_i \approx 0.53$, which place electrons well within the strongly magnetized regime and ions outside it.
The numerical collisions responsible for the thermalization mentioned above can also enhance the effective collision frequency, reducing $\omega_{c0}/\nu$ from the MCC value, and this will be discussed further in the next section.

Figure~\ref{fig.4_8}(i) shows the plasma potential, $\phi$, which is negative throughout the bulk, with the rods and outer walls held at $\phi = 0$ by the boundary condition.
Figure~\ref{fig.4_8}(j) shows the electric field magnitude, $E$, with the unit vector $\bvec{E}/E$ overlaid as white arrows.
The unit vectors point in the cross-field direction throughout the domain, with $E_\zeta$ directed toward the rods and the parallel component $E_\eta$ negligible.
This structure arises from a negative sheath effect.
The magnetic field is approximately parallel to the rod surfaces, so electrons reach the rods only by slow cross-field transport.
Electron loss to the rods is therefore suppressed more strongly than ion loss, which causes electron accumulation in the magnetized region surrounding them and generates a negative sheath.
The magnetic field is also approximately parallel to portions of the outer walls, but its magnitude is much weaker there, so the magnetized sheath effect is weaker than near the rods.
The smallness of $E_\eta$ indicates that $\phi$ varies weakly along the field lines, and $\phi$ decreases from the outer walls toward the rods in the cross-field direction.
In the ROI, $\phi(\zeta=-0.2) \approx \SI{-0.039}{V}$ and $\phi(\zeta=+0.2) \approx \SI{-0.032}{V}$, which corresponds to a potential increase of approximately $\SI{0.007}{V}$.

The relatively small magnitude of $\phi$ in Fig.~\ref{fig.4_8}(i), together with the resulting small magnitude of $E$ in Fig.~\ref{fig.4_8}(j), is a consequence of the low simulated density.
In practice, the injection rate of primary electrons from the rods is adjusted to reach the steady-state density target.
The density target is chosen specifically to keep the Debye length larger than the cell size at an affordable grid resolution, satisfying the standard PIC stability requirement $\Delta x \lesssim \lambda_D$.\cite{Birdsall1991}
Within the ROI, using representative values $n \sim \SI{2e11}{\per\cubic\meter}$ and $T_e \sim \SI{0.05}{eV}$, the Debye length is $\lambda_D \approx \SI{4}{mm}$, giving $\Delta x \approx \lambda_D / 4$ and satisfying the requirement.
Physically typical densities for low pressure plasmas ($\gtrsim \SI{e16}{\per\cubic\meter}$) would require sub-millimeter cells; combined with the slow cross-field transport by magnetization, this makes the simulation computationally intractable in the present configuration.

The two-dimensional PIC profiles in Fig.~\ref{fig.4_8} display the qualitative features anticipated by the reduced models: a density plateau structure near the separatrix in the strongly magnetized regime.
In the next section, we compare the $\eta$-averaged one-dimensional profiles of the PIC results with the reduced models and analyze the discrepancies.



\section{Quantitative comparison with the reduced models}\label{sec.4.5}

The PIC profiles presented in Sec.~\ref{sec.4.4} exhibit qualitative features consistent with the reduced model predictions of Sec.~\ref{sec.4.3}.
In this section, we make the comparison explicit by reducing the PIC data to $\eta$-averaged one-dimensional profiles, testing the closure assumption that underlies the reduction, and overlaying the reduced model profiles calculated from the PIC plasma quantities.
The comparison confirms that the reduced models capture the geometric mechanism of plateau formation, while revealing quantitative discrepancies attributable to the combined effects of the limitations of the PIC simulation and of the drift-diffusion framework near the null.

The PIC profiles in Sec.~\ref{sec.4.4} were obtained by depositing the moments of the particle distribution onto the Cartesian $\norm{x}-\norm{y}$ grid.
To show the cross-field profiles in $\zeta$, the moments are instead first deposited onto the $\zeta-\eta$ grid using the coordinate transformation, Eq.~\eqref{eq.2_zetaeta}, and at each $\zeta$, the data are averaged over $\eta \in (-\pi, +\pi]$ following Eq.~\eqref{eq.4_eta_avg}.
The chosen $\zeta-\eta$ grid has resolutions $\Delta\zeta = 0.004$ and $\Delta\eta = \pi/90$~($=2^\circ$).
Throughout this section, in order to suppress residual PIC sampling noise without affecting the physical features of interest, discrete $\zeta$-derivatives are computed by first smoothing the input profile.
The smoothing convolves the profile with a Gaussian kernel of standard deviation $2$ grid points before computing the discrete gradient.
Other quantities, such as $\langle n_s \rangle_\eta$ and $\epsilon_\mathcal{N}$, are plotted directly without smoothing.

\begin{figure}
    \centering
    \includegraphics[width=10cm]{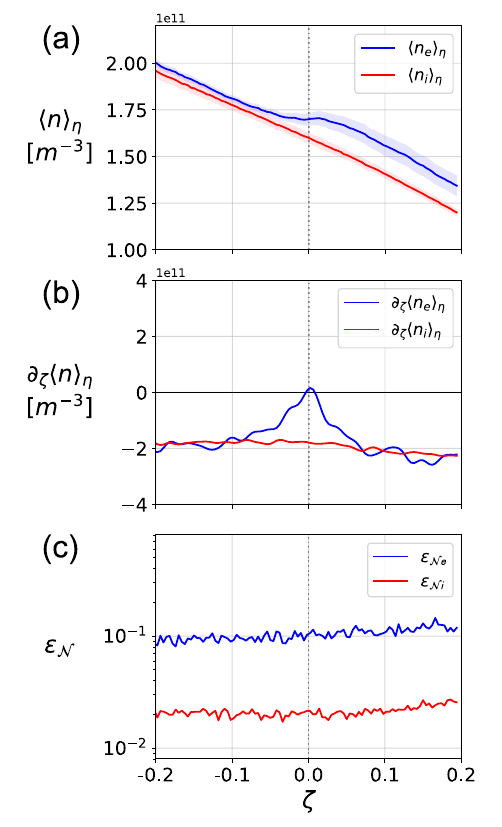}
    \caption[
    $\eta$-averaged PIC profiles and the closure diagnostic in the ROI
    ]{
    $\eta$-averaged PIC profiles in the ROI. Electrons are shown in blue and ions in red in all panels.
    (a) $\langle n_s \rangle_\eta(\zeta)$, with shaded bands indicating the standard deviation of $n_s$ along $\eta$ at each $\zeta$.
    (b) The corresponding gradient profiles $\partial_\zeta \langle n_s \rangle_\eta(\zeta)$.
    (c) The fluctuation diagnostic $\epsilon_\mathcal{N}(\zeta)$ defined in Eq.~\eqref{eq.4_epsilon}.
    The gray dotted line indicates the separatrix.
    }\label{fig.4_9}
\end{figure}

Figure~\ref{fig.4_9}(a) shows the $\eta$-averaged density profiles for the two species across the ROI, $\zeta \in [-0.2, +0.2]$.
The shaded bands represent the standard deviation of $n_s$ taken along $\eta$ at each $\zeta$.
Both densities decrease from the rod side toward the outer side, but their behaviors near the separatrix differ.
For electrons, $\langle n_e \rangle_\eta$ shows a visible inflection at $\zeta = 0$, with the slope flattening near the separatrix and steepening again on the outer side.
For ions, $\langle n_i \rangle_\eta$ remains close to linear across the entire ROI, with no inflection near the separatrix.
This is the density plateau formation for electrons, and its absence for ions, predicted by the reduced models for the respective magnetization regimes.
Due to this species-asymmetric plateau behavior, the electron-to-ion density ratio in Fig.~\ref{fig.4_9}(a) increases near the separatrix.
The self-consistent electric field acts to restore quasineutrality, but its magnitude in the present simulation is too weak for this restoration within the ROI, as discussed in Sec.~\ref{sec.4.4}.

The corresponding gradient profiles in Fig.~\ref{fig.4_9}(b) sharpen the species contrast.
For electrons, $\partial_\zeta \langle n_e \rangle_\eta$ shows a broad peak centered at $\zeta = 0$, with the gradient briefly crossing zero at the separatrix and returning to approximately $\SI{-2e11}{\per\cubic\meter}$ away from the separatrix.
This is the gradient signature of the density plateau: a localized relaxation of the gradient near the separatrix corresponds to the relative flattening of $\langle n_e \rangle_\eta$ visible in Fig.~\ref{fig.4_9}(a).
For ions, $\partial_\zeta \langle n_i \rangle_\eta$ remains approximately constant near $\SI{-2e11}{\per\cubic\meter}$ across the ROI, with no such feature.
The contrast between electrons and ions reflects their respective magnetization regimes in the ROI, established in Sec.~\ref{sec.4.4}.
The PIC simulation thus reproduces the central qualitative prediction of the reduced model analysis.

The shaded bands in Fig.~\ref{fig.4_9}(a) indicate that $n_s$ varies weakly along the field lines.
To rigorously test the fast parallel equilibration discussed in Sec.~\ref{sec.4.2}, we use the $\mathcal{N}$ fluctuation diagnostic, $\epsilon_\mathcal{N}(\zeta)$ (Eq.~\eqref{eq.4_epsilon}).
A small $\epsilon_\mathcal{N}$ indicates that the closure is satisfied.
Figure~\ref{fig.4_9}(c) shows $\epsilon_\mathcal{N}(\zeta)$ for both species.
Throughout the ROI, electrons satisfy $\epsilon_{\mathcal{N},e} \approx 0.1$, while ions satisfy $\epsilon_{\mathcal{N},i} \approx 0.02$.
Both values are small compared to unity, confirming that the closure underlying the reduced models holds across the ROI for both species; the reduced models may therefore be applied for direct quantitative comparison.

For the quantitative comparison of the density profiles, we evaluate the general reduced model, Eq.~\eqref{eq.4_reduced_model_gen}, and the diffusion-only reduced model, Eq.~\eqref{eq.4_reduced_model_diff}, using the PIC plasma quantities.
The general model retains the Boltzmann response of the density to $\phi$, and the diffusion-only model omits the Boltzmann factor and corresponds to the $|q_s\phi/T_s| \to 0$ limit of the general model.
The PIC quantities $\nu_s$ and $T_s$ remain approximately uniform across the ROI as noted in Sec.~\ref{sec.4.4}, while $B$ and $\phi$ are the spatially varying quantities of interest.
The conserved fluxes, $G_s$, are evaluated to be $G_e \approx \SI{5.01e10}{\per\meter\per\second}$ and $G_i \approx \SI{1.15e11}{\per\meter\per\second}$, and are verified by the particle absorption data at the outer walls; the difference in the conserved flux values is attributed to the asymmetric loss rates of the two species, explained in Sec.~\ref{sec.4.4}.

\begin{figure}
    \centering
    \includegraphics[width=15cm]{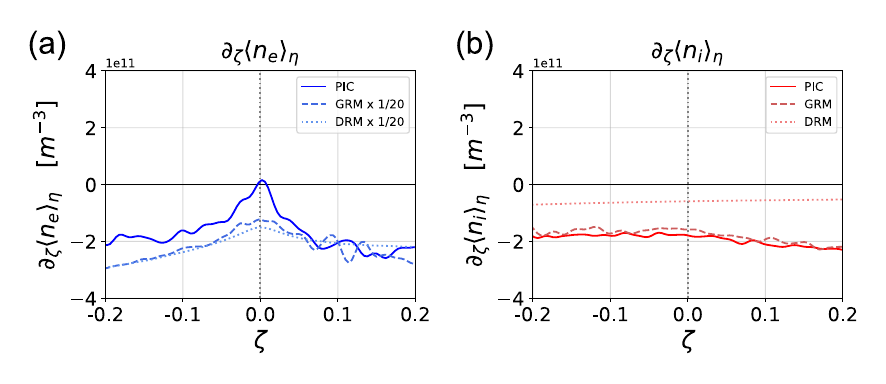}
    \caption[
    Comparison of the density gradient between PIC and the reduced models
    ]{
    Comparison of $\partial_\zeta \langle n_s \rangle_\eta(\zeta)$ between PIC (solid), the general reduced model (GRM, dashed), and the diffusion-only reduced model (DRM, dotted), for (a) electrons (blue) and (b) ions (red), with different shades indicating the reduced models.
    The GRM and DRM curves in panel (a) are shown rescaled by $1/20$ to permit shape comparison on common axes; panel (b) curves are unscaled reduced model results.
    The gray dotted line indicates the separatrix in both panels.
    }\label{fig.4_10}
\end{figure}

Figure~\ref{fig.4_10} shows the comparison of $\partial_\zeta \langle n_s \rangle_\eta$ between PIC and the two reduced models for each species.
For ions in Fig.~\ref{fig.4_10}(b), the general model agrees well with the PIC measurement, both curves remaining near $\SI{-2e11}{\per\cubic\meter}$ across the ROI with no localized feature.
The diffusion-only model, by contrast, predicts a gradient magnitude approximately a factor of three smaller than the PIC measurement.
This contrast indicates that the electric drift contribution in the ROI is not negligible for ions, so the Boltzmann response retained in the general model is essential to reproduce the ion gradient, and the diffusion-only limit is not sufficient.
For electrons in Fig.~\ref{fig.4_10}(a), the general and diffusion-only models are nearly indistinguishable from each other, which reflects the negligibility of the electric drift contribution for electrons, and both reproduce the shape of the PIC gradient including the peak at $\zeta = 0$ and the modest asymmetry between the gradient recovery at $\zeta = -0.2$ and $+0.2$.
However, both models also predict a gradient magnitude approximately a factor of $20$ larger than that measured by PIC, and the reduced model curves in panel (a) are therefore divided by $20$ to permit shape comparison on common axes.
This discrepancy is the important feature of Fig.~\ref{fig.4_10} and is interpreted in the following discussion.
We note two additional points about Fig.~\ref{fig.4_10}(a).
First, the general model curve shows small fluctuations absent from the diffusion-only model, because residual noise in the potential data enters the general model pointwise through the Boltzmann factor while the diffusion-only model neglects the potential.
Second, the gradient peak at $\zeta = 0$ appears less sharp than in the model demonstrations of Sec.~\ref{sec.4.3}; this smoothing reflects the finite resolution of the PIC diagnostics, as well as the finite-Larmor-radius (FLR) effect discussed below.

The magnitude discrepancy for electrons in Fig.~\ref{fig.4_10}(a) reflects the limitations of both the PIC simulation and the drift-diffusion framework under the present conditions.
Two distinct effects contribute: the enhancement of the effective collision frequency by numerical collisions in the PIC simulation, and the limited validity of the local fluid description of the drift-diffusion framework in the vicinity of the magnetic null.

The first effect is the enhancement of the effective collision frequency by numerical collisions, a well known feature of two-dimensional PIC quantified for unmagnetized configurations by Jubin~\textit{et al.}\cite{Jubin2024}
This effect was noted briefly in connection with the low electron temperature in Sec.~\ref{sec.4.4}, and we examine it here in the context of the collision frequency.
Numerical collisions originate from the discreteness of macroparticle interactions and act on the electron velocity distribution in a manner analogous to Coulomb collisions.
A directly applicable quantitative estimate is not available for the present configuration, since the existing analyses are derived for either an unmagnetized plasma or one with the magnetic field perpendicular to the simulation plane, whereas the applied TWM field is purely in-plane in this work.
The implication of the numerical effect is nonetheless clear: in the strongly magnetized regime, the predicted gradient magnitude scales as $|dn_e / d\zeta| \propto 1/(D_e \langle f_e \rangle_\eta) \propto 1/\nu_e$, and a numerically enhanced $\nu_e$ implies that the models evaluated with the MCC-based $\nu_e$ overpredict the gradient, in the direction of the observed discrepancy.
Numerical collisions therefore degrade the quantitative agreement between the models and PIC, contributing to the order-of-magnitude gap; however, the gap cannot be inverted to estimate the effective $\nu_e$, because the contribution of numerical collisions is compounded with that of the FLR effect addressed in the following.

The second effect is the limited validity of the local fluid description of the drift-diffusion framework in the vicinity of the magnetic null.
The standard FLR correction\cite{Chen2016} is not directly applicable here because $\rho_L = m v_\perp / (|q|B)$ diverges as $B \to 0$ at the X-point, so the relevant orbit excursion must be characterized differently.
The collisionless single particle analysis of Chapter~\ref{chap.3} provides such a characterization in the TWM system.
Applying the analysis method of Eq.~\eqref{eq.3_zeta_maxmin} in Chapter~\ref{chap.3}, we find that the orbit excursion bound for a thermal electron in the present simulation is $\Delta\zeta_\text{bound} = \zeta_\text{max} - \zeta_\text{min} \approx 0.02$, which is non-negligible relative to the characteristic width of the $f$ enhancement near the X-point (Fig.~\ref{fig.4_2}).
Consequently, the cross-field flux is no longer set by the local value of $f(\zeta,\eta)$ but by an average over the orbit excursion, which smooths the sharp $f \to 1$ enhancement that the reduced models capture locally at the X-point.
In addition, within its accessible range, a particle mixes freely along its trajectory, so the density at a given $\zeta$ receives contributions from all particles whose orbits reach that field line.
These FLR effects cannot be captured by the local drift-diffusion description.
The compounded influence of numerical collisions and orbit excursion mixing is presumably responsible for the magnitude discrepancy between the PIC results and the reduced models.
The quantitative agreement for ion profiles is consistent with this picture: numerical collisions affect ions far less than electrons due to the much larger ion mass, and the magnetization reduction factor of weakly magnetized ions has no sharp feature near the X-point to be smoothed by orbit excursion, so neither limitation is expected to be significant for the ion case.

\begin{figure}
    \centering
    \includegraphics[width=12cm]{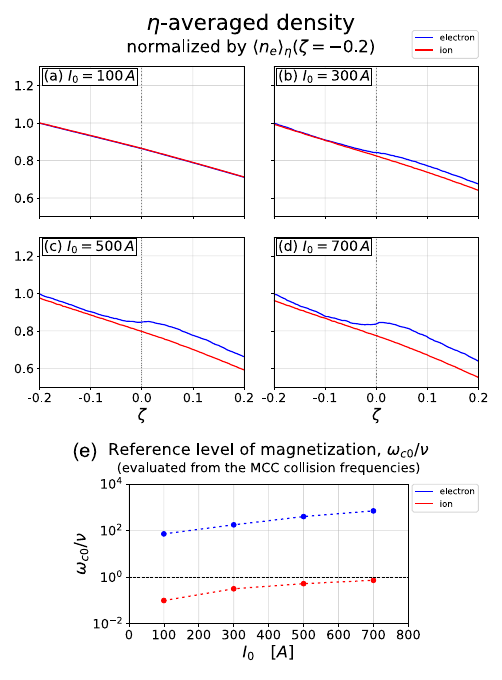}
    \caption[
    Wire current scan of the $\eta$-averaged density profiles
    ]{
    $\eta$-averaged density profiles in the ROI for wire currents (a) $I_0 = \SI{100}{A}$, (b) $\SI{300}{A}$, (c) $\SI{500}{A}$, and (d) $\SI{700}{A}$, with all other simulation parameters identical to Table~\ref{tab.4_simparams}.
    Electrons are shown in blue and ions in red in all panels.
    The gray dotted line indicates the separatrix.
    Each profile is normalized by $\langle n_e \rangle_\eta(\zeta = -0.2)$ of the corresponding case to allow a direct comparison of the profile shapes.
    (e) The reference levels of magnetization, $\omega_{ce0}/\nu_e$ and $\omega_{ci0}/\nu_i$, of the four cases, evaluated from the MCC collision frequencies; the black dashed line marks $\omega_{c0}/\nu = 1$.
    }\label{fig.4_11}
\end{figure}

To further examine the magnetization dependence of the plateau, we repeat the simulation with wire currents $I_0 = \SI{100}{A}$, $\SI{300}{A}$, and $\SI{700}{A}$, keeping all other parameters of Table~\ref{tab.4_simparams} fixed, and compare them with the $\SI{500}{A}$ case analyzed above.
Since $B_0 \propto I_0$, this directly scans the reference magnetization: from the MCC collision frequencies, the electron magnetization ratios are evaluated to be $\omega_{ce0}/\nu_e \approx 73$, $178$, $403$, and $707$ for $I_0 = \SI{100}{A}$, $\SI{300}{A}$, $\SI{500}{A}$, and $\SI{700}{A}$, respectively, while the ions remain weakly magnetized throughout, with $\omega_{ci0}/\nu_i \approx 0.10$, $0.32$, $0.53$, and $0.74$ (Fig.~\ref{fig.4_11}(e)).
Figures~\ref{fig.4_11}(a--d) show the $\eta$-averaged density profiles for the four cases, normalized by the electron density at the inner boundary, $\langle n_e \rangle_\eta(\zeta = -0.2)$, of each case.
At $\SI{100}{A}$, the electron profile is linear with no plateau, identical to the ion profile; at $\SI{300}{A}$, a slight flattening of the electron profile emerges near the separatrix; and at $\SI{500}{A}$ and $\SI{700}{A}$, a more pronounced electron density plateau is present.
The ion profile remains linear at every current.
The scan thus reproduces the predicted dependence of the plateau on the magnetization as a trend, though with a caveat: within the idealized reduced models, $\omega_{ce0}/\nu_e \approx 73$ for $\SI{100}{A}$ already lies in the strongly magnetized regime, where the predicted plateau shape is nearly saturated (Fig.~\ref{fig.4_4}), so an electron density plateau would be expected at all four currents.
Its absence at $\SI{100}{A}$ is attributed to the combined effect of the two limitations identified above.
In particular, the orbit excursion bound grows as the field weakens, which further limits the validity of the local fluid description near the X-point and smooths out the plateau at lower currents.
The density plateau therefore becomes visible only at a higher magnetization than the idealized reduced models predict.

The numerical and theoretical limitations, however, do not alter the central qualitative result of this work.
The density plateau near the separatrix is a geometric consequence of the magnetization reduction factor structure in the TWM, arising from the field magnitude weakening near the X-point and being communicated along the closed field lines by parallel equilibration.
This mechanism is captured by the analytic reduced models and qualitatively reproduced by the self-consistent kinetic PIC simulation.
The comparison thus supports the central claim that the magnetic X-point produces a density plateau on the closed field lines near the separatrix in strongly magnetized low temperature plasmas confined in a TWM field, with the plateau strengthening as the magnetization increases.



\section{Summary}\label{sec.4.6}

In this chapter, we have investigated how a true magnetic null shapes classical cross-field transport and the steady-state density profile of a magnetized low temperature plasma confined in the TWM configuration.
We formulated a drift-diffusion model in the field-aligned TWM coordinate system, in which the cross-field flux is reduced by the magnetization reduction factor $f = 1/(1+\beta^2)$.

By $\eta$-averaging the model along the closed field lines and applying a fast parallel equilibration closure, we derived one-dimensional reduced models for the cross-field density profile.
These models predict density plateau formation near the separatrix in the strongly magnetized regime: as the field magnitude weakens toward the X-point, the reduction factor rises sharply toward unity, so the conserved cross-field flux is carried by a relaxed density gradient, and fast parallel equilibration communicates this relaxation along the closed field line to flatten the density profile around the separatrix.

Self-consistent two-dimensional electrostatic PIC simulations with the EDIPIC-2D code reproduce this prediction, and the parallel equilibration closure is confirmed directly through the fluctuation diagnostic.
The strongly magnetized electrons develop the density plateau while the weakly magnetized ions do not, and a wire current scan confirms that the plateau strengthens with the magnetization.
Relative to the PIC results, the reduced models overpredict the electron density gradient by about an order of magnitude under the present conditions, which we attribute to numerical collisions in the PIC method and the FLR effect near the null.
These limitations, however, do not alter the central qualitative result of this chapter.
The density plateau near the separatrix is therefore a geometric consequence of the magnetization reduction factor structure in the TWM, and these results establish a classical baseline for cross-field transport near a magnetic null.


\chapter{Discussion and Conclusion}\label{chap.5}

This work has investigated plasma dynamics near a true magnetic null, using the guide-field-free two-wire model (TWM) as a clean theoretical setting in which the role of the X-point is isolated from the many other effects present in real devices.
The investigation proceeded in two complementary regimes, the collisionless dynamics of individual charged particles and the collisional transport of a magnetized low temperature plasma, both developed within the TWM field and its coordinate system.

Chapter~\ref{chap.2} established the analytic and geometric foundation for the work.
The TWM field, generated by two parallel current-carrying wires, features a magnetic X-point and a separatrix, and is expressed by a closed analytic form that allows an exact treatment of the geometry.
From this field, a field-aligned conformal curvilinear coordinate system $(\zeta, \eta)$ was constructed, in which $\zeta$ labels the closed field lines, and $\eta$ parametrizes position along them.
The coordinate system and the corresponding differential operators provided the framework used throughout the subsequent chapters.

Chapter~\ref{chap.3} investigated the collisionless dynamics of single charged particles, in which the motion is governed solely by the externally applied TWM field.
A Lagrangian analysis identified two constants of motion: the total kinetic energy $W_\text{total}$, conserved because the magnetic field does no work, and the base field line value $\zeta_\text{base}$, derived from the axial canonical momentum and conserved by the translational symmetry along the wires.
Near the X-point, where the field magnitude weakens and its spatial gradient becomes large, the magnetic moment is no longer conserved and undergoes abrupt shifts.
These shifts are chaotic and sensitive to initial conditions, yet their long time statistics are determined entirely by the invariant pair.
When a shift produces a Larmor radius large enough to cross the separatrix, the particle migrates to the corresponding branch of its base field line on the opposite side of the null.
An effective potential analysis yielded a threshold energy for migration, $\norm{W}_\text{total} > \zeta_\text{base}^2$.
The time between consecutive migrations was found to be exponentially distributed, so that migration is a memoryless process characterized by a migration confinement time that depends only on the two invariants.
An empirical expression for this confinement time was formulated from simulations spanning a wide range of invariant pairs, and a tokamak-like demonstration showed that migration preferentially removes high energy particles from inside the separatrix.

Chapter~\ref{chap.4} turned to the collisional regime, in which a magnetized low temperature plasma fills the TWM configuration, and collisions and self-consistent electric fields shape the dynamics.
A drift-diffusion model (DD model) was formulated in the TWM coordinate system, in which the cross-field flux is reduced by the magnetization reduction factor $f = 1/(1+\beta^2)$, with $\beta = \omega_c/\nu$.
By $\eta$-averaging the model along the closed field lines under a fast parallel equilibration closure, one-dimensional reduced models for the cross-field density profile were derived.
These models predict the formation of a density plateau near the separatrix in the strongly magnetized regime: the density gradient required to carry the conserved cross-field flux through a closed field line is inversely proportional to $\langle f \rangle_\eta$, which is sharply enhanced on the field lines passing close to the null, so that the null acts as a localized leak in the magnetic confinement and fast parallel equilibration communicates its effect along the entire field line, flattening the density profile around the separatrix.
Self-consistent two-dimensional electrostatic particle-in-cell (PIC) simulations with the EDIPIC-2D code reproduced this prediction with a clear species contrast, a density plateau for the strongly magnetized electrons and no pronounced feature for the weakly magnetized ions, and a wire current scan confirmed that the plateau strengthens with the magnetization.
The closure underlying the reduced models was validated directly through the fluctuation diagnostic $\epsilon_\mathcal{N}$.
Quantitatively, the reduced models overpredict the electron density gradient by about an order of magnitude under the present conditions, a discrepancy attributed to the compounded effects of numerical collisions in the PIC method and the finite-Larmor-radius (FLR) effect near the null.

Although the two regimes were analyzed with different methods, single particle orbit integration in Chapter~\ref{chap.3} and fluid drift-diffusion with kinetic simulation in Chapter~\ref{chap.4}, the resulting phenomena are two manifestations of the same underlying geometric fact.
At a true magnetic null, the field magnitude vanishes and the local magnetization breaks down.
In the collisionless regime, this breakdown causes the loss of magnetic moment conservation, which drives the chaotic magnetic moment shifts and the cross-separatrix migration.
In the collisional regime, it causes the rise of the magnetization reduction factor $f$ toward unity, which leads to density plateau formation.
The migration and the density plateau are therefore the kinetic and fluid signatures of one common cause, the local weak magnetization at the X-point.

This connection between the two regimes also appears quantitatively, in the comparison between the reduced models and the PIC results near the null.
The standard FLR correction of fluid theory cannot be applied there, because the Larmor radius diverges as the field magnitude vanishes at the X-point.
The collisionless analysis of Chapter~\ref{chap.3} instead provides a way to estimate the relevant orbit excursion, which indicates where the local fluid description becomes unreliable.
More broadly, the two regimes bracket the same physical system from opposite limits of collisionality, with a real plasma near a true null lying between them.

Taken together, the two regimes establish a classical baseline for plasma behavior near a true magnetic null, isolated from the guide field, curvature, turbulence, and boundary effects of real devices, against which each such effect can be reintroduced as a controlled extension.

The most direct application is to the MAXIMUS device,\cite{Lim2020} which realizes a near-exact guide-field-free TWM field.
Measured profiles can be mapped into the TWM coordinates for field-aligned analysis, the closure can be tested on the measurements through $\epsilon_\mathcal{N}$, and the reduced models provide the classical profile predictions against which the measurements can be compared.
In particular, the mechanism identified in Chapter~\ref{chap.4} offers a candidate classical explanation for the locally flattened pressure profile and the transport barrier observed near the separatrix in that device.

With a superposed guide field, as in tokamak divertor regions, the field magnitude retains a finite floor and the true null is replaced by a field minimum, so that the enhancement of the reduction factor near the X-point is capped at a finite value set by the guide field magnetization.
A weakened form of the density plateau can still be expected near the separatrix, and the guide-field-free result presented here provides an upper bound on this X-point transport enhancement.
The collisionless migration is likewise modified, since a guide field makes the Larmor radius finite everywhere and would slow the cross-separatrix transport, while collisions, absent in Chapter~\ref{chap.3}, would enhance transport.
The guide-field-free TWM therefore brackets the tokamak-relevant case from the strong-X-point side in both regimes.

A truly guide-field-free null is realized in the closed field line region of the field-reversed configuration (FRC),\cite{Steinhauer2011} whose purely poloidal field vanishes at the X-points on the geometric axis.
The closed field lines near the FRC separatrix pass close to these nulls, producing a field structure similar to the one analyzed here, so that the same mechanism would relax the cross-field density gradient near the FRC separatrix in collisional regimes.
Because FRC plasmas are often weakly collisional, with large particle orbits near the nulls, the collisional result serves as a classical baseline to which the orbit-excursion corrections of the collisionless analysis in Chapter~\ref{chap.3} must be added, a combination that the connection discussed above makes natural.
The betatron character of the migrating orbits found in Chapter~\ref{chap.3} resembles the particle orbits near FRC nulls.\cite{Steinhauer2011}

Beyond these magnetic confinement devices, the invariant quantities and magnetic moment statistics of Chapter~\ref{chap.3} provide a kinetic baseline for single particle orbit dynamics near a static magnetic null, the idealized limit of a reconnection site prior to the addition of its driving electric field.

Several concrete directions follow from this work.
First, a quantitative comparison with MAXIMUS measurements, using the field-aligned profile mapping described above, would test both the migration predictions and the reduced models against experiment and quantify any anomalous contribution to X-point transport.
Second, the extension of the analysis to a finite guide field would connect the present limiting case to fusion-relevant geometry, and would determine whether the weakened plateau and the slowed migration anticipated above survive with a superposed field.
Third, the quantitative discrepancy identified in Chapter~\ref{chap.4} motivates further work on both of its sources: a characterization of numerical collisions in PIC simulations with an in-plane magnetic field, extending the analysis of Jubin~\textit{et al.},\cite{Jubin2024} and a kinetic transport description near the null that builds directly on the orbit analysis of Chapter~\ref{chap.3}.
Finally, higher density simulations, closer to physically typical low temperature plasmas, would sharpen the quantitative comparison.

The investigation of plasma dynamics in the TWM field has revealed the essential physics of a true magnetic null.
By treating the collisionless and collisional regimes within a single analytic geometry, this work has shown that the local breakdown of magnetization at the X-point shapes both the orbits of individual particles and the steady-state profile of a confined plasma, and that the cross-separatrix migration and the density plateau share a common geometric origin.
The result is a concrete classical baseline for transport near a magnetic X-point, together with a transferable framework that applies to the broader class of configurations containing magnetic nulls, from linear multipole devices to field-reversed configurations and the divertor regions of magnetic confinement experiments.
The complicating effects of real devices can each be reintroduced as a controlled extension of this baseline, against which their separate contributions to X-point transport can be assessed.


\appendix


\chapter{Two-Wire Model Electric Field}\label{chap.A}

In the main text, the TWM field was treated as purely magnetic, generated by the currents in the two wires.
This appendix provides a supplementary analysis of the TWM electric field, which is produced when the wires carry a uniform linear charge density.
We derive the electric field, show that it is perpendicular to the TWM magnetic field, and that the coordinate $\zeta$, which is proportional to the TWM magnetic vector potential, is also proportional to the electric scalar potential.
We then present several quantities that follow from the TWM electric and magnetic fields together.


\section{TWM electric field}\label{sec.A.1}

For brevity, we refer to the TWM magnetic and electric fields as the TWM B field and the TWM E field, respectively.
The TWM B field is generated by the currents $I_0$ in the two wires, as derived in Chapter~\ref{chap.2}.
The TWM E field is the electrostatic field produced when the same two wires each carry a uniform linear charge density $\lambda_0$.

By analogy with the characteristic magnetic vector potential $A_0 = I_0\mu_0/2\pi$, we define the characteristic electric scalar potential
\begin{equation}\label{eq.appx_phi0}
    \phi_0 = \frac{\lambda_0}{2\pi\epsilon_0},
\end{equation}
and the characteristic electric field magnitude $E_0 = \phi_0/\ell_0$, in parallel with $B_0 = A_0/\ell_0$.

A single infinite line charge at the origin produces the radial field $\bvec{E} = (\phi_0/r)\,\uvec{r}$, the electrostatic analogue of the azimuthal field $\bvec{B} = (A_0/r)\,\uvecg{\theta}$ of a single current-carrying wire.
Superposing the fields of the two line charges at $(x,y) = (0,\pm\ell_0)$ gives the TWM E field
\begin{equation}\label{eq.appx_E_field}
    \bvec{E}(x,y)
    = E_x \uvec{x} + E_y \uvec{y}
    = \phi_0 \left[ \left( \frac{x}{r_+^2} + \frac{x}{r_-^2} \right) \uvec{x}
    + \left( \frac{y_+}{r_+^2} + \frac{y_-}{r_-^2} \right) \uvec{y} \right].
\end{equation}
Comparing these components with those of the TWM B field in Eq.~\eqref{eq.2_vecB} yields
\begin{equation}\label{eq.appx_E_dual}
    \bvec{E} = \frac{\phi_0}{A_0} \left( B_y \uvec{x} - B_x \uvec{y} \right).
\end{equation}
The map $(B_x, B_y) \mapsto (B_y, -B_x)$ is a clockwise rotation by $90^\circ$, so the TWM E field is everywhere perpendicular (in the plane) to the TWM B field and rescaled by the constant factor $\phi_0/A_0$.

In the TWM coordinate system, since $\bvec{B} \parallel \uvecg{\eta}$ and a clockwise rotation by $90^\circ$ of $\uvecg{\eta}$ gives $\uvecg{\zeta}$ (from $\uvecg{\zeta} \times \uvecg{\eta} = \uvec{z}$), the two fields are
\begin{subequations}\label{eq.appx_EB_zetaeta}
\begin{align}
    \bvec{B} & = \uvec{z}\times (A_0\nabla\zeta) = A_0 \nabla\eta \parallel \uvecg{\eta},
    \\
    \bvec{E} & = (\phi_0\nabla\eta)\times\uvec{z} = \phi_0 \nabla\zeta  \parallel \uvecg{\zeta},
\end{align}
\end{subequations}
and in normalized form,
\begin{subequations}\label{eq.appx_EB_zetaeta_norm}
\begin{align}
    \norm{\bvec{B}} & = \uvec{z}\times \norm{\nabla}\zeta = \norm{\nabla}\eta \parallel \uvecg{\eta},
    \\
    \norm{\bvec{E}} & = \norm{\nabla}\eta\times\uvec{z} = \norm{\nabla}\zeta \parallel \uvecg{\zeta}.
\end{align}
\end{subequations}
Thus, the TWM E field is directed across the TWM B field lines, and the two fields have the same normalized magnitude,
\begin{equation}\label{eq.appx_Etil}
    \norm{E} = \frac{E}{E_0} = \frac{2\norm{r}}{s^2} = \norm{B}.
\end{equation}

We now turn to the scalar potential.
The TWM E field is electrostatic, so it derives from a scalar potential, $\bvec{E} = -\nabla\phi$.
Comparing with $\bvec{E} = \phi_0\nabla\zeta$ from Eq.~\eqref{eq.appx_EB_zetaeta} gives
\begin{equation}\label{eq.appx_phi}
    \phi = -\phi_0 \zeta.
\end{equation}
This is the exact electrostatic counterpart of the magnetic relation $A_z = -A_0\zeta$ derived in Sec.~\ref{sec.2.2}.
Both potentials are therefore proportional to the $\zeta$-coordinate:
\begin{equation}\label{eq.appx_dual_potential}
    \zeta = -\frac{\phi}{\phi_0} = -\frac{A_z}{A_0},
\end{equation}
so $-\zeta$ is simultaneously the normalized electric scalar potential and the normalized magnetic vector potential.
The contours of $\zeta$, the Cassini ovals, are therefore simultaneously equipotentials of the TWM E field and field lines of the TWM B field.


\section{Electromagnetic energy and drift}\label{sec.A.2}

This section presents the $\bvec{E}\times\bvec{B}$ drift, the Poynting vector, and the electromagnetic energy density that arise when both the TWM E and B fields are present.

The $\bvec{E}\times\bvec{B}$ drift is spatially uniform despite the strong spatial variation of both fields.
Because the two fields are everywhere perpendicular with the fixed ratio $E/B = \phi_0/A_0$, the drift velocity is
\begin{equation}\label{eq.appx_ExB}
    \bvec{v}_{\bvec{E}\times\bvec{B}}
    = \frac{\bvec{E}\times\bvec{B}}{B^2}
    = \frac{\phi_0}{A_0}\,\uvec{z}
    = \frac{1}{\epsilon_0\mu_0}\frac{\lambda_0}{I_0}\,\uvec{z}
    = c^2\,\frac{\lambda_0}{I_0}\,\uvec{z}.
\end{equation}
For $E > cB$, the formula would imply a superluminal drift, but the explanation of this is beyond the scope of this work.

The Poynting vector, giving the electromagnetic energy flux, is
\begin{equation}\label{eq.appx_poynting}
    \bvec{S} = \frac{1}{\mu_0}\bvec{E}\times\bvec{B}
    = \frac{EB}{\mu_0}\,\uvec{z}
    = \frac{\lambda_0 I_0}{\pi^2\epsilon_0}\,\frac{r^2}{r_+^2 r_-^2}\,\uvec{z},
\end{equation}
directed along the wire axis and carrying the product $\lambda_0 I_0$ of the two sources.

The electromagnetic energy density is
\begin{equation}\label{eq.appx_energy}
    u = \frac{1}{2}\left( \epsilon_0 E^2 + \frac{B^2}{\mu_0} \right)
    = \frac{1}{2\pi^2}\,\frac{r^2}{r_+^2 r_-^2}\left( \frac{\lambda_0^2}{\epsilon_0} + \mu_0 I_0^2 \right).
\end{equation}
The shared spatial factor $r^2/(r_+^2 r_-^2)$ reflects the common normalized magnitude $\norm{E} = \norm{B} = 2\norm{r}/s^2$, while the electric and magnetic contributions split into the two terms.


\chapter{Supplementary Analyses for the Reduced Drift-Diffusion Model}\label{chap.B}

This appendix presents two supplementary analyses that support the reduced DD models developed in Chapter~\ref{chap.4}.
Appendix~\ref{sec.B.1} derives an analytic validity criterion for the fast parallel equilibration closure underlying the reduced models, expressed as an order-of-magnitude inequality between the cross-field gradient length of the $\eta$-averaged quantity and a critical length set by the magnetization reduction factor and the Boltzmann factor.
Appendix~\ref{sec.B.2} derives the ambipolar fluxes in the TWM coordinate system by imposing quasineutrality and ambipolarity on the two-species drift-diffusion fluxes, and shows that the resulting ambipolar reduced model takes the same form as the single-species diffusion-only model with $f$ and $D$ replaced by ambipolar counterparts $f_\text{A}$ and $D_\text{A}$.


\section{Validity criterion for the closure}\label{sec.B.1}

This appendix derives an analytic validity criterion for the closure assumed in the reduced models of Sec.~\ref{sec.4.2}.
The criterion is obtained by estimating the magnitude of the fluctuation $\mathcal{N}'$ via order-of-magnitude scaling, and takes the form of a scalar inequality comparing the cross-field gradient length of $\mathcal{N}_0 = \langle \mathcal{N} \rangle_\eta$ to a critical length scale set by the magnetization reduction factor and the Boltzmann factor.

Since $\eta$ is periodic on $(-\pi,+\pi]$, any scalar quantity $a(\zeta,\eta)$ can be written as a Fourier series in $\eta$, equivalent to the average-fluctuation decomposition of Eq.~\eqref{eq.4_eta_avg_fluc}:
\begin{equation}\label{eq.appx_gen_decomp}
\begin{aligned}
    a(\zeta,\eta)
    & = a_0(\zeta) + \sum_{m \ne 0} a_m(\zeta) e^{im\eta}
    \\
    & = \langle a \rangle_\eta(\zeta) + a'(\zeta,\eta),
\end{aligned}
\end{equation}
where $a_0 = \langle a \rangle_\eta$ is the zeroth Fourier mode and $a'$ is the sum of non-zero modes.

We make use of the commutation identities $\langle \partial_\zeta a \rangle_\eta = \partial_\zeta \langle a \rangle_\eta$ and $\langle \partial_\eta a \rangle_\eta = 0$ from Eq.~\eqref{eq.4_eta_avg_partial_zetaeta}.
For any two scalar quantities, the average-fluctuation decomposition also gives the identities:
\begin{equation}\label{eq.appx_eta_avg_id_1}
    \langle ab \rangle_\eta = a_0 b_0 + \langle a'b' \rangle_\eta,
\end{equation}
\begin{equation}\label{eq.appx_eta_avg_id_2}
    (ab)' = ab - \langle ab \rangle_\eta = a' b_0 + a_0 b' + (a' b')'.
\end{equation}

For compactness, abbreviate the Boltzmann factor as
\begin{equation}\label{eq.appx_w_def}
    w(\zeta,\eta) = \exp\left(-\frac{q\phi(\zeta,\eta)}{T}\right),
\end{equation}
so that $fw = f e^{-q\phi/T}$.
Applying the decomposition Eq.~\eqref{eq.appx_gen_decomp} to $\mathcal{N}$ and $fw$,
\begin{equation}\label{eq.appx_N_decomp}
\begin{aligned}
    \mathcal{N}(\zeta,\eta)
    & = \mathcal{N}_0(\zeta) + \sum_{m \ne 0} \mathcal{N}_m(\zeta) e^{im\eta}
    \\
    & = \langle \mathcal{N} \rangle_\eta(\zeta) + \mathcal{N}'(\zeta,\eta),
\end{aligned}
\end{equation}
\begin{equation}\label{eq.appx_fw_decomp}
\begin{aligned}
    (fw)(\zeta,\eta)
    & = \langle fw \rangle_\eta(\zeta) + \sum_{m \ne 0} (fw)_m(\zeta) e^{im\eta}
    \\
    & = \langle fw \rangle_\eta(\zeta) + (fw)'(\zeta,\eta).
\end{aligned}
\end{equation}
For clarity in this appendix, we use $\mathcal{N}_0$ and $\langle fw \rangle_\eta$ for the $\eta$-averages (the zeroth Fourier modes).

Using $w$, the cross-field and parallel fluxes in Eqs.~\eqref{eq.4_Gamma_zeta} and~\eqref{eq.4_Gamma_eta} are
\begin{subequations}\label{eq.appx_flux_N_form}
\begin{align}
    \label{eq.appx_flux_N_form_zeta}
    h\Gamma_\zeta & = - D fw \, \partial_\zeta \mathcal{N},
    \\
    \label{eq.appx_flux_N_form_eta}
    h\Gamma_\eta & = - D w \, \partial_\eta \mathcal{N}.
\end{align}
\end{subequations}

Taking the $\eta$-average of Eq.~\eqref{eq.appx_flux_N_form_zeta} and using $G = \langle h\Gamma_\zeta \rangle_\eta$ from Eq.~\eqref{eq.4_G},
\begin{equation}\label{eq.appx_mean_balance}
    \langle fw \, \partial_\zeta\mathcal{N}\rangle_\eta = -\frac{G}{D},
\end{equation}
and applying Eq.~\eqref{eq.appx_eta_avg_id_1} to the left-hand side,
\begin{equation}\label{eq.appx_mean_balance_decomp}
    \langle fw \rangle_\eta \partial_\zeta \mathcal{N}_0
    + \langle (fw)' \, \partial_\zeta\mathcal{N}' \rangle_\eta
    = -\frac{G}{D}.
\end{equation}
The general reduced model, Eq.~\eqref{eq.4_reduced_model_gen}, is recovered by neglecting the correlation term here.
This neglect is implied by the closure of Sec.~\ref{sec.4.2}, which in Fourier form states that all non-zero modes are negligible, $|\mathcal{N}_0| \gg |\mathcal{N}'|$, as in Eq.~\eqref{eq.4_closure_form}:
\begin{equation}\label{eq.appx_closure_gen}
    \mathcal{N}(\zeta,\eta) \approx \mathcal{N}_0(\zeta).
\end{equation}
To quantify when the closure is valid, we estimate the magnitude of $\mathcal{N}'$, which requires an equation for the fluctuation.

Substituting Eqs.~\eqref{eq.appx_flux_N_form} into the steady-state continuity equation, Eq.~\eqref{eq.4_continuity}, gives
\begin{equation}\label{eq.appx_continuity_N}
    \partial_\zeta(fw \, \partial_\zeta\mathcal{N}) + \partial_\eta(w \, \partial_\eta\mathcal{N}) = 0.
\end{equation}
Taking the $\eta$-average,
\begin{equation}\label{eq.appx_continuity_N_etaavg}
    \partial_\zeta \langle fw \, \partial_\zeta\mathcal{N} \rangle_\eta = 0,
\end{equation}
which is the differential form of Eq.~\eqref{eq.appx_mean_balance}.
Both equations are exact under the assumptions of Sec.~\ref{sec.4.1}, and no closure has been applied.
Subtracting Eq.~\eqref{eq.appx_continuity_N_etaavg} from Eq.~\eqref{eq.appx_continuity_N} gives the fluctuation equation,
\begin{equation}\label{eq.appx_fluc_exact}
    \partial_\zeta\left[(fw)' \, \partial_\zeta\mathcal{N}_0\right]
    + \partial_\zeta\left[(fw \, \partial_\zeta\mathcal{N}')'\right]
    + \partial_\eta(w \, \partial_\eta\mathcal{N}') = 0,
\end{equation}
where the identities introduced above have been used.
This is a linear partial differential equation for $\mathcal{N}'$ with $\mathcal{N}_0$ entering as a forcing term.

We seek an order-of-magnitude bound instead of an exact solution of Eq.~\eqref{eq.appx_fluc_exact}.
Two simplifications are applied.
First, the cross-field term of $\mathcal{N}'$ carries an additional factor $f$, so it is subdominant in the strongly magnetized regime and is dropped at leading order.
Second, $w$ is taken to be weakly varying along the field line, $|w'| \ll \langle w \rangle_\eta$, so $w$ is replaced by $\langle w \rangle_\eta$ in the parallel operator.
Equation~\eqref{eq.appx_fluc_exact} then reduces to
\begin{equation}\label{eq.appx_fluc_lin}
    \partial_\zeta \left[(fw)' \, \partial_\zeta\mathcal{N}_0\right] + \langle w \rangle_\eta \partial_\eta^{2}\mathcal{N}' \approx 0.
\end{equation}
Expanding $\mathcal{N}'$ and $(fw)'$ in Fourier modes via Eqs.~\eqref{eq.appx_N_decomp} and~\eqref{eq.appx_fw_decomp} and substituting in Eq.~\eqref{eq.appx_fluc_lin}, the modes decouple and each non-zero mode satisfies
\begin{equation}\label{eq.appx_mode_balance}
    \partial_\zeta \left[(fw)_m \, \partial_\zeta\mathcal{N}_0\right] - m^2 \langle w \rangle_\eta \mathcal{N}_m = 0,
\end{equation}
or equivalently
\begin{equation}\label{eq.appx_mode_amplitude}
    \mathcal{N}_m(\zeta) = \frac{1}{m^2 \langle w \rangle_\eta} \partial_\zeta \left[(fw)_m \, \partial_\zeta\mathcal{N}_0\right].
\end{equation}
The factor $m^{-2}$ suppresses higher modes, so we focus on $|m|=1$, which dominates $\mathcal{N}'$.

The natural scale on which $\mathcal{N}_0$ varies in $\zeta$ is its cross-field gradient length,
\begin{equation}\label{eq.appx_delta_zeta_N0}
    \Delta\zeta_{\mathcal{N}_0} = \left| \partial_\zeta \ln \mathcal{N}_0 \right|^{-1}.
\end{equation}
Note that this is a coordinate length, not a physical length.
Assuming $(fw)_1$ varies slower in $\zeta$ than $\mathcal{N}_0$, the forcing in Eq.~\eqref{eq.appx_mode_amplitude} is bounded by
\begin{equation}\label{eq.appx_forcing_OOM}
    \left| \partial_\zeta \left[(fw)_1 \, \partial_\zeta\mathcal{N}_0\right] \right|
    \lesssim \frac{|(fw)_1| \, |\mathcal{N}_0|}{\Delta\zeta_{\mathcal{N}_0}^{2}}.
\end{equation}
Bounding the Fourier amplitude of $fw$ by its $\eta$-average,
\begin{equation}\label{eq.appx_weight_OOM}
    |(fw)_1| \lesssim \langle fw \rangle_\eta,
\end{equation}
and combining with Eq.~\eqref{eq.appx_mode_amplitude} gives
\begin{equation}\label{eq.appx_amp_bound}
    |\mathcal{N}_1| \lesssim
    \frac{\langle fw \rangle_\eta}{\langle w \rangle_\eta}
    \frac{|\mathcal{N}_0|}{\Delta\zeta_{\mathcal{N}_0}^{2}}.
\end{equation}

The closure requires $|\mathcal{N}_0| \gg |\mathcal{N}'|$.
Since the $|m|=1$ mode dominates $\mathcal{N}'$, this becomes $|\mathcal{N}_0| \gg |\mathcal{N}_1|$.
Substituting the bound from Eq.~\eqref{eq.appx_amp_bound} and rearranging,
\begin{equation}\label{eq.appx_validity_gen_crit}
    \Delta\zeta_{\mathcal{N}_0} \gg \Delta\zeta_{\text{c,g}}
    = \sqrt{\frac{\langle fw \rangle_\eta}{\langle w \rangle_\eta}}
    = \sqrt{\frac{\langle f e^{-q\phi/T} \rangle_\eta}{\langle e^{-q\phi/T} \rangle_\eta}},
\end{equation}
where $\Delta\zeta_{\text{c,g}}$ is the critical cross-field length.
This is the validity criterion for the general reduced model:
under the order-of-magnitude assumptions used in the derivation, the criterion is a sufficient condition for the closure of the general reduced model.
When the gradient length of $\mathcal{N}_0$ is much longer than the critical length set by the magnetization reduction factor $f$ and the Boltzmann factor $e^{-q\phi/T}$, the estimated fluctuation amplitude $|\mathcal{N}_1|$ is small compared to $|\mathcal{N}_0|$, and the closure is expected to hold.

In the diffusion-only limit, $\phi$ is spatially uniform, so $w$ is constant and $\langle fw \rangle_\eta / \langle w \rangle_\eta = \langle f \rangle_\eta$.
Since $\mathcal{N}_0$ differs from $n_0 = \langle n \rangle_\eta$ only by the constant $w$, $\Delta\zeta_{\mathcal{N}_0} = \Delta\zeta_{n_0}$ with $\Delta\zeta_{n_0} = |\partial_\zeta \ln n_0|^{-1}$.
The criterion reduces to
\begin{equation}\label{eq.appx_validity_diff}
    \Delta\zeta_{n_0} \gg \Delta\zeta_{\text{c,d}}
    = \sqrt{\langle f \rangle_\eta},
\end{equation}
which depends only on the magnetization reduction factor.
Physically, this criterion gives a quantitative form of the closure condition.
The nonuniform $f$ converts the cross-field gradient into a fluctuating cross-field flux that varies along the field line, and parallel diffusion works to flatten that variation.
Since the whole field line contributes to this balance, $f$ enters the criterion through its $\eta$-average; in the general case, the average is additionally weighted by the Boltzmann factor.

The criterion is order-of-magnitude in nature and rests on three working assumptions used in the derivation: (i) the cross-field term involving $\mathcal{N}'$ in Eq.~\eqref{eq.appx_fluc_exact} is subdominant in the strongly magnetized regime, (ii) the Boltzmann factor varies weakly along the field line, $|w'| \ll \langle w \rangle_\eta$, and (iii) $(fw)_1$ varies more slowly in $\zeta$ than $\mathcal{N}_0$.
Under these assumptions, the criterion is sufficient but not necessary; the bound $|(fw)_1| \lesssim \langle fw \rangle_\eta$ used in the derivation can be loose, so closure may still hold when the criterion is violated, for instance when $fw$ is weakly $\eta$-varying.
The approximations are also least justified near the X-point, where $f$ approaches unity and the strongly magnetized ordering breaks down locally, and where the sharp $\zeta$-variation of $(fw)_1$ violates assumption~(iii) within a narrow layer around the separatrix.
Numerical factors are not tracked, and the criterion should be read as a guide to the regime of validity rather than a sharp threshold.
For sharper testing of the closure, the direct fluctuation diagnostic $\epsilon_\mathcal{N}$ introduced in Sec.~\ref{sec.4.3} is recommended, which can be applied to simulation or experimental data.


\section{Ambipolar flux derivation}\label{sec.B.2}

The DD model in Sec.~\ref{sec.4.1} was formulated for a single plasma species under a prescribed electric field.
In a quasineutral ambipolar plasma, however, the electron and ion densities are approximately equal in the bulk, while the electric field is determined self-consistently so that electron and ion fluxes match, which makes the system current-free.
This appendix derives the ambipolar fluxes in the TWM coordinate system by applying quasineutrality and ambipolarity to the two-species drift-diffusion fluxes.
It shows that the ambipolar fluxes reduce to an effective diffusive form, with the magnetization reduction factor and diffusion coefficient replaced by ambipolar counterparts $f_{\text{A}}$ and $D_{\text{A}}$.

We start by writing the in-plane fluxes from Eqs.~\eqref{eq.4_flux_comp} with $\mathbf{E}$ in place of $-\nabla\phi$, and with species subscript $s$,
\begin{subequations}\label{eq.appx_species_flux}
\begin{align}
    \Gamma_{\zeta,s} & = - f_s D_s (\nabla n_s)_\zeta + \sigma_s f_s \mu_s n_s E_\zeta,
    \\
    \Gamma_{\eta,s} & = - D_s (\nabla n_s)_\eta + \sigma_s \mu_s n_s E_\eta.
\end{align}
\end{subequations}
We consider electrons ($s = e$, $\sigma_e = -1$) and singly-charged ions ($s = i$, $\sigma_i = +1$), and assume $T_s$ and $\nu_s$ to be spatially uniform, as in Sec.~\ref{sec.4.1}.
The derivation rests on two assumptions for the plasma: quasineutrality, $n_e = n_i = n$, and ambipolarity, $\bvecg{\Gamma}_e = \bvecg{\Gamma}_i = \bvecg{\Gamma}$.

Imposing these two conditions in Eqs.~\eqref{eq.appx_species_flux} yields the ambipolar electric field,
\begin{subequations}\label{eq.appx_E_amb}
\begin{align}
    E_\zeta & = \frac{f_i D_i - f_e D_e}{f_e \mu_e + f_i \mu_i} \frac{(\nabla n)_\zeta}{n},
    \\
    E_\eta & = \frac{D_i - D_e}{\mu_e + \mu_i} \frac{(\nabla n)_\eta}{n}.
\end{align}
\end{subequations}
Substituting Eqs.~\eqref{eq.appx_E_amb} back into Eqs.~\eqref{eq.appx_species_flux} gives
\begin{subequations}\label{eq.appx_amb_flux_pre}
\begin{align}
    \Gamma_\zeta & = - D_{\zeta\text{A}} (\nabla n)_\zeta,
    \\
    \Gamma_\eta & = - D_{\eta\text{A}} (\nabla n)_\eta,
\end{align}
\end{subequations}
with
\begin{subequations}\label{eq.appx_D_amb}
\begin{align}
    D_{\zeta\text{A}} & = f_e f_i \, \frac{D_e \mu_i + D_i \mu_e}{f_e \mu_e + f_i \mu_i},
    \\
    D_{\eta\text{A}} & = \frac{D_e \mu_i + D_i \mu_e}{\mu_e + \mu_i}.
\end{align}
\end{subequations}

The parallel coefficient $D_{\eta\text{A}}$ is the standard ambipolar diffusion coefficient,\cite{Chen2016} unaffected by magnetization.
We therefore simply write
\begin{equation}\label{eq.appx_DA}
    D_\text{A} = D_{\eta\text{A}},
\end{equation}
and introduce a reduction factor $f_\text{A}$ such that $D_{\zeta\text{A}} = f_\text{A} D_\text{A}$:
\begin{equation}\label{eq.appx_fA}
\begin{aligned}
    f_\text{A} = \frac{D_{\zeta\text{A}}}{D_\text{A}}
    & = \frac{f_e f_i (\mu_e + \mu_i)}{f_e \mu_e + f_i \mu_i}
    \\
    & = \frac{f_e f_i (\mu_e + \mu_i)}{f_e f_i (\mu_e + \mu_i)(1 + \beta_e \beta_i)}
    \\
    & = \frac{1}{1 + \beta_e \beta_i},
\end{aligned}
\end{equation}
where we used $\beta_s = \mu_s B$ and $f_s = 1/(1+\beta_s^2)$.
We refer to $f_\text{A}$ as the ambipolar magnetization reduction factor.
The ambipolar fluxes then take the same form as the single-species expressions, Eqs.~\eqref{eq.4_flux_comp_zeta} and~\eqref{eq.4_flux_comp_eta},
\begin{subequations}\label{eq.appx_amb_flux_final}
\begin{align}
    \Gamma_\zeta & = - f_\text{A} D_\text{A} (\nabla n)_\zeta,
    \\
    \Gamma_\eta & = - D_\text{A} (\nabla n)_\eta.
\end{align}
\end{subequations}

The ambipolar reduced model can be obtained by applying the substitutions $f \to f_\text{A}$ and $D \to D_\text{A}$ to the diffusion-only reduced model in Sec.~\ref{sec.4.2}.
Since $\phi$ has been eliminated in this setting, the drift contribution does not appear separately, and the resulting model takes the same diffusion-only form as Eq.~\eqref{eq.4_reduced_model_diff},
\begin{equation}\label{eq.appx_reduced_model_amb}
    \frac{dn}{d\zeta} = -\frac{G}{D_\text{A}} \frac{1}{\langle f_\text{A}\rangle_\eta},
\end{equation}
where $G = \langle h \Gamma_\zeta \rangle_\eta$ is the conserved cross-field flux, the same for both species, and $n = n_e = n_i$ is the common density.
The ambipolar reduced model can be used to predict the density gradient profile for a quasineutral ambipolar plasma.
Here, the ambipolar reduction factor $f_\text{A}$ is governed by the product $\beta_e \beta_i$, rather than $\beta_s^2$ as in the single-species case.
Since $\beta_e \gg \beta_i$ in typical low temperature plasmas, $f_\text{A}$ suppresses cross-field transport much less than $f_e$ does.


\bibliographystyle{unsrt}
\bibliography{twm_bib}

\end{document}